\documentclass[preprint,authoryear,12pt]{elsarticle}
\makeatletter
\def\ps@pprintTitle{%
 \let\@oddhead\@empty
 \let\@evenhead\@empty
 \def\@oddfoot{\centerline{\thepage}}%
 \let\@evenfoot\@oddfoot}
\makeatother
\usepackage{wrapfig, amsmath, amssymb, multirow, subfig, epsfig, graphicx}
\usepackage{mathtools, paralist, caption, tabularx, wasysym, hyperref, natbib, comment}
\usepackage{float}
\usepackage[table]{xcolor}
\usepackage[margin=1in]{geometry}
\usepackage{algorithm, algorithmic, verbatim, amsthm}
\usepackage[english]{babel}
\usepackage[utf8]{inputenc}
\usepackage{color, colortbl}
\definecolor{Gray}{gray}{0.9}
\definecolor{LightCyan}{rgb}{0.88,1,1}
\usepackage{booktabs}

\graphicspath{{figures-final/}}
\newcommand{\omitme}[1]{}

\newcommand{\m}[1]{$\matr{#1}$}

\newcommand{\X}{\ensuremath{\mathbf{X}}} \newcommand{\T}{\ensuremath{\mathbf{T}}} \newcommand{\bT}{\ensuremath{\hat{\mathbf{T}}}}   
\newcommand{\tX}{\ensuremath{\tilde{\mathbf{X}}}} \newcommand{\h}{\ensuremath{\mathrm{h}}} 
\newcommand{\U}{\ensuremath{\mathbf{U}}} \newcommand{\W}{\ensuremath{\mathbf{W}}} \newcommand{\w}{\ensuremath{\mathbf{w}}}
\renewcommand{\S}{\ensuremath{\mathbf{S}}} 
\newcommand{\s}{\ensuremath{\mathbf{s}}}

\theoremstyle{definition}

\newcolumntype{C}{>{\centering\arraybackslash}p{1em}}

\definecolor{darkgreen}{RGB}{0,0,205}

\newcommand{\rnote}[1]{{\begin{color}{red} #1 \end{color}}}
\DeclarePairedDelimiter{\ceil}{\lceil}{\rceil}

\begin{document}

\begin{frontmatter}
\date{}

\title{Accelerating Permutation Testing in Voxel-wise Analysis \\ through Subspace Tracking: A new plugin for SnPM}
\author[a]{Felipe Gutierrez-Barragan}
\author[a]{Vamsi K. Ithapu}
\author[a]{Chris Hinrichs}
\author[e]{Camille Maumet}
\author[c]{\\Sterling C. Johnson}
\author[e]{Thomas E. Nichols}
\author[b,a]{Vikas Singh}
\author{\\for the Alzheimer's Disease Neuroimaging Initiative \footnote{Data used in preparation of this article was obtained from the Alzheimer’s Disease
Neuroimaging Initiative (ADNI) database (adni.loni.ucla.edu). 
As such, the investigators within the ADNI contributed to the design and implementation of ADNI and/or provided data but did not participate in analysis or writing of this report. 
A complete listing of ADNI investigators can be found at: \url{http://adni.loni.ucla.edu/wp-content/uploads/how\_to\_apply/ADNI\_Acknowledgement\_List.pdf}}
}

\address[a]{Department of Computer Sciences, University of Wisconsin-Madison}
\address[b]{Department of Biostatistics \& Med. Informatics, University of Wisconsin-Madison}
\address[c]{Department of Medicine, University of Wisconsin-Madison and William S. Middleton Veteran's Hospital}
\address[e]{Department of Statistics, The University of Warwick\\
{\tt \href{http://felipegb94.github.io/RapidPT/}{http://felipegb94.github.io/RapidPT/}
}

}
\begin{abstract} Permutation testing is a non-parametric method for obtaining the max null distribution used to compute corrected $p$-values that provide strong control of false positives. In neuroimaging, however, the computational burden of running such an algorithm can be significant. We find that by viewing the permutation testing procedure as the construction of a very large permutation testing matrix, $\T$, one can exploit structural properties derived from the data and the test statistics  to reduce the runtime under certain conditions. In particular, we see that $\T$ is low-rank plus a low-variance residual. This makes $\T$ a good candidate for low-rank matrix completion, where only a very small number of entries of $\T$ ($\sim0.35\%$ of all entries in our experiments) have to be computed to obtain a good estimate. Based on this observation, we present RapidPT, an algorithm that efficiently recovers the max null distribution commonly obtained through regular permutation testing in voxel-wise analysis. We present an extensive validation on a synthetic dataset and four varying sized datasets against two baselines: Statistical NonParametric Mapping (SnPM13) and a standard permutation testing implementation (referred as NaivePT). We find that RapidPT achieves its best runtime performance on medium sized datasets ($50 \leq n \leq 200$), with speedups of 1.5x - 38x (vs. SnPM13) and 20x-1000x (vs. NaivePT). For larger datasets ($n \geq 200$) RapidPT outperforms NaivePT (6x - 200x) on all datasets, and provides large speedups over SnPM13 when more than 10000 permutations (2x - 15x) are needed. The implementation is a standalone toolbox and also integrated within SnPM13, able to leverage multi-core architectures when available.
 \end{abstract}
\begin{keyword} Voxel-wise analysis, Hypothesis test, Permutation test, Matrix completion \end{keyword}
\end{frontmatter}

{\footnotesize \noindent Corresponding Author(s): Felipe Gutierrez Barragan (fgutierrez3@wisc.edu)}


\section{Introduction} \label{sec:intro}

Nonparametric voxel-wise analysis, e.g., via permutation tests, are widely used in the brain image analysis literature. 
Permutation tests are often utilized to control the family-wise error rate (FWER) in voxel-wise hypothesis testing.
As opposed to parametric hypothesis testing schemes \cite{Friston94, Worsley92, Worsley96}, nonparametric permutation tests \cite{Holmes96,Nichols02} 
can provide exact control of false positives while making minimal assumptions on the data. 
Further, despite the additional computational cost, permutation tests have been widely adopted in image analysis \cite{Arndt96, Halber97, Holmes96, Nichols02, Nichols03} via implementations in broadly used software libraries available in the community \cite{SnPM,FSL,Winkler14}.

{\em Running time aspects of Permutation Testing.} Despite the varied advantages of permutation tests, 
there is a general consensus that the computational cost of performing permutation tests in neuroimaging analysis can often be quite high. 
As we will describe in more detail shortly, high dimensional imaging datasets essentially mean that {\em for each permutation}, 
hundreds of thousands of test statistics need to be computed. 
Further, as imaging technologies continue to get better (leading to higher resolution imaging data) and the concurrent 
slowdown in the predicted increase of processor speeds (Moore's law), 
it is reasonable to assume that the associated runtime will continue to be a problem in the short to medium term. 
To alleviate these runtime costs, ideas that rely on code optimization and parallel computing have been explored \cite{Eklund11,Eklund12,Eklund13}. 
These are interesting strategies but any hardware-based approach will be limited by the amount of resources at hand.
Clearly, significant gains may be possible if {\em more efficient schemes} that exploit the underlying structure of the imaging data were available. It seems likely  
that such algorithms can better exploit the resources (e.g., cloud or compute cluster) one has available as part of a study
and may also gain from hardware/code improvements that are being reported in the literature. 

Data acquired in many scientific studies, especially imaging and genomic data, are highly structured. Individual genes and/or individual voxels share a great 
deal of commonality with other genes and voxels. 
It seems reasonable that such correlation can be exploited towards better (or more efficient) statistical algorithms. 
For example, in genomics, \cite{Cheverud01} and \cite{Li05} used correlations in the data to estimate the effective number of 
independent tests in a genome sequence to appropriately threshold the test statistics. 
Also motivated by bioinformatics problems, \cite{knijnenburg09} approached 
the question of estimating the tail of the distribution of permutation values via an approximation by a generalized Pareto distribution (using 
fewer permutations).
In the context of more general statistical analysis, the authors in \cite{subramanian05} 
proposed Gene Set Enrichment Analysis (GSEA) 
which exploits the underlying structure among the genes, to analyze gene-sets (e.g., where sets were obtained from biological pathways) 
instead of individual genes. If the genes within a gene-set have similar expression pattern, one may see improvements in statistical power. 
This idea of exploiting the ``structure'' towards efficiency (broadly construed) was more rigorously studied in \cite{efron07} and a nice non-parametric Bayesian perspective was offered in \cite{dahl07}. 
Within neuroimaging, a similar intuition drives Random Field theory based analysis \cite{taylor08}, albeit the objective there is to obtain a less conservative 
correction, rather than computational efficiency. 
Recently, motivated by neuroimaging applications and computational issues, \cite{Gaonkar13} 
derived an analytical approximation of statistical significance maps to reduce the computational burden imposed by permutation tests commonly used 
to identify which brain regions contribute to a Support Vector Machines (SVM) model. 
In summary, exploiting the structure of the data to obtain alternative efficient solutions is not new, but we find that in the context 
of permutation testing on {\em imaging data}, there is a great deal of voxel-to-voxel correlations that if leveraged properly can, in principle, yield 
interesting new algorithms.

For permutation testing tasks in neuroimaging in particular, several groups have recently investigated ideas to make use of the underlying structure of the data to accelerate the procedure.
In a preliminary conference paper (\cite{Hinrichs13}), we introduced the notion of exploiting correlations in neuroimaging data via the underlying low-rank structure of the permutation testing procedure. A few years later, \cite{Winkler16} presented the first thorough evaluation of the accuracy and runtime gains of six approaches that leverage the problem structure to accelerate permutation testing for neuroimage analysis. Among these approaches \cite{Winkler16} presented an algorithm which relied on some of the ideas introduced by \cite{Hinrichs13} to accelerate permutation testing through low-rank matrix completion (LRMC). Overall, algorithms that exploit the underlying structure of permutation testing in neuroimaging have provided substantial computational speedups. 

\subsection{Main Idea and Contributions}\label{sec:mainideas} 

The starting point of our formulation is to analyze the entire permutation testing procedure via numerical linear algebra. 
In particular, the object of interest is the permutation testing matrix, $\T$. 
Each row of $\T$ corresponds to the voxel-wise statistics, and each column is a specific permutation of the labels of the data. 
This perspective is not commonly used because a typical permutation test in neuroimaging rarely instantiates or operates on this matrix of statistics.
Apart from the fact that $\T$, in neuroimaging, contains millions of entries, the reason for not working {\it directly} with it is because the goal is to derive the maximum null distribution. 
The central aspect of this work is to {\it exploit} the structure in $\T$ -- the spatial correlation across different voxel-statistics.
Such correlations are not atypical because the statistics are computed from anatomically correlated regions in the brain. 
Even far apart voxel neighbourhoods are inherently correlated because of the underlying biological structures. 
This idea drives the proposed novel permutation testing procedure. 
We describe the contributions of this paper based on the observation that the permutation testing matrix is filled with related entries. 

\begin{itemize}

\item {\bf Theoretical Guarantees.}
The basic premise of this paper is that {\it 
permutation testing in high-dimensions (especially, imaging data) is extremely redundant}. We show how we can model $\T$ as a low-rank plus a low-variance residual. We provide two theorems that support this claim and demonstrate its practical implications. Our first result justifies this modeling assumption and several of the components involved in recovering $\T$. The second result shows that the error in the {\it global} maximum null distribution 
obtained from the estimate of $\T$ is quite small.


\item {\bf A novel, fast and robust, multiple-hypothesis testing procedure. }
Building upon the theoretical development, we propose a fast and accurate algorithm for permutation testing involving high-dimensional imaging data. 
The algorithm achieves state of the art runtime performance by estimating (or recovering) the statistics in $\T$ rather than ``explicitly'' computing them.
We refer to the algorithm as {\it RapidPT}, and we show that compared to existing state-of-the-art libraries for non-parametric testing, 
the proposed model achieves approximately $20\times$ speed up over existing procedures. 
We further identify regimes where the speed up is even higher.
RapidPT also is able to leverage serial and parallel computing environments seamlessly.
\item {\bf A plugin in SnPM (with stand-alone libraries). }
Given the importance and the wide usage of permutation testing in neuroimaging (and other studies involving high-dimensional and multimodal data), 
we introduce a heavily tested implementation of RapidPT integrated as a plugin within the current development version of SnPM --- a widely used non-parametric testing toolbox. 
Users can invoke RapidPT directly from within the SnPM graphical user interface and benefit from SnPM's familiar pre-processing and post-processing capabilities. This final contribution, without a separate installation, brings the performance promised by the theorems to the neuroimaging community.
Our documentation \cite{RapidPT} gives an overview of how to use RapidPT within SnPM.
\end{itemize}

  Although the present work shares some of the goals and motivation of \cite{Winkler16} -- specifically, utilizing the algebraic structure of $\T$ -- there are substantial technical differences in the present approach, which we outline further below.
  First, unlike \cite{Winkler16}, we directly study permutation testing for images at a more fundamental level and seek to characterize 
  mathematical properties of relabeling (i.e., permutation) procedures operating on high-dimensional imaging data. This is 
  different from assessing whether the underlying operations of classical statistical testing procedures can be reformulated (based on the correlations) 
  to reduce computational burden as in \cite{Winkler16}.
  Second, by exploiting celebrated technical results in random matrix theory, we provide theoretical guarantees for estimation and 
  recovery of $\T$. Few such results were known. 
  Note that empirically, our machinery avoids a large majority of the operations performed in \cite{Winkler16}.
  Third, some speed-up strategies proposed in \cite{Winkler16} can be considered as special cases of our proposed algorithm --- 
  interestingly, if we were to increase the number 
  `actual' operations performed by RapidPT (from 
  $\approx 1\%$, suggested by our experiments, to $50\%$), the computational workload begins approaching what is described in 
  \cite{Winkler16}.

\section{Permutation Testing in Neuroimaging} \label{sec:permtest} 

In this section, we first introduce some notations and basic concepts. Then, we give additional background on permutation testing for hypothesis testing in neuroimaging
to motivate our formulation.
Matrices and vectors will be denoted by bold upper-case and lower-case letters respectively, and scalars will be represented using non-bold letters.
For a matrix $\X$, $\X[i,:]$ denotes the $i^{th}$ row and $\X[i,j]$ denotes the element in $i^{th}$ row and $j^{th}$ column.

Permutation testing is a nonparametric procedure for estimating the empirical distribution of the global null \cite{Edgington69a,Edgington69b,Edgington07}.
For a two-sample univariate statistical test, a permutation test computes an {\it unbiased} estimate of the null distribution 
of the relevant univariate statistic (e.g., $t$ or $\chi^2$).
Although univariate null distributions are, in general, well characterized, 
the sample maximum of the voxel-wise statistics usually does not have an analytical form due to
strong correlations across voxels, as discussed in Section \ref{sec:intro}.
Permutation testing is appropriate in this high-dimensional setting because it is non-parametric 
and does not impose any restriction on the correlations across the voxel-wise statistics.
Indeed, when the test corresponds to group differences between samples based on a stratification variable, 
under the null hypothesis $\mathcal{H}_0$, the grouping labels given to the samples are artificial, i.e., they
correspond to the set of all possible {\it relabellings} of the samples. 
In neuroimaging studies, typically the groups correspond to the presense or absence of an underlying disease condition (e.g., controls and diseased).
Whenever $\mathcal{H}_0$ is true, the data sample representing a healthy subject is, roughly speaking, `similar' to a diseased subject. 
Under this setting, in principle, 
{\it interchanging} the labels of the two instances will have no effect on the distribution of the resulting voxel-wise statistics across all the dimensions (or covariates or features). 
So, if we randomly permute the labels of the data instances from both groups, 
under $\mathcal{H}_0$ the recomputed sets of voxel-wise statistics correspond to the same global null distribution.
We denote the number of such relabellings (or permutations) by $L$. 
The histogram of all $L$ maximum statistics i.e., the maximum across all voxel-wise statistics for a given permutation, 
is the empirical estimate of the {\it exact} maximum null distribution under $\mathcal{H}_0$.
When given the true/real labeling, 
to test for significant group-wise differences,
one can simply compute the fraction of the max null that is more extreme than the maximum statistic computed across all voxels for this real labeling.


\noindent \paragraph{\bf The case for strong null}
Observe that when testing {\em multiple} sets of hypotheses there are two different types of control for the Type $1$ error rate (FWER): weak and strong control \cite{Hochberg87}. 
A test is referred to as weak control for FWER whenever the Type $1$ error rate is controlled only when all the hypotheses involved 
(here, the number of voxels being tested) are true. That is, $\mathcal{H}_0$ is true for (all) voxels.
On the other hand, a test provides strong control for FWER whenever Type $1$ error rate is controlled under any combination/proportion of the true hypotheses.
It is known that the procedure described above (i.e., using the max null distribution calculated across all voxel-wise statistics) 
provides strong control of FWER \cite{Holmes96}.
This is easy to verify because the maximum of all voxel-wise statistics is used to compute the corrected $p$-value, and so, the exact proportion of which hypotheses are
true does not matter.
Further, testing based on strong control will classify non-activated voxels as activated with a probability upper bounded by $\alpha$, 
i.e., it has localizing power \cite{Holmes96}, a desirable property in neuroimaging studies in particular.
For the remainder of this paper, we will focus on such strong control and restrict our presentation to the case of group difference analysis for two groups. 


\subsection{\textsc{NaivePT}: The exhaustive Permutation Testing procedure} \label{sec:naivept}

\begin{figure}[H]
\centerline{%
\includegraphics[width=0.8\textwidth]{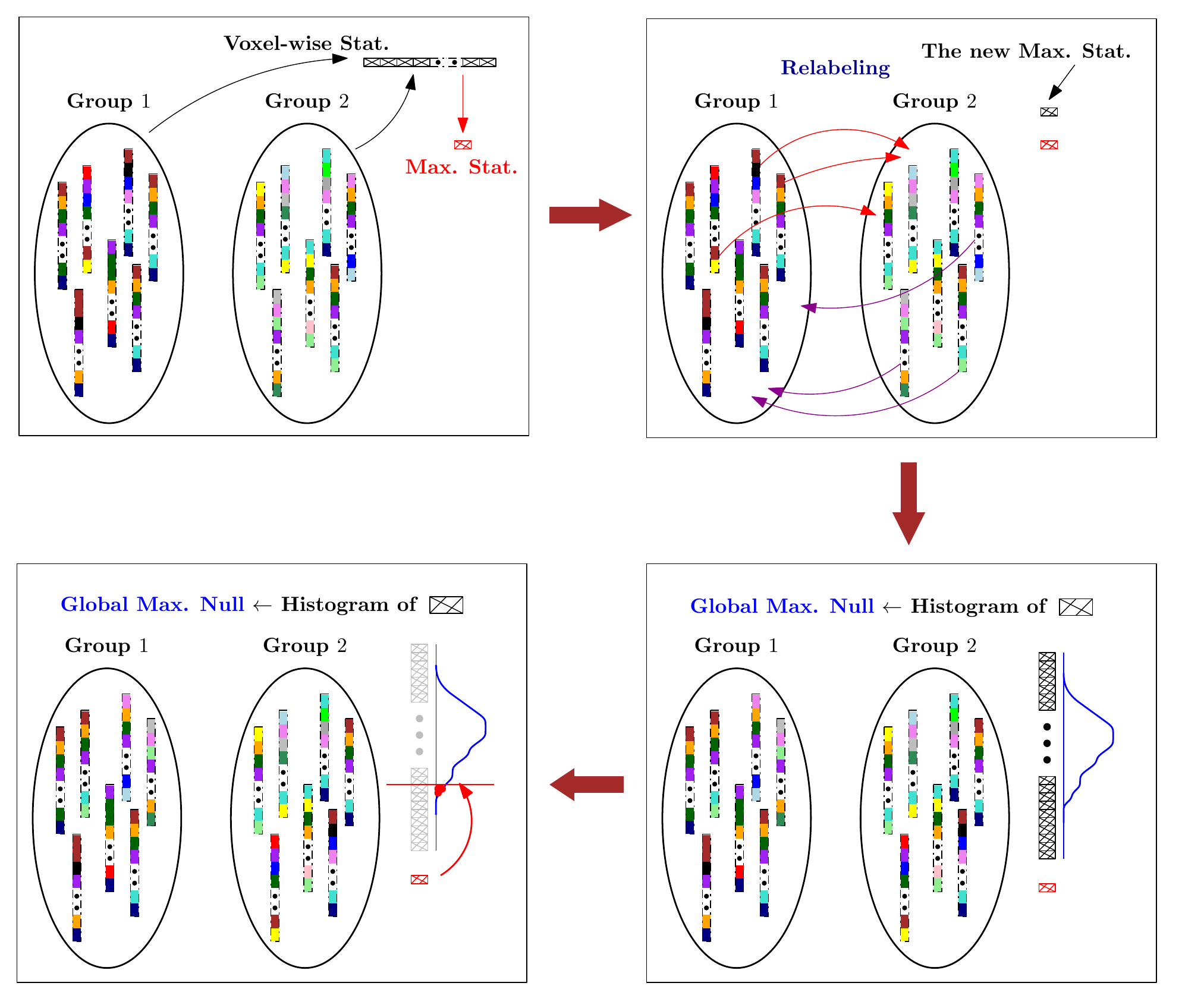}%
}%
\caption{Flow diagram of the permutation testing procedure described in algorithm \ref{alg:naivept}. Group 1 and Group 2 correspond to $\X^1$ and $\X^2$, each new max stat corresponds to each $m_i$, and the global max null corresponds to $h^L$. }
\label{fig:naiveptFlow}
\end{figure}

Figure \ref{fig:naiveptFlow} and algorithm \ref{alg:naivept} illustrate the permutation testing procedure. 
Given the data instances from two groups, $\X^1 \in \mathbb{R}^{v \times n_1}$ and $\X^2 \in \mathbb{R}^{v \times n_2}$, 
where $n_1$ and $n_2$ denote the number of subjects from each group respectively.  
Also, $v$ denotes the number of voxels in the brain image.
Let $n = n_1 + n_2$ and $\X = [\X^1; \X^2]$ give the (row-wise) stacked data matrix ($\X \in \mathbb{R}^{v \times n}$).
Note that a permutation of the columns of $\X$ corresponds to a group relabelling.
The $v$ distinct voxel-wise statistics are then computed for $L$ such permutations, and used to construct the
{\it permutation testing matrix} $\T \in \mathbb{R}^{v \times L}$. 
The empirical estimate of the max null is simply the histogram of the maximum of each of the columns of $\T$ -- denoted by $\h^L$. 
Algorithm \ref{alg:naivept} is occasionally referred to as Monte Carlo permutation tests in the literature because of the random sampling
involved in generating the statistics. This standard description of a permutation test will be used in the following sections to describe our
proposed testing algorithm. 

\begin{algorithm}[h!]
\caption{\label{alg:naivept}~~\textsc{NaivePT} The exhaustive permutation testing procedure.} 
\begin{algorithmic}
\REQUIRE $\X^1$, $\X^2$, $L$
\ENSURE $\T$, $\h^L$
\STATE $\X = [\X^1; \X^2]$, $n = n_1 + n_2$
\STATE $m_1\ldots,m_L \leftarrow [\varnothing]$
\FOR {$i \in 1,\ldots,L$}
\STATE $j_1\ldots,j_{n} \sim \textsc{Permute}[1,n]$
\STATE $\tX^1 \leftarrow \X[:,j_1,\ldots,j_{n_1}]$, $\tX^2 \leftarrow \X[:,j_{n_1+1},\ldots,j_{n}]$
\STATE $\T[:,i] \leftarrow {\rm test}(\tX^1, \tX^2)$
\STATE $m_i \leftarrow \textsc{Max}(\T[:,i])$
\ENDFOR
\STATE $\h^L \leftarrow \textsc{Histogram}({m_1,\ldots,m_L})$
\end{algorithmic}
\end{algorithm}


\subsection{The situation when $L$ is large} \label{sec:large-r}

Evidently as the number of permutations ($L$) increases Algorithm \ref{alg:naivept} becomes extremely expensive. 
Nonetheless, performing a large number of permutations is essential in neuroimaging for various reasons.
We discuss these reasons in some detail next. 

\begin{compactenum}[\bfseries 1)] 

\item Random sampling methods draw many samples from near the mode(s) of the distribution of interest, but fewer samples from the tail. 
  To characterize the threshold for a small portion of the tail of a distribution, we must invariably 
  draw a very large number of samples just so that the estimate converges. 
  So, if we want an $\alpha = 0.01$ threshold from the max null distribution, we require many thousands of permutation samples --- a computationally expensive procedure in neuroimaging as previously discussed. 
%
\item Ideally we want to obtain a precise threshold for any $\alpha$, in particular small $\alpha$. However, the smallest possible $p$-value that can be obtained from the empirical null is $\frac{1}{L}$. Therefore, to calculate very low $p$-values (essential in many applications), $L$ must be very large.
%
\item A typical characteristic of brain imaging disorders, for instance in the {\em early} (e.g., preclinical) stages of 
  Alzheimer's disease (AD) and other forms of dementia, is that the disease signature is subtle --- 
  for instance, in AD, the deposition of Amyloid load estimated via positron emission tomographic (PET) images or 
  atrophy captured in longitudinal structural magnetic resonance images (MRI) image scans in the asymptomatic stage of the disease. 
  The signal is weak in this setting and 
  requires large sample size studies (i.e., large $n$) and a need for estimating the Type 1 error threshold with high confidence. 
  The necessity for high confidence tail estimation implies that we need to sample many more relabelings, 
  requiring a large number of permutations $L$.
%
\end{compactenum}

\section{A Convex Formulation to characterize the structure of $T$} \label{sec:fastperm}

It turns out that the computational burden of algorithm \ref{alg:naivept} can be mitigated by exploiting the structural properties of 
the permutation testing matrix $\T$. 
Our strategy uses ideas from LRMC, subspace tracking, and random matrix theory, to exploit the correlated structure of $\T$ and model it in an alternative form.
In this section, we first introduce LRMC, followed by the overview of the eigen-spectrum propoerites of $\T$, which then leads to our proposed model of $\T$. 

\subsection{Low-Rank Matrix Completion} \label{sec:matcomp}

Given only a small fraction of the entries in a matrix, the problem of low-rank matrix completion (LRMC) \cite{Candes10} seeks to {\it recover} the missing entries of the 
entire matrix.  
Clearly, with no assumption on the properties of the matrix, such a recovery is ill-posed. 
Instead, if we {\it assume} that the column space of the matrix is low-rank and the observed entries 
are randomly sampled, 
then the authors of \cite{Candes10} and others have shown that, 
with sufficiently small number of entries, one can recover the orthogonal basis of the row space as well as the expansion coefficients for each column --- 
that is, fully recover the missing entries of the matrix. 
Specifically, the number of entries required is roughly $r \log(d)$ where $r$ is the column space's rank and $d$ is the ambient dimension.
By placing an $\ell_1$-norm penalty on the eigenvalues of the recovered matrix, i.e., the nuclear norm \cite{fazel04,recht10}, 
one optimizes a convex relaxation of an (non-convex) objective function which explicitly minimizes the rank. 
Alternatively, we can specify a rank $r$ ahead of time, and estimate an orthogonal basis of that rank by following a gradient 
along the Grassmannian manifold \cite{Balzano10, He12}. 
The LRMC problem has received a great deal of attention in the period after the Netflix Prize \cite{bennett07}, and
numerous applications in machine learning and computer vision have been investigated \cite{ji10}. 
Details regarding existing algorithms and their analyses including strong recovery guarantees are available in \cite{candes09,recht11}. 

{\bf LRMC Formulation.} Let us consider a matrix $\T \in \mathbb{R}^{v \times L}$. Denote the set of randomly subsampled entries of this matrix 
as $\Omega$.
This means that we have access to $\T_{\Omega}$, and our recovery task corresponds to estimating $\T_{\Omega^C}$, where 
$\Omega^C$ corresponds to the complement of the set $\Omega$. 
Let us denote the estimate of the complete matrix be $\bT$. The completion problem can be written as the following optimization task,
\begin{align} \label{eq:completion1}
  \min_{\bT} & \quad \|  \T_{\Omega} - \bT_{\Omega}\|^2_{\rm Frob} \\
  \text{s.t.} & \quad \bT = \U\W \\
  \ & \quad \U \hspace{2mm}\text{is orthogonal}.
\end{align}
where $\U \in \mathbb{R}^{v \times r}$ is the low-rank basis of $\T$, i.e., the columns of $\U$ correspond to the orthogonal basis vectors of the column space of $\T$. 
Here, $\Omega$ gives the measured entries and $\W$ is the matrix of coefficients that lets us reconstruct $\bT$.

\subsection{Low-rank plus a long tail in $T$} \label{sec:longtail}

Most datasets encountered in the real world (and especially in neuroimaging) have a dominant low-rank component. 
While the data may not be {\em exactly} characterized by a low-rank basis, the residual will not significantly alter the 
eigen-spectrum of the sample covariance in general. 
Strong correlations nearly always imply a skewed eigen-spectrum, because as the the eigen-spectrum becomes flat, 
the resulting covariance matrix tends to become sparser (the ``uncertainty principle'' between low-rank and sparse matrices \cite{chandrasekaran11}).
Low-rank structure in the data is encountered even more frequently in neuroimaging --- unlike natural images in computer vision, there is much stronger
voxel-to-voxel homogeneity in a brain image. 

\begin{figure}[h!]

 \centerline{
 	\includegraphics[width=0.4\textwidth]{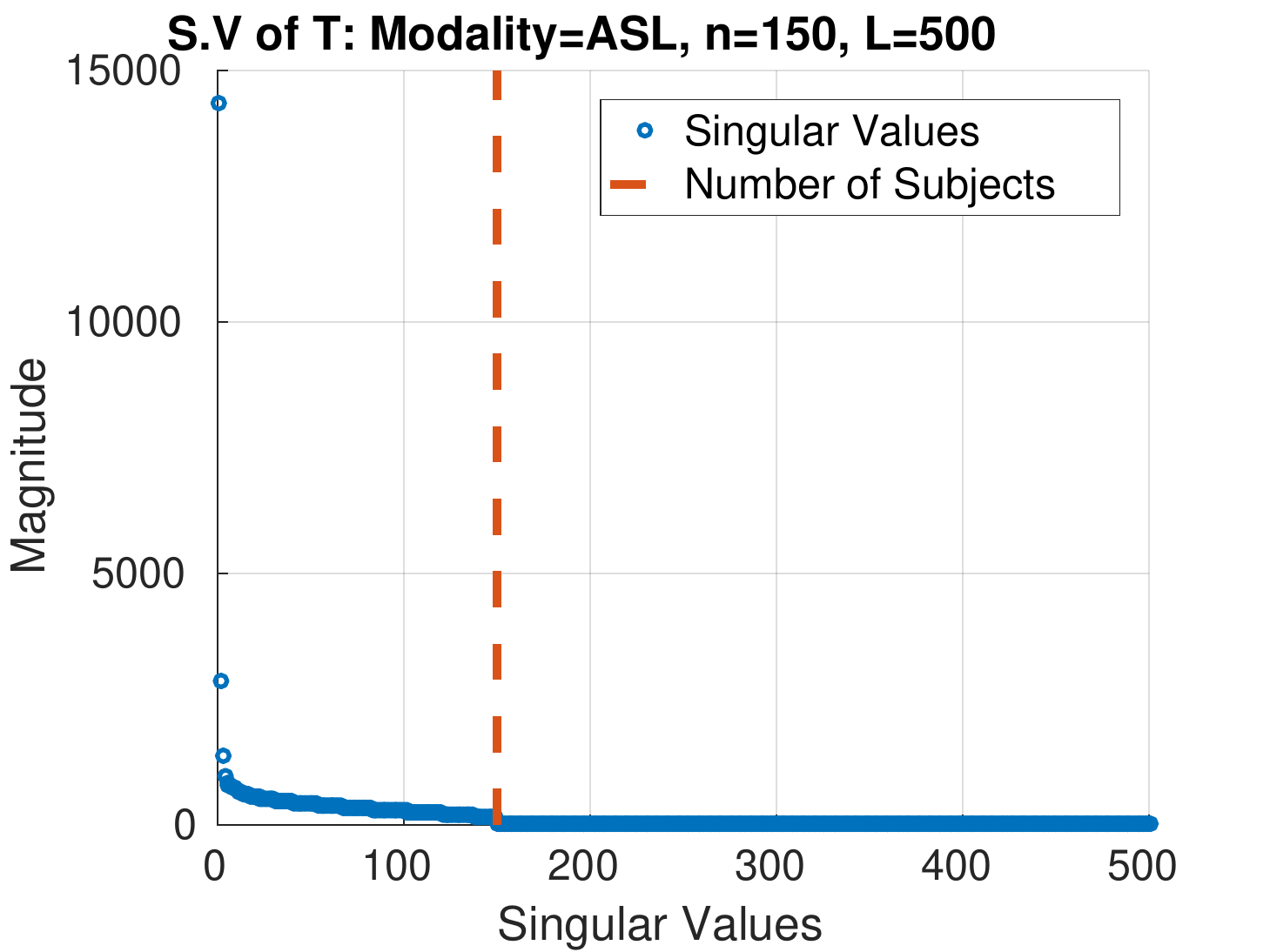}
 	\includegraphics[width=0.4\textwidth]{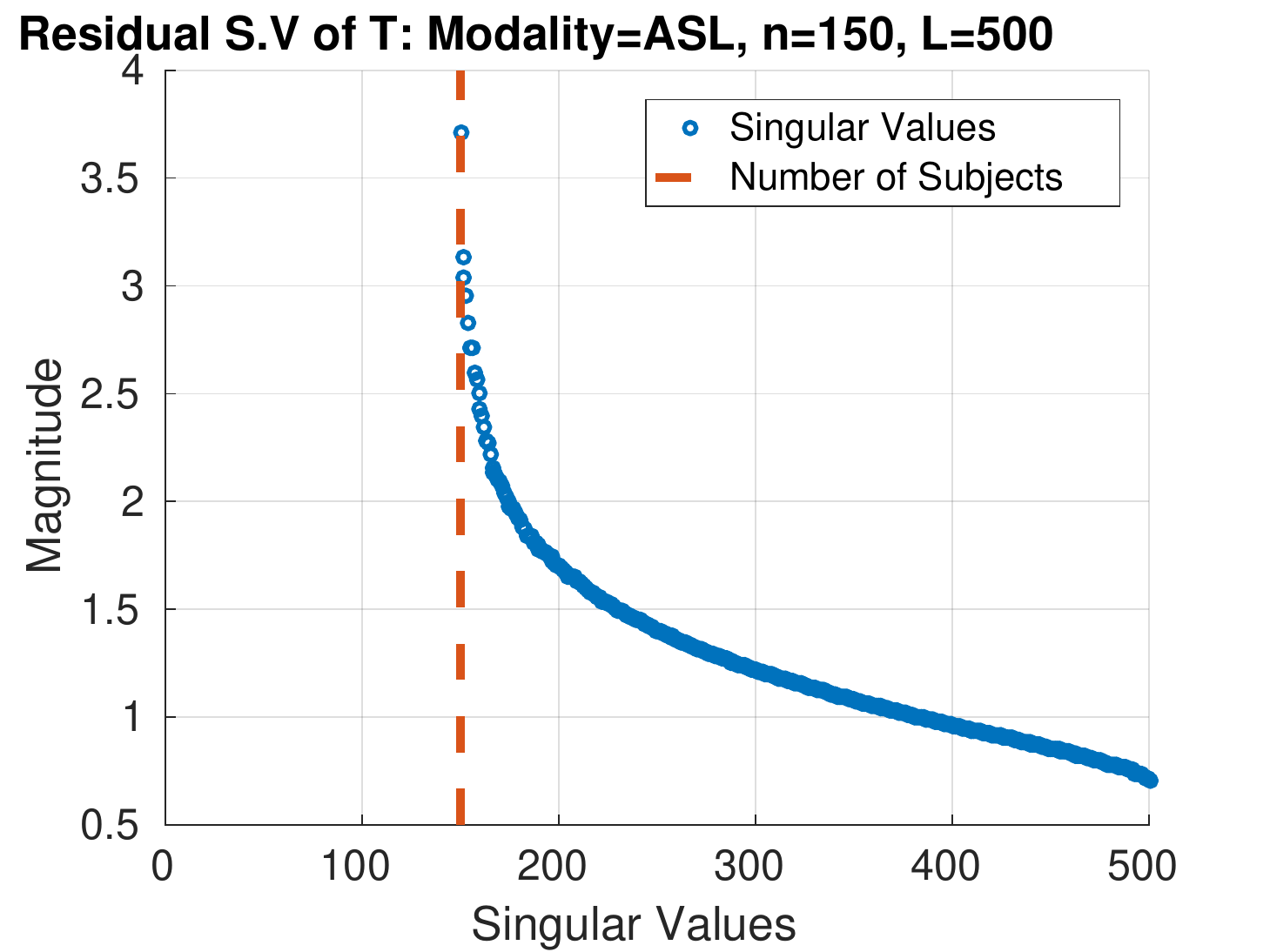}
 }
 \centerline{
 	\includegraphics[width=0.4\textwidth]{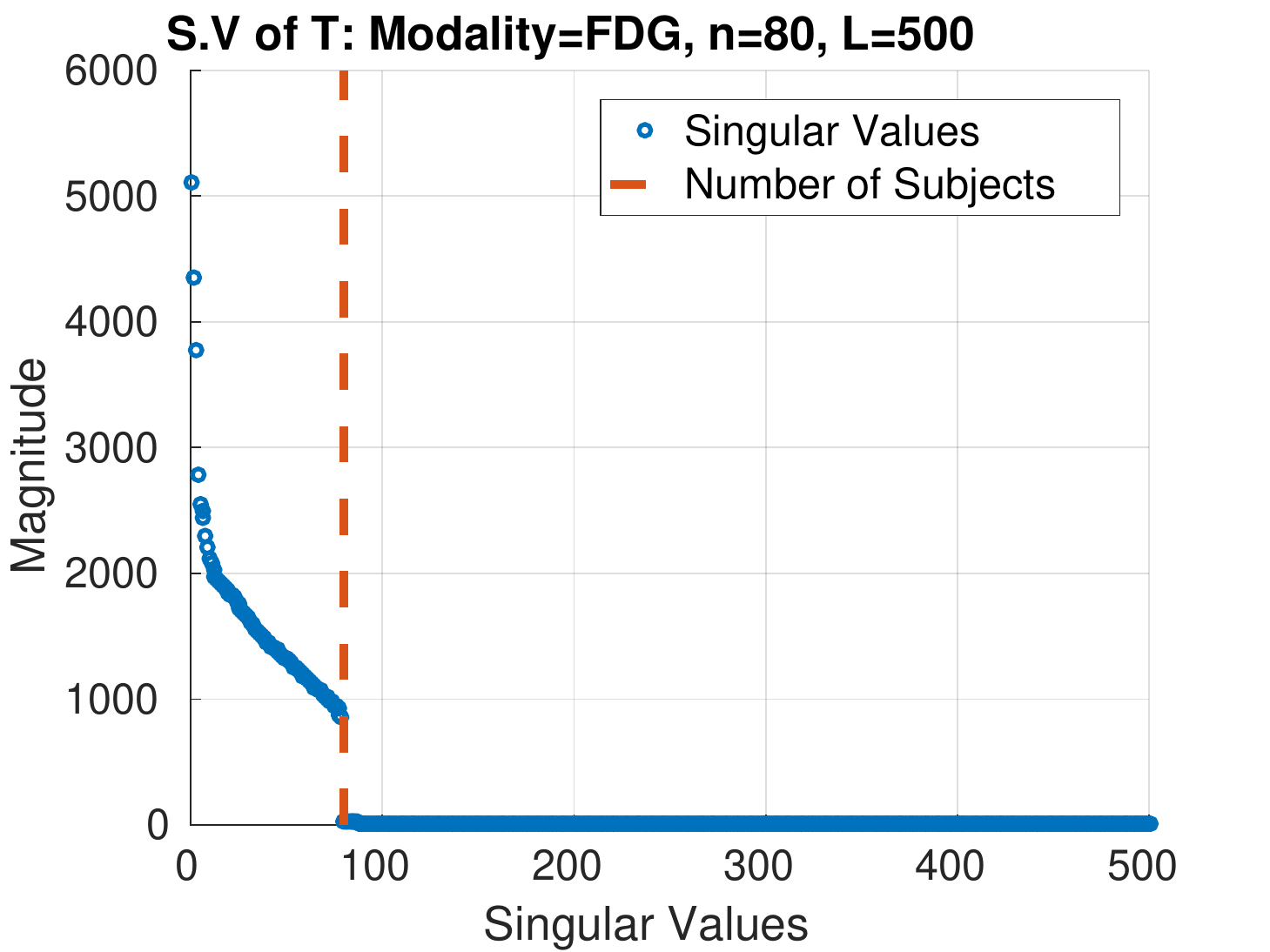}
 	\includegraphics[width=0.4\textwidth]{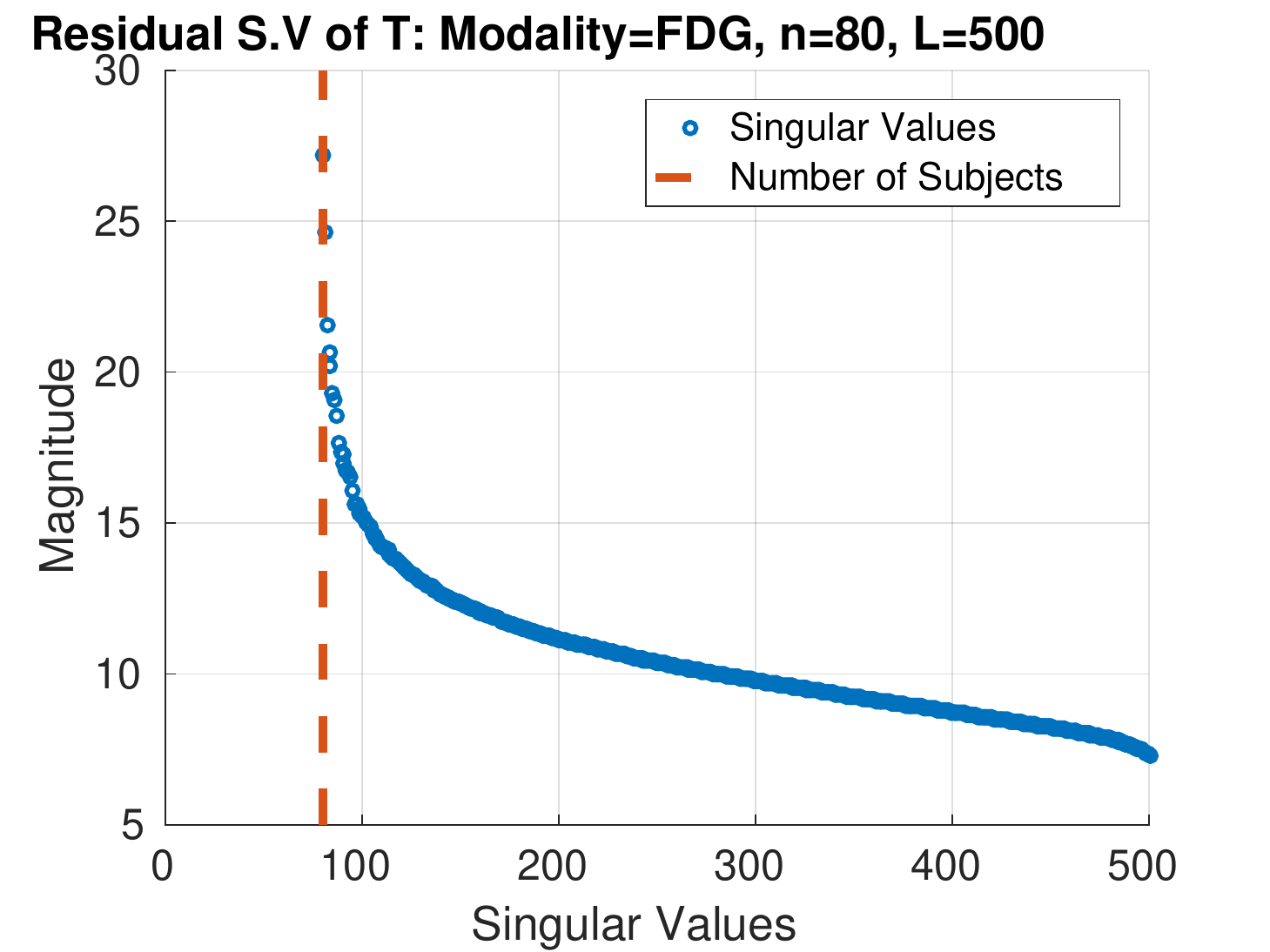}
 }
 \centerline{
 	\includegraphics[width=0.4\textwidth]{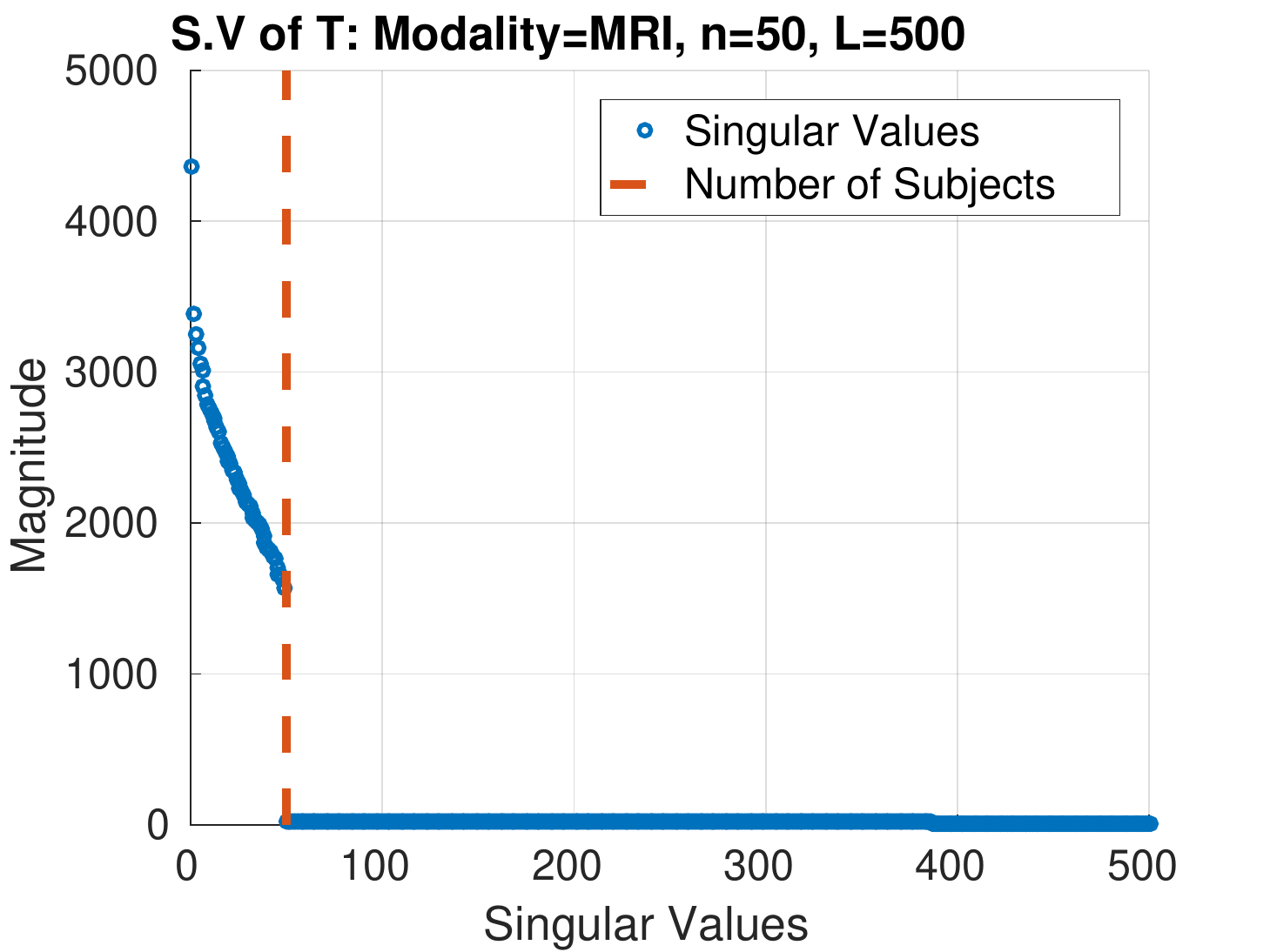}
 	\includegraphics[width=0.4\textwidth]{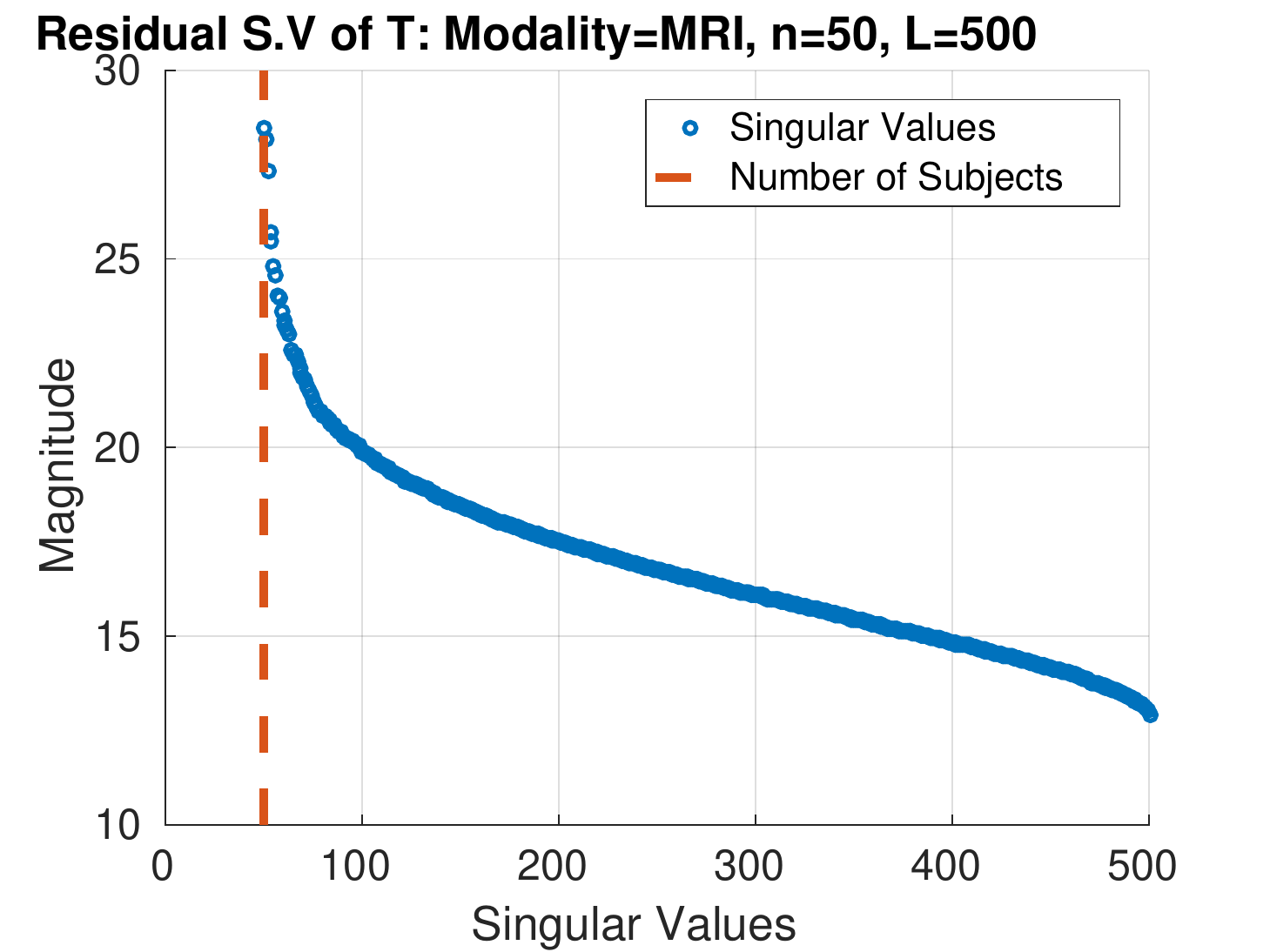}
 }
 \centerline{
 	\includegraphics[width=0.4\textwidth]{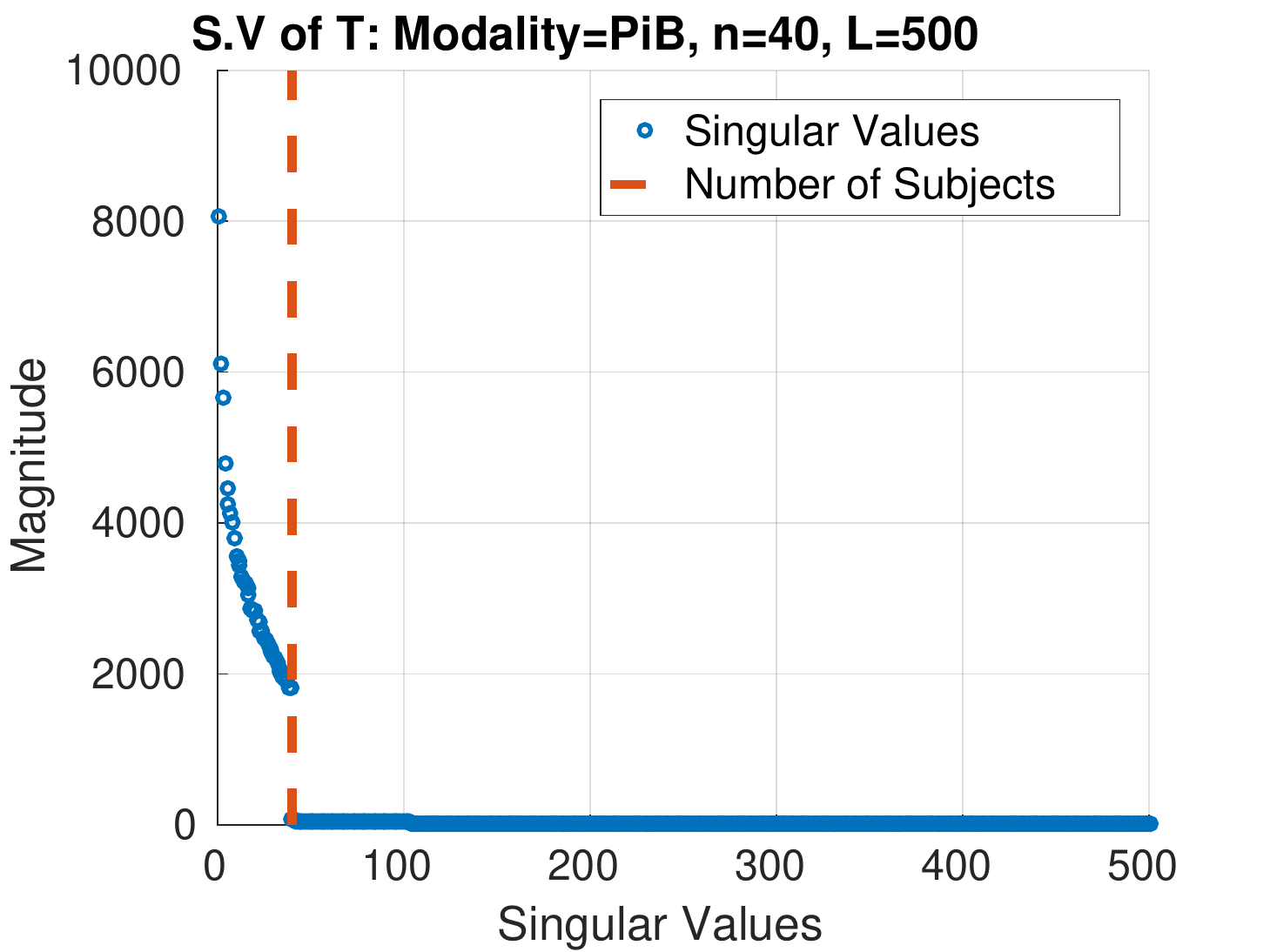}
 	\includegraphics[width=0.4\textwidth]{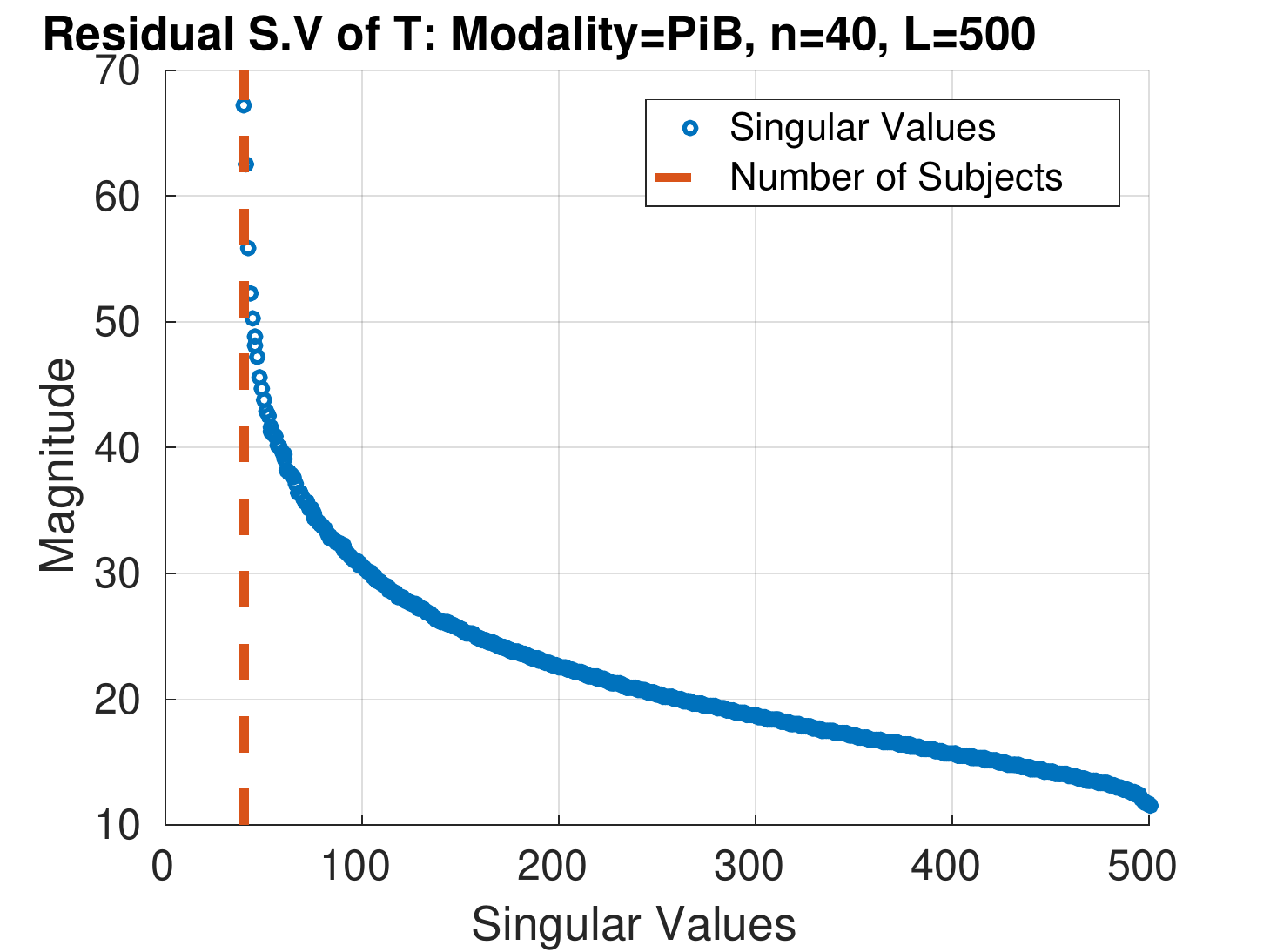}
 }

 \caption{Singular value spectrums for permutation testing matrices with dimensions generated from the imaging modalities: ASL, FDG PET, MRI, and PiB PET. \textit{Left:} Full spectrum of $\T$ with $L$ rows and $v$ columns. \textit{Right:} Residual singular values of $\T$. }
   
\label{fig:SingVals}
\end{figure}

While performing statistical hypothesis testing on these images, the low-rank structure described above 
carries through to $\T$ for purely linear statistics such as sample means, mean differences and so on. 
However, non-linearities in the test statistic, e.g., normalizing by pooled variances, 
will perturb the eigen-spectrum of the original data, contributing a long tail of eigenvalues (see Figure \ref{fig:SingVals}).
This large number of significant singular values needs to be accounted for, if one 
intends to model $\T$ using low-rank structure.
Ideally, we require that this long tail should either decay rapidly, or that it does not overlap with the dominant eigenvalues. 
This is equivalent to asking that the resulting non-linearities do not {\it decorrelate} the test statistics, 
to the point that the matrix $\T$ cannot be approximated by a low-rank matrix with high fidelity.
For $t$-statistics, the non-linearities come from normalization by pooled variances, see for example a two-sample $t$-test shown in \eqref{eq:twosample}. Here ($\mu_1$, $\sigma_1$) and ($\mu_2$, $\sigma_2$) are the mean and standard deviations for the two groups respectively. 
Since the pooled variances calculated from correlated data $X$ are unlikely to change very much from one permutation sample to another (except outliers), 
we expect that the spectrum of $\T$ will resemble that of the data (or sample) covariance, {\it with the addition of a long, exponentially decaying tail}. 
More generally, if the non-linearity does not decorrelate the test statistics too much, it will almost certainly preserve the low-rank structure.
\begin{equation} \label{eq:twosample}
t = \frac{\mu_1 - \mu_2}{\sqrt{\frac{\sigma_1}{n_1} + \frac{\sigma_2}{n_2}}}
\end{equation}

\noindent
\textbf{Does low-rank plus a long tail assumption hold for other image modalities?}

The underlying thesis of our proposed framework is that the permutation testing matrix $\T$, in general, has this low-rank plus long tail structure. In Figure \ref{fig:SingVals}, we show evidence that this is in fact the case for a variety of imaging modalities that are commonly used in medical studies -- Arterial Spin Labeling (ASL), Magnetic Resonance Imaging (MRI), and two Positron Emission Tomography (PET) modalities including fluorodeoxyglucose (FDG) and Pittsburgh compound B (PiB). Using several images coming from each of these modalities four different $\T$s are constructed. Each row in Figure \ref{fig:SingVals} shows the decay of the singular value spectrum of these four different $\T$s. The low-rank (left column) and the remaining long tail (right column) is clearly seen in these spectrum plots which suggests that the core modeling assumptions are satisfied. We note that the MRI data that was used in Figure \ref{fig:SingVals} was in fact the de facto dataset for our evaluation presented in section \ref{sec:results}. The underlying construct of such high correlations across multiple covariates or predictors is common to most biological datasets beyond brain scans, like genetic datasets. 


\subsection{Overview of Proposed method} \label{sec:overview}

If the low-rank structure dominates the long tail described above, then its contribution to $\T$ can be modeled as a low variance Gaussian \textsc{i.i.d.} residual. 
A Central Limit argument appeals to the number of independent eigenfunctions that contribute to this residual, and, 
the orthogonality of eigenfunctions implies that as more of them meaningfully contribute to each entry in the residual, the more independent those entries become. 
In other words, if this long tail begins at a low magnitude and decays slowly, then we can treat it as a Gaussian \textsc{i.i.d.} residual; 
and if it decays rapidly, then the residual will perhaps be less Gaussian, but also more negligible. 
Thus, our algorithm makes no direct assumption about these eigenvalues themselves, 
but rather that the residual corresponds to a low-variance \textsc{i.i.d.} Gaussian random matrix -- 
its contribution to the covariance of test statistics will be {\it Wishart} distributed, and from this property, we can characterize its eigenvalues.

\medskip
{\bf Why should we expect runtime improvements?} The low-rank + long tail structure of the permutation testing matrix then translates to the following identity, 
\begin{equation} \label{eq:model} \T = \U\W + \S \end{equation}
where $\S$ is the {\it unknown} \textsc{i.i.d.} Gaussian random matrix. 
We do not restrict ourselves to one-sided tests here, and so, $\S$ is modeled to be zero-mean. Later in Section \ref{sec:rapidpt}, 
we show that this apparent zero-mean assumption is addressed because of a post-processing step.
The low-rank portion of $\T$ can be reconstructed by sub-sampling the matrix at $\Omega$ using the LRMC optimization from \eqref{eq:completion1}.
Recall from the discussion in Section \ref{sec:matcomp} that $\Omega$ corresponds to a subset of indices of the entries  in $\T$, 
i.e., instead of computing all voxel-wise statistics for a given relabeling (a column of $\T$), only a small fraction $\eta$, 
referred to as the sub-sampling rate, are computed. 
Later in Sections \ref{sec:experiments} and \ref{sec:results}, we will show that $\eta$ is very small (on the orders of $<1\%$).
Therefore, the overall number of entries in $\Omega$ --- {\em the number of statistics actually calculated to recover $\T$} -- is $\eta vL$ as opposed to $vL$ for $\eta \ll 1$.

Since the core of the proposed method is to model $\T$ by accessing only a small subset of its entries $\Omega$,
we refer to it as a {\it rapid permutation testing} procedure -- \textsc{RapidPT}.
Observe that a large contributor to the running time of online subspace tracking algorithms, including the LRMC optimization from \eqref{eq:completion1}, 
is the module which updates the basis set $\U$; but once a good estimate for $\U$ has been found, this additional calculation is no longer needed.
Second, the eventual goal of the testing procedure is to recover the max null as discussed earlier in Section \ref{sec:permtest}, 
which then implies that the residual $\S$ should also be recovered with high fidelity. 
Exact recovery of $\S$ is not possible.
Although, for our purposes, we only need its effect on the {\it distribution of the maximum} per permutation test. 
An estimate of the mean and variance of $\S$ then provides reasonably good estimates of the max null.
We therefore divide the entire process into two steps: {\it training}, and {\it recovery} which is described in detail in the next section.  


\section{Rapid Permutation Testing -- \textsc{RapidPT}} \label{sec:rapidpt}

In this section, we discuss the specifics of the training and recovery stages of \textsc{RapidPT}, 
and then present the complete algorithm, followed by some theoretical guarantees regarding consistency and recovery of $\T$. Figure \ref{fig:rapidptFlow} shows a graphical representation of the RapidPT algorithm \ref{alg:rapidpt}.

\subsection{The training phase} \label{sec:training}

The goal of the training phase is to estimate the basis $\U$. 
Here, we perform a small number of fully sampled permutation tests, i.e., 
for approximately a few hundred of the columns of $\T$ (each of which corresponds to a permutation) denoted by $\ell$,
all the $v$ voxel-wise statistics are computed.
This $v \times \ell$ ``sub-matrix'' of $\T$ is referred to as the {\it training set}, denoted by $\T_{ex}$.
In our experiments, $\ell$ was selected to be either a fraction or a multiple of the total number of subjects $n$ as described in section \ref{sec:hyperparameters}. 
From $\T_{ex}$, we estimate the basis $\U$ using sub-sampled matrix completion methods \cite{Balzano10, He12}, 
making {\it multiple} passes over the training set with the (given) sub-sampling rate $\eta$, until convergence.
This corresponds to initializing $\U$ as a random orthogonal matrix of a pre-determined rank $r$, 
and using the columns of $\T_{ex}$ repeatedly to iteratively update it until convergence 
(see \cite{Balzano10,He12} for details regarding subspace tracking). 
Once $\U$ is obtained in this manner, $\W_{ex}$ is obtained by running a simple least-squares procedure on $\T_{ex}$ and $\U$.
The histogram of $\T_{ex} - \U\W_{ex}$ will then be an estimate of the empirical distribution of the residual $\S$ over the training set. 
We denote the standard deviation of these `left over' entries as $\sigma$. We now discuss a few relevant aspects of this training phase. 

Notice that in principle, one can estimate $\U$ directly from $\T_{ex}$ by simply computing the leading $r$ principal components. 
This involves a brute-force approximation of $\U$ by computing the singular-value decomposition of a dense $v \times \ell$ matrix. 
Even for reasonably small $v$, this is a costly operation. 
Second, 
$\bT$, by definition, contains a non-trivial residual. We have no direct control on the structure of $\S$ except that it is $\textsc{i.i.d}$ Gaussian.
Clearly, the variance of entries of $\S$ will depend on the fidelity of the approximation provided by $\U$.
Since the sub-sampling rate $\eta$ (the size of the set $\Omega$ compared to $vL$) is known ahead of time, 
estimating $\U$ via a subspace-tracking procedure using $\eta$ fraction of the entries of $\T_{ex}$ 
(where each column of $\T_{ex}$ modifies an existing estimate of $\U$, one-by-one, without requiring to store all the entries of $\T_{ex}$) 
directly provides an estimate of $\S$. 

\medskip
\noindent {\bf Bias-Variance Trade-off.} 
When using a very sparse subsampling method i.e., sampling with small $\eta$, there is a bias-variance trade-off in estimating $\S$. 
Clearly, if we use the entire matrix $\T$ to estimate $\U$, $\W$ and $\S$, we will obtain reliable estimates of $\S$. 
But, there is an overfitting problem: the least-squares objective used in fitting $\W$ (in getting a good estimate of the max null) 
to such a small sample of entries is likely to grossly underestimate the variance of $\S$ compared to when we use the entire matrix; 
the sub-sampling problem is not nearly as over-constrained as it is for the full matrix. 
This sampling artifact reduces the apparent variance of $\S$, and induces a bias in the distribution of the sample maximum, 
because extreme values are found less frequently. 
This sampling artifact has the effect of ``shifting'' the distribution of the sample maximum towards zero.
We refer to this as a bias-variance trade-off because, 
we can think of the need for shift as an outcome of the sub-optimality of the estimate of $\sigma$ 
versus the deviation of the true max null from the estimated max null, 
We correct for this bias by estimating the amount of the shift during the training phase, 
and then shifting the recovered sample max distribution by this estimated amount.
This shift is denoted by $\mu$.

\subsection{The recovery phase} \label{sec:recovery}

In the recovery phase, we sub-sample a small fraction of the entries of each column of $\T$ successively, i.e., for each new relabeling, 
the voxel-wise statistics are computed over a small fraction of all the voxels to populate $\T_{\Omega}$. 
Using this $\T_{\Omega}$, and the pre-estimated $\U$, the reconstruction coefficients for this column $\w \in \mathbb{R}^{r \times 1}$ are computed.
After adding the random residuals -- $\textsc{i.i.d}$ Gaussian with mean $\mu$ and standard deviation $\sigma$, to this $\U\w$, 
we have our estimate $\bT$ for this specific relabeling/permutation. 
Recall that $\S$ was originally modeled to be zero-mean (see \eqref{eq:model}), 
but the presence of the shift $\mu$ suggests a $\mathcal{N}(\mu,\sigma^2)$ distribution instead.
Overall, this entails recovering a total of $v$ voxel-wise statistics from $\eta v$ of such entries where $\eta \ll 1$. 
This process repeats for all the remaining $L-\ell$ columns of $\T$, eventually providing $\bT$.
Once $\bT$ has been estimated, we proceed exactly as in the $\textsc{NaivePT}$, to compute the max null and test for the significance of the true labeling. 

\begin{figure}[h!]
\centerline{%
\includegraphics[width=0.8\textwidth]{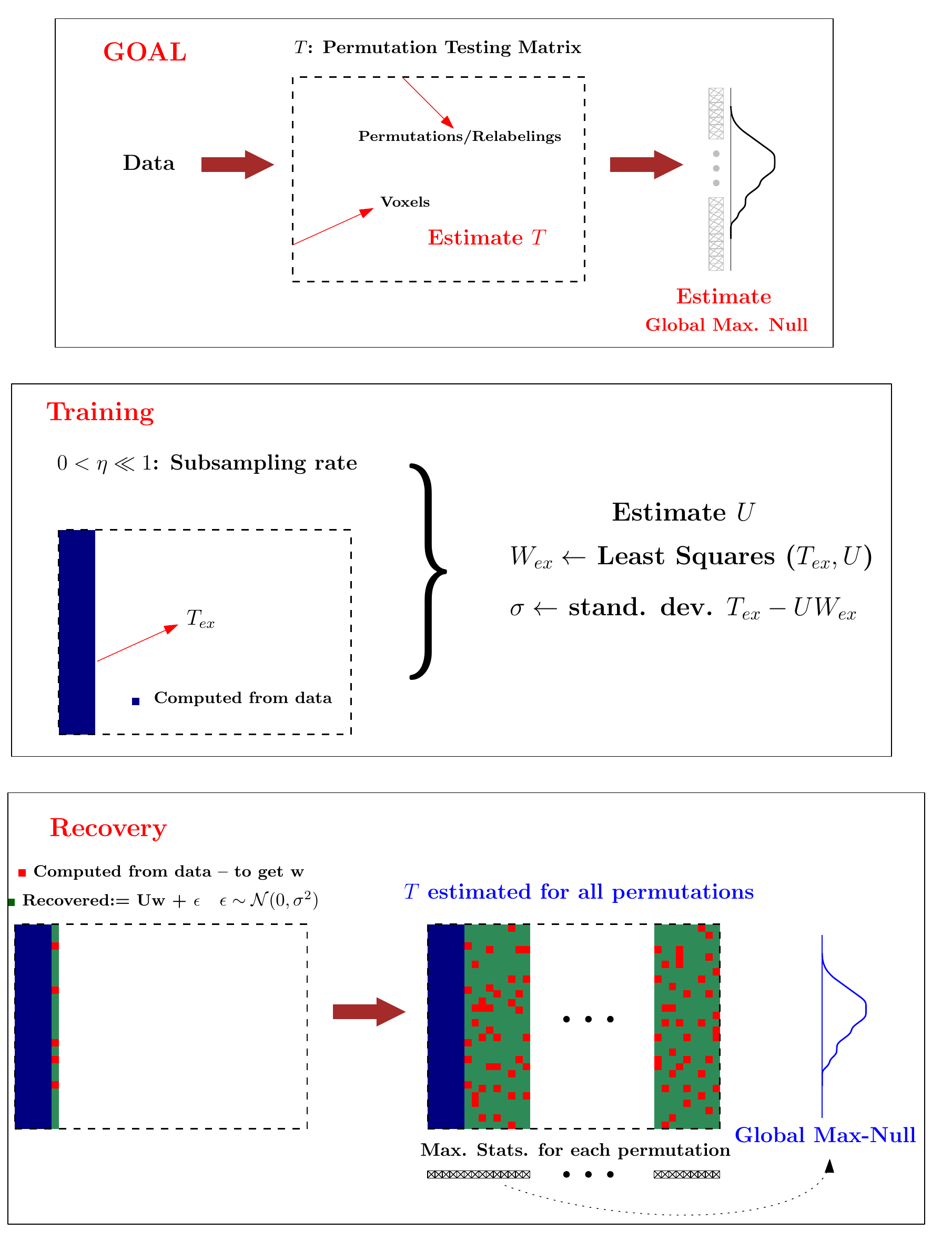}%
}%
\caption{Flow diagram of the training and recovery steps in RapidPT. The number of training samples ($l$) and the rank of \U ($r$) is the number of columns computed in the training phase (the blue area). The sub-sampling rate, $\eta$, is the fraction of red over green entries computed per column. The global max null is $h^L$ in the algorithm.}
\label{fig:rapidptFlow}
\end{figure}


\subsection{The Algorithm} \label{sec:algo}

Algorithm \ref{alg:rapidpt} and Figure \ref{fig:rapidptFlow} summarizes $\textsc{RapidPT}$.
The algorithm takes in the input data $\X$, the rank of the basis $r$, the sub-sampling rate $\eta$, 
the number of training columns $\ell$ and the total number of columns $L$ as inputs, 
and returns the estimated permutation testing matrix $\T$ and the max null distribution $\h^L$. 
As described earlier in Sections \ref{sec:training} and \ref{sec:recovery}, Algorithm \ref{alg:rapidpt} first estimates $\U$, 
$\sigma$ and the shift $\mu$, which are then used to compute $\W$ and $\S$ for the $L$ number of permutations.

\begin{algorithm}[h!]
\caption{\label{alg:rapidpt}~~The \textsc{RapidPT} algorithm for permutation testing.}
\begin{algorithmic}
\REQUIRE $\X^1$, $X^2$, $r$, $\eta$, $L$, $\ell$, {\rm stat}
\ENSURE $\bT$, $\h^L$
\STATE $\X = [\X^1; \X^2]$, $n = n_1 + n_2$
\STATE \textsc{TRAINING}
\STATE $\U \leftarrow \textsc{Rand. Orth.}$, $\W_{ex} = [\varnothing]$
\FOR {$i \in 1,\ldots,\ell$}
\STATE $j_1\ldots,j_{n} \sim \textsc{Permute}[1,n]$
\STATE $\tX^1 \leftarrow \X[:,j_1,\ldots,j_{n_1}]$
\STATE $\tX^2 \leftarrow \X[:,j_{n_1+1},\ldots,j_{n}]$
\STATE $\T_{ex}[:,i] \leftarrow {\rm test}(\tX^1, \tX^2)$ 
\STATE $k_1,\ldots,k_{\ceil{\eta v}} \sim \textsc{UNIF}[1,v]$
\STATE $\tilde{\T} \leftarrow \T_{ex}[k_1,\ldots,k_{\ceil{\eta v}},i]$
\STATE $\U, \W_{ex}[:,i] \leftarrow \textsc{Subspace-Tracking}(r)$
\ENDFOR
\STATE $\sigma \leftarrow \textsc{Standard Deviation}\{ \T_{ex} - \U\W_{ex}\}_{\Omega}$
\STATE $\mu \leftarrow \sup_i \textsc{Max}\{\T_{ex}[:,i] - \U\W_{ex}[:,i]\}$
\FOR {$i \in 1,\ldots,\ell$}
\STATE $\bT[:,i] \leftarrow \T[:,i]$
\ENDFOR 
\STATE \textsc{RECOVERY}
\FOR {$i \in \ell+1,\ldots,L$}
\STATE $k_1,\ldots,k_{\ceil{\eta v}} \sim \textsc{UNIF}[1,v]$
\STATE $j_1\ldots,j_{n} \sim \textsc{Permute}[1,n]$
\STATE $\tX^1 \leftarrow \X[k_1,\ldots,k_{\ceil{\eta v}},j_1,\ldots,j_{n_1}]$
\STATE $\tX^2 \leftarrow \X[k_1,\ldots,k_{\ceil{\eta v}},j_{n_1+1},\ldots,j_{n}]$
\STATE $\tilde{\T} \leftarrow {\rm test}(\tX^1, \tX^2)$%
\STATE $\W[:,i] \leftarrow \textsc{Complete}(\U,\tilde{\T},k_1,\ldots,k_{\ceil{\eta v}})$
\STATE $\s \leftarrow {\rm i.i.d} \mathcal{N}^v(0,\sigma^2)$
\STATE $\bT[:,i] \leftarrow \U\W[:,i] + \s$
\ENDFOR
\FOR {$i \in 1,\ldots,L$}
\IF {$i \leq \ell$}
\STATE $m_i \leftarrow \textsc{Max}(\bT[:,i])$
\ELSE 
\STATE $m_i \leftarrow \textsc{Max}(\bT[:,i]) + \mu$
\ENDIF
\ENDFOR
\STATE $\h^L \leftarrow \textsc{Histogram}({m_1,\ldots,m_L})$
\end{algorithmic}
\end{algorithm}

\subsection{Summary of theoretical guarantees} \label{sec:algo}

Algorithm \ref{alg:rapidpt} shows an efficient way to recover the max null by modeling $\T$ as a low-rank ($\U\W$) plus a low-variance residual ($\S$).
While this is useful, one may ask if this modeling assumption is reasonable and whether such a procedure will, in fact, recover the true statistics. 
Our two technical results answer these questions, and for brevity, we present them in the supplementary material accompanying the main paper. 
The informal summary of the results is (1) the basic model (i.e., low-rank and low-variance residual) is indeed meaningful for the setting we are interested in, 
and (2) Recovering the low-rank and the residual by Algorithm \ref{alg:rapidpt} guarantees a high fidelity estimate of the max null, 
and shows that the error is small.

\section{Experimental Setup} \label{sec:experiments}

We evaluate RapidPT in multiple phases.
First we perform a simulation study where the goal is to empirically demonstrate the requirements on the input hyperparameters that will guarantee an accurate recovery of the max null. These empirical results are compared to the analytical bounds governed by the theory (and further discussed in the supplement). 
The purpose of these evaluations is to walk the reader through the choices of hyperparameters, and how the algorithm is very robust to them.
Next, we perform another simulation study where our goal is to evaluate the performance of RapidPT on multiple synthetic datasets
  generated by changing the strength of group-wise differences and the sparsity of the signal (e.g., how many voxels are different).
We then conduct an extensive experiments to evaluate RapidPT against competitive methods on real brain imaging datasets.
These include comparisons of RapidPT's accuracy, runtime speedups and overall performance gains against two baselines.
The first baseline we used was the latest release of the widely used MATLAB toolbox for nonparametric permutation testing in neuroimaging, Statistical NonParametric Mapping (SnPM) (\cite{SnPM, Nichols02}). The second baseline was a standard MATLAB implementation of algorithm \ref{alg:naivept}, which we will call NaivePT. Both baselines serve to evaluate RapidPT's accuracy.
Further, the very small differences between the results provided by SnPM and NaivePT offer a secondary reference point that tells us an acceptable range for RapidPT's results (in terms of
differences).
For runtime performance, SnPM acts as a state of the art baseline. On the other hand, NaivePT is used to evaluate how an unoptimized permutation testing implementation will perform on
the datasets we use in our experiments. In the next section, we describe the experimental data, the hyperparameters space evaluated, the methods used to quantify accuracy,
and the environment where all experiments were run. 


\subsection{Simulation Data I} \label{sec:simData1}

The dataset consisted of $n=30$ synthetic images composed of $v=20000$ voxels. The signal in each voxel is derived from one of the following two normal distributions: $N(\mu=0,\sigma^{2}=1)$ and $N(\mu=1,\sigma^{2}=1)$. Two groups were then constructed with 15 images in each and letting $1\%$ ($200$ voxels) exhibit voxel-wise group differences. The signal in the remaining $99\%$ of
the voxels was assumed to come from $N(\mu=0,\sigma^{2}=1)$.

\subsection{Simulation Data II} \label{sec:simData2}

The dataset consisted of a total of $48$ synthetically generated datasets, each with $v=20000$ voxels.
The datasets were generated by varying: the number of images ($n$) in the dataset, the strength of the signal (i.e., deviation of $\mu$ in $N(\mu,1)$ from $N(0,1)$)
and the sparsity of the signal (percentage of voxels showing group differences). The dataset sizes were $n=60$, $n=150$, $n=600$.
Each dataset was split into two \textit{equally sized} groups for which various degrees of signal and sparsity of the signal were used to generate the different datasets.
The first ``group'' (for group-difference analysis) in all datasets was generated from a standard normal distribution, $N(\mu=0,\sigma^{2}=1)$.
In the second ``group'', we chose $\{1\%, 5\%, 10\%, 25\%\}$ of the voxels from one of four normal distributions ($N(\mu=1,\sigma^{2}=1)$, $N(\mu=5,\sigma^{2}=1)$,
$N(\mu=10,\sigma^{2}=1)$, $N(\mu=25,\sigma^{2}=1)$). The signal in the remaining voxels in the second group was also obtained from $N(\mu=0,\sigma^{2}=1)$.

To summarize the simulation setup, Simulation Data I fixes the dataset and changes the algorithmic hyperparameters whereas Simulation Data II fixes the algorithmic hyperparameters
and generates different datasets. 

\subsection{Data} \label{sec:data}

The data used to evaluate RapidPT comes from the Alzheimer’s disease Neuroimaging Initiative-II (ADNI2) dataset. The ADNI project
was launched in 2003 by the National Institute on Aging, the National Institute of Biomedical Imaging and Bioengineering, the Food and Drug Administration,
private pharmaceutical companies, and nonprofit organizations, as a \$60 million, 5-year public-private partnership.
The overall goal of ADNI is to test whether serial MRI, positron emission tomography (PET), other biological markers, and clinical and neuropsychological assessment
can be combined to measure the progression of MCI and early AD. 
Determination of sensitive and specific markers of very early AD progression is intended to aid researchers and clinicians to develop new treatments and monitor their effectiveness, as well as lessen the time and cost of clinical trials. The principal investigator of this initiative is Michael W. Weiner, M.D., VA Medical Center and University of California — San Francisco. ADNI is the result of the efforts of many co-investigators from a broad range of academic institutions and private corporations, and subjects have been recruited from over 50 sites across the U.S. and Canada.
  The initial aim of ADNI was to recruit 800 adults, ages 55 to 90, to participate in the research — approximately 200 cognitively normal older individuals to be followed for 3 years, 400 people with MCI to be followed for 3 years, and 200 people with early AD to be followed for 2 years.

For the experiments presented in this paper, we used gray matter tissue probability maps derived from T1-weighted magnetic resonance imaging (MRI) data.
From this data, we constructed four varying sized datasets. We sampled $n_1$ and $n_2$ subjects from the CN and AD groups in the cohort, respectively.
Table \ref{Datasets} shows a summary of the datasets used for our evaluations.

\begin{table}[h!]
\centering
	\begin{tabular}{| c | c | c | c |}
    \hline
    \rowcolor{Gray}
 		\multicolumn{4}{|c|}{\textbf{Dataset Size: $n$ ($n_1$,$n_2$)}} \\
 		\hline
 		50 (25,25) & 100 (50,50) & 200 (100,100) & 400 (200,200)  \\
    \hline
	\end{tabular}
	\caption{\label{Datasets} Dataset sizes used in our experiments. The table lists the total number of subjects ($n$) and how many of the participants were sampled
          from the CN  ($n_1$) and AD groups ($n_2$).}
\end{table}

\subsubsection{Data Pre-processing} \label{sec:preprocessing}

All images were pre-processed using voxel-based morphometry (VBM) toolbox in Statistical Parametric Mapping software (SPM, http://www.fil.ion.ucl.ac.uk/spm).
After pre-processing, we obtain a data matrix $\X$ composed of $n$ rows and $v$ columns for each dataset shown in Table \ref{Datasets}.
Each row in $\X$ corresponds to a subject and each column is associated to a voxel that denotes approximately the same anatomical location across subjects (since the images
are already co-registered). This pre-processing is commonly used in the literature and not specialized to our experiments. 


\subsection{Hyperparameters} \label{sec:hyperparameters}

As outlined in Algorithm \ref{alg:rapidpt}, there are three high-level input parameters that will impact the performance of the procedure: the \textit{number of training samples} ($l$),
the \textit{sub-sampling rate} ($\eta$), and the \textit{number of permutations} ($L$).
To demonstrate the robustness of the algorithm to these parameter settings, 
we explored and report on hundreds of combinations of these hyperparameters on each dataset.
This also helps us identify the general scenarios under which RapidPT will be a much superior alternative to regular permutation testing.
The baselines for a given combination of these hyperparameters are given by the max null distribution constructed by SnPM and NaivePT.
The number of permutations used for the max null distributions of the baselines is the same as the number of permutations used by RapidPT for a given combination of hyperparameters. 

\begin{itemize}
\item \textit{Number of Training Samples: }The number of training samples, $l$, determines how many columns of $\T$ are calculated to estimate the basis of the subspace, 
  and also how many training passes are performed to estimate the shift that corrects for the bias-variance tradeoff discussed in Section \ref{sec:training}. 
  We decided to use the total number of subjects $n$ as a guide to pick a sensible $l$, the rationale is that 
  the maximum possible rank of $\T$ is $n$ (as discussed in Section \ref{sec:fastperm}). 
  Further, $l$ is also used to determine the number of passes performed to calculate the shift of the max null distribution. Calculating the shift is 
  a cheap step, therefore it makes sense to use all the information available in $\T_{ex}$ to calculate the shift. Table \ref{TrainingSamples} shows a summary of the different values for $l$ used in our evaluations.

\begin{table}[h!]
\centering
	\begin{tabular}{c|c|c|c|c}
 		\rowcolor{Gray}
    \multicolumn{5}{c}{\textbf{Number of Training Samples:} $l$}\\
    \cline{2-5}
 		\cellcolor{Gray}\textbf{Simulations} & $\frac{n}{3}$ & $n$ & $2n$\\
    \cline{2-5}
    \cellcolor{Gray}\textbf{Experiments} & $\frac{n}{2}$ & $\frac{3n}{4}$ & $n$ & $2n$\\
	\end{tabular}
	\caption{\label{TrainingSamples}Number of training samples used to evaluate RapidPT. $n$ corresponds to the total number of subjects in the dataset. For instance, for the $400$ subject dataset the values for $l$ used were $100$, $200$, $400$, and $800$.}
\end{table}

\item \textit{Sub-sampling rate: }The sub-sampling rate, $\eta$, is the percentage of all the entries of $\T$ that we will calculate (i.e., sample) 
  when recovering the max null distribution. In the recovery phase, $\eta$ determines how many voxel-wise test statistics will be calculated at each 
  permutation to recover a column of $\bar{\T}$. For instance, if the data matrix has $v$ columns (number of voxels) then instead of calculating $v$ 
  test statistics, we will sample only $\eta v$ (where $\eta \ll 1$) random columns and calculate test statistics only for those columns. Table \ref{SamplingRates} shows a summary of the different values for $\eta$ used in our evaluations.

\begin{table}[h!]
\centering
	\begin{tabular}{ c | c | c | c | c | c | c | c | c | c }
 		\rowcolor{Gray}
    \multicolumn{10}{c}{\textbf{Sub-sampling rate:} $\eta$} \\
    \cline{2-10}
    \cellcolor{Gray} \textbf{Simulations} & $0.5\%$ & $1\%$ & $1.6\%$ & $2\%$ & $4\%$ & $8\%$ & $16\%$ & $32\%$ & $64\%$  \\
    \cline{2-10}
 		\cellcolor{Gray} \textbf{Experiments} & $0.1\%$ & $0.35\%$ & $0.5\%$ & $0.7\%$ & $1.5\%$ & $5\%$ \\
	\end{tabular}
	\caption{\label{SamplingRates}Sub-sampling rates used to evaluate RapidPT. $\eta$ is the percentage of the total number of entries in $\T$ that will be calculated during the recovery phase.}
\end{table}

\item \textit{Number of Permutations}
  The number of permutations, $L$, determines the total number of columns in $\T$. By varying $L$ we are able to see how the size of $\T$ affects the 
  accuracy of the algorithm and also how it scales compared to a standard permutation testing implementation with the same number 
  of permutations (e.g., NaivePT or SnPM). Table \ref{NumPermutations} shows a summary of the different values for $L$ used in our evaluations.

\begin{table}[h!]
\centering
	\begin{tabular}{ c | c | c | c | c | c | c | c }
 		\rowcolor{Gray}
    \multicolumn{8}{c}{\textbf{Number of Permutations:} $L$} \\
    \cline{2-8}
    \cellcolor{Gray} \textbf{Simulations} & $5,000$ & $10,000$ & $20,000$ & $40,000$ & $50,000$ & $100,000$ \\
    \cline{2-8}
 		\cellcolor{Gray} \textbf{Experiments} & $2,000$ & $5,000$ & $10,000$ & $20,000$ & $40,000$ & $80,000$ & $160,000$ \\
	\end{tabular}
	\caption{\label{NumPermutations}Number of permutations done to evaluate RapidPT.}
\end{table}

\end{itemize}


\subsection{Accuracy Benchmarks} \label{sec:benchmarks}

In order to assess the accuracy and overall usefulness of the recovered max null distribution by RapidPT we used three different 
measures: Kullback-Leibler Divergence (KL-Divergence), $t$-thresholds/$p$-values and the resampling risk.

\textit{Kullback-Leibler Divergence}: The KL-Divergence provides a measure of the difference between two probability distributions. One of the distributions represents the 
ground truth (SnPM or NaivePT) and the other an ``approximation'' (obtained via RapidPT). In this case, 
the distributions are the max null distributions ($h^{L}$). We use the KL-Divergence to identify under which circumstances (i.e., hyperparameters) 
RapidPT provides a good estimate of the overall max null distribution and if there are cases where the results are unsatisfactory.  

\textit{$T$-Thresholds/$p$-values}: Once we have evaluated whether all methods recover a similar max null distribution, we analyze if 
$t$-thresholds associated to a given $p$-value calculated from each max null distribution are also similar.  
	
\textit{Resampling Risk}: Two methods can recover a similar max null distribution and $p$-values, and yet partially disagree in {\em which voxels} 
should be classified as statistically significant (e.g., within a group difference analysis). 
The resampling risk is the probability that the decision of accepting/rejecting the null hypothesis differs between two methods (\cite{Jockel84}). 
Let $v_{1}$ and $v_{2}$ be the number of voxels whose null hypothesis was rejected according to the max null derived from Method 1 and 2, respectively. 
Further, let $v_{c}$ be the number of voxels that are the (set) intersection of $v_1$ and $v_2$. The resampling risk can then be calculated 
as shown in \eqref{eq:resamplingRisk}.

\begin{equation} \label{eq:resamplingRisk}
	\text{risk} = \frac{\frac{v_{1} - v_{c}}{v_{1}} + \frac{v_{2} - v_{c}}{v_{2}}}{2}
\end{equation}


\subsection{Implementation Environment and other details} \label{sec:environment}

All evaluation runs reported in this paper were performed on multiple machines with the same hardware configuration. 
The setup consisted of multiple 16 core machine with two Intel(R) Xeon(R) CPU E5-2670, each with 8 cores. This means that any MATLAB application will 
be able to run a maximum of 16 threads. To evaluate the runtime performance of RapidPT, we performed all experiments on two different setups
which forced MATLAB to use a specific number of threads. First, we forced MATLAB to only use a single thread (single core) when running SnPM, RapidPT, and NaivePT. 
The performance results on a single threaded environment attempt to emulate a scenario where the application was running on an older laptop/workstation serially. 
In the second setup, we allow MATLAB to use all 16 threads available to demonstrate how RapidPT is also able to leverage a parallel computing environment 
to reduce its overall runtime.

Although all machines had the same hardware setup, to further ensure that we were making a fair runtime performance comparison, all measurements of a 
given Figure shown in Section \ref{sec:results} were obtained from the same machine.


\section{Results} \label{sec:results}

The hyperparameter space explored through hundreds of runs allowed identifying specific scenarios where RapidPT is most effective. 
To demonstrate accuracy, we first show the impact of the hyperparameters on the recovery of the max null distribution by analyzing KL-Divergence. 
Then we focus on the comparison of the corrected $p$-values across methods and the resampling risk associated with those $p$-values. To demonstrate the 
runtime performance gains of RapidPT, we first calculate the speedup results across hyperparameters. We then focus on the hyperparameters that a user would 
use and look at how RapidPT, SnPM, and NaivePT scale with respect to the dataset and the number of permutations. Overall, the large hyperparameter space 
that was explored in these experiments produced hundreds of figures. In this section, we summarize the results of all figures within each subsection,
but only present the figures that we believe will convey the most important information about RapidPT. 
An extended results section is presented in the supplementary materials.
We point out that following the results and the corresponding discussion of the plots and figures, we discuss the open-source toolbox version of RapidPT that is mde available online. 


\subsection{Accuracy} \label{sec:accuracy}
\noindent

\textbf{Simulations}

\begin{figure}[H]
\centerline{%
	\includegraphics[width=0.33\textwidth]{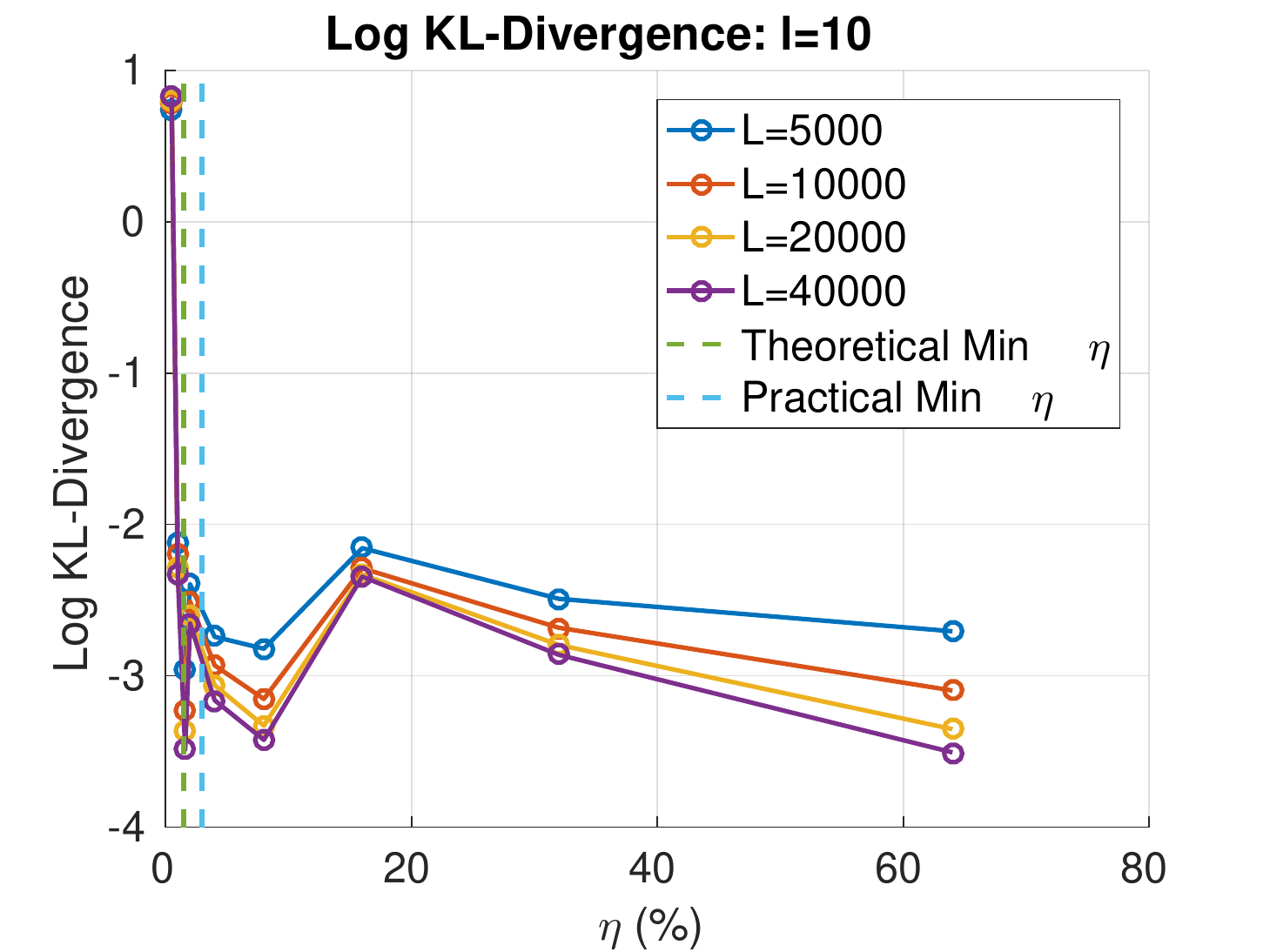}
	\includegraphics[width=0.33\textwidth]{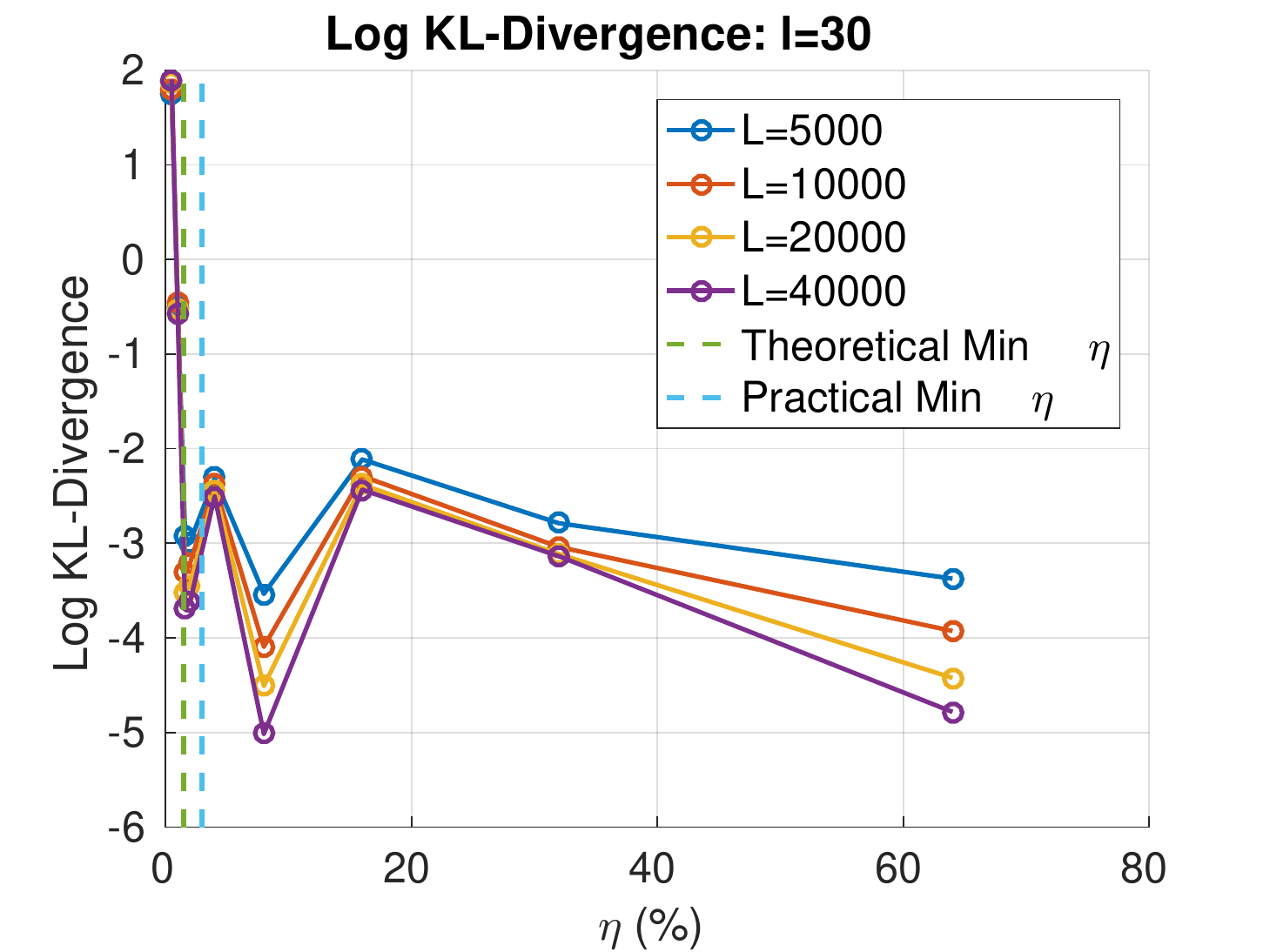}
	\includegraphics[width=0.33\textwidth]{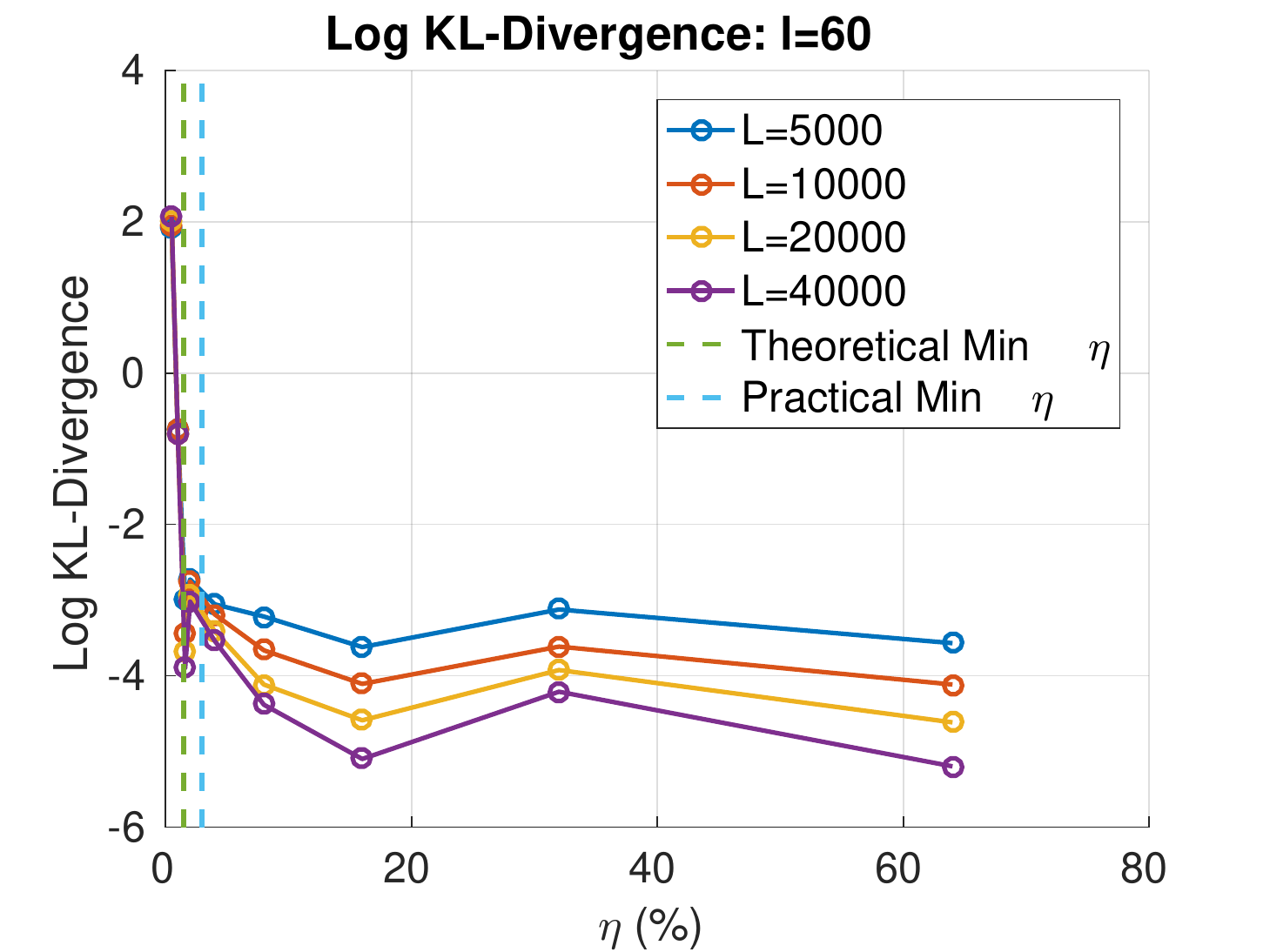}
}%

\caption{KL-Divergence between the true max null and the one recovered by RapidPT. Each line corresponds to a different number of permutations. The dotted lines are the theoretical minimum sub-sampling rate and the "practical" one, i.e., the one the toolbox will set it to automatically if none is specified.}
\label{fig:KLDivSimData}
\end{figure}

Figure \ref{fig:KLDivSimData} shows the log KL-Divergence between the max null recovered by regular permutation testing and the max null recovered by RapidPT. We can observe that once the sub-sampling rate, $\eta$, exceeds the minimum value established by LRMC theory, RapidPT is able to accurately recover the max null with a KL-Divergence $< 10^{-2}$.
Furthermore, increasing $l$ can lead to slightly lower KL-Divergence as seen in the middle and right most plots. Finally, increasing the number of permutation improves the accuracy.
A through discussion of how to choose the important hyperparameters is in Section \ref{sec:hyperreco}.

\begin{figure}[H]
\centerline{%
	\includegraphics[width=0.33\textwidth]{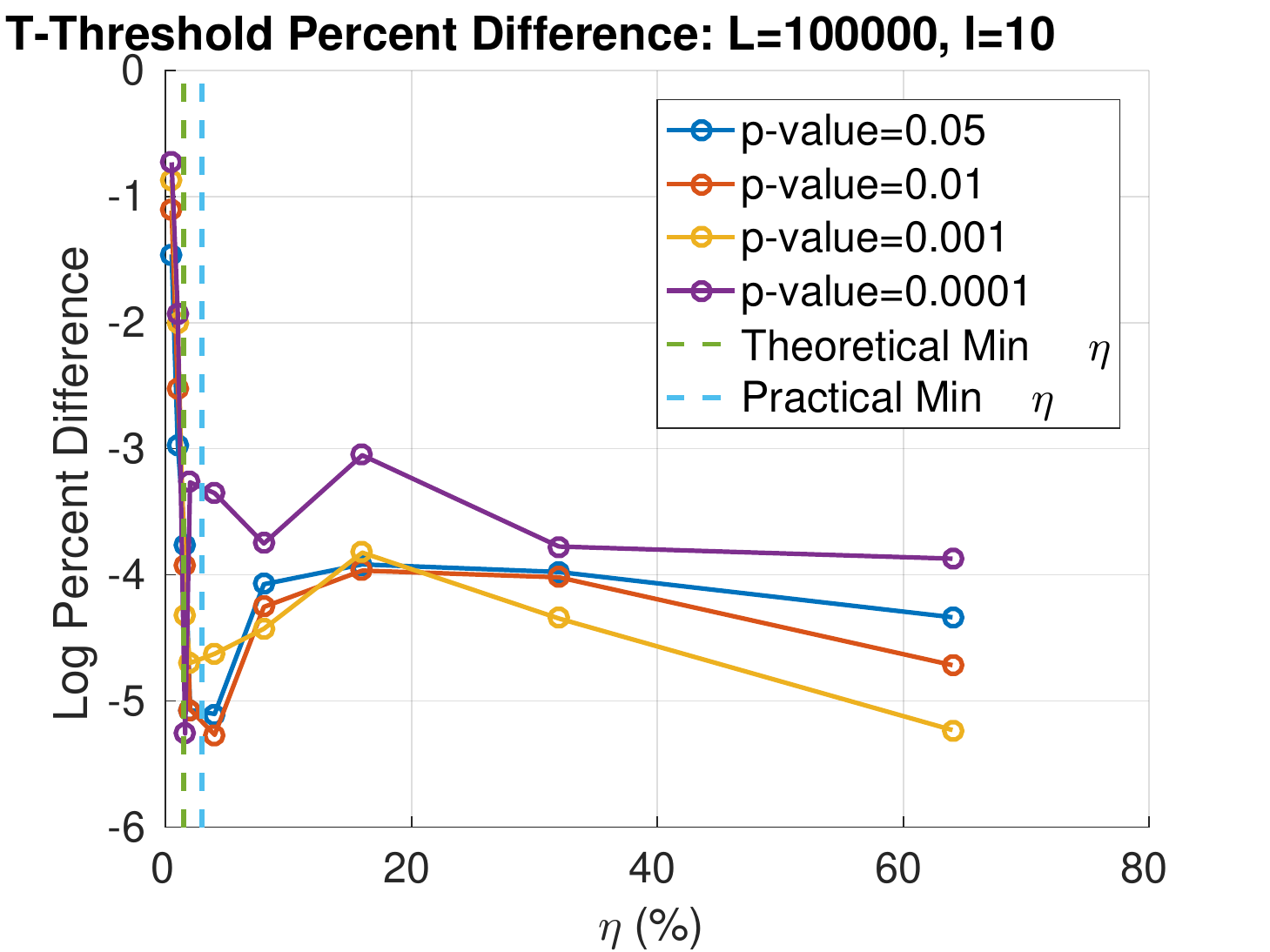}
	\includegraphics[width=0.33\textwidth]{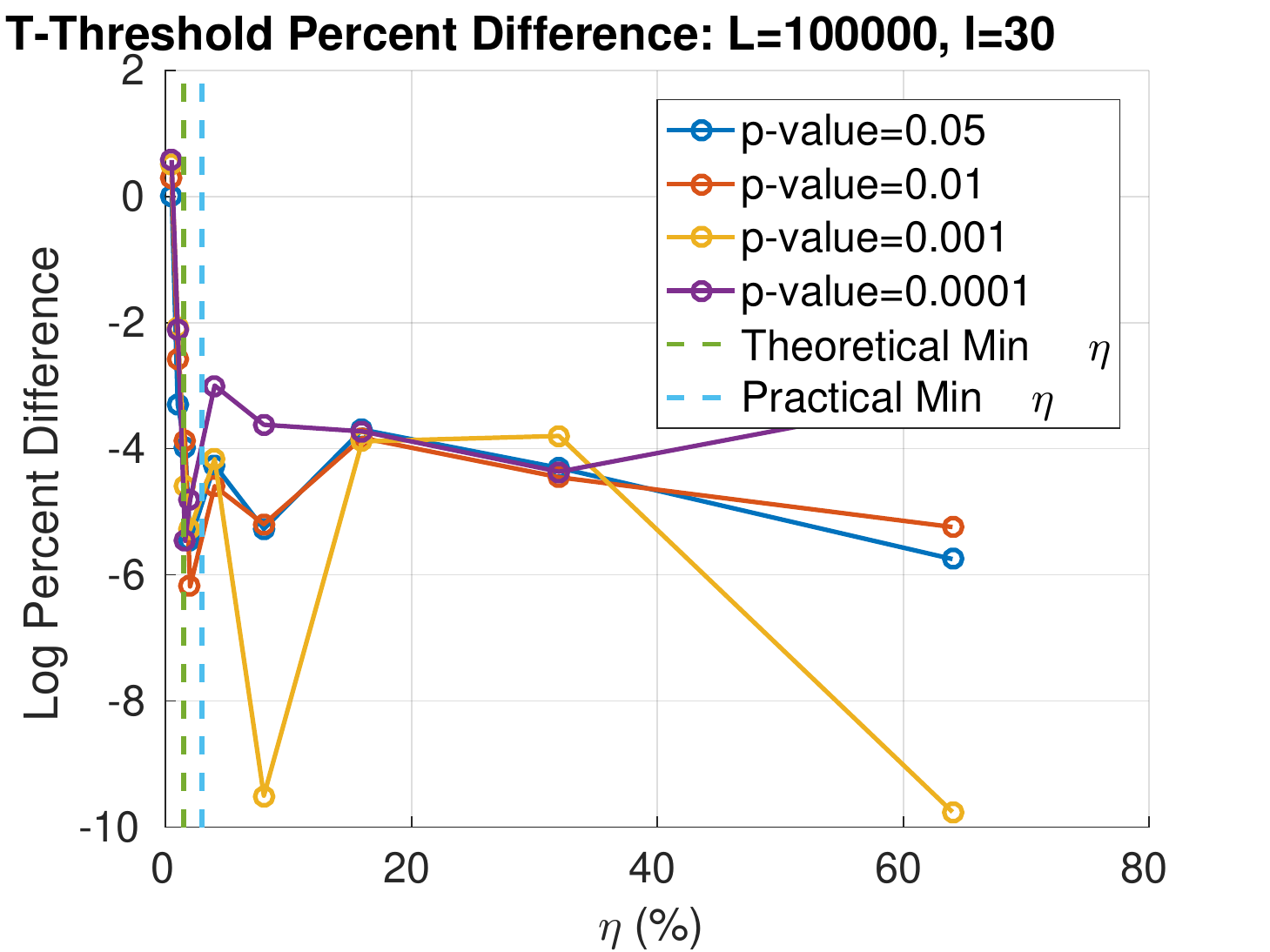}
	\includegraphics[width=0.33\textwidth]{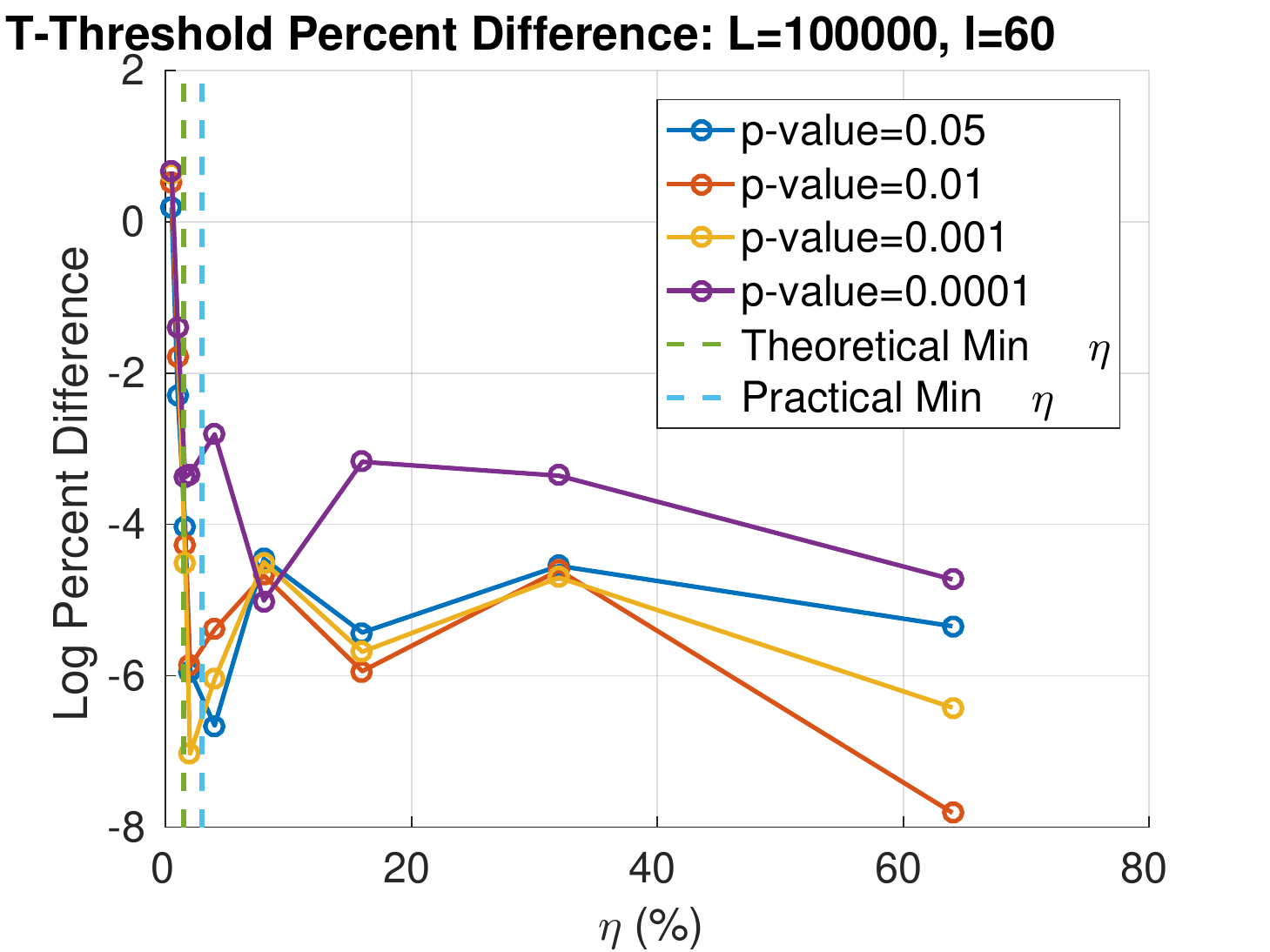}
}%

\caption{Percent difference between the $t$-threshold (for different $p$-values) obtained from the true max null and the one recovered by RapidPT. The dotted lines are the theoretical minimum sub-sampling rate and the "practical" one, i.e., the one the toolbox will set it to automatically if none is specified.}
\label{fig:TThreshSimData}
\end{figure}

Figure \ref{fig:TThreshSimData} shows the log percent difference between the $t$-thresholds for different $p$-values obtained from the true max null and the one recovered by RapidPT. Similar to Figure \ref{fig:KLDivSimData}, it is evident that once the strict requirement on the minimum value of $\eta$ is achieved, we obtain a reliable $t$-threshold i.e., percent difference $< 10^{-3}$. Additionally, increasing the number of training samples (progression of plots from left to right) gives a improvement in the accuracy, however, not incredibly significant since we are already at negligible percent differences. Overall we see that RapidPT is able to estimate accurate thresholds even at extremely low $p$-value regimes. 

\begin{figure}[H]
\centerline{%
	\includegraphics[width=0.33\textwidth]{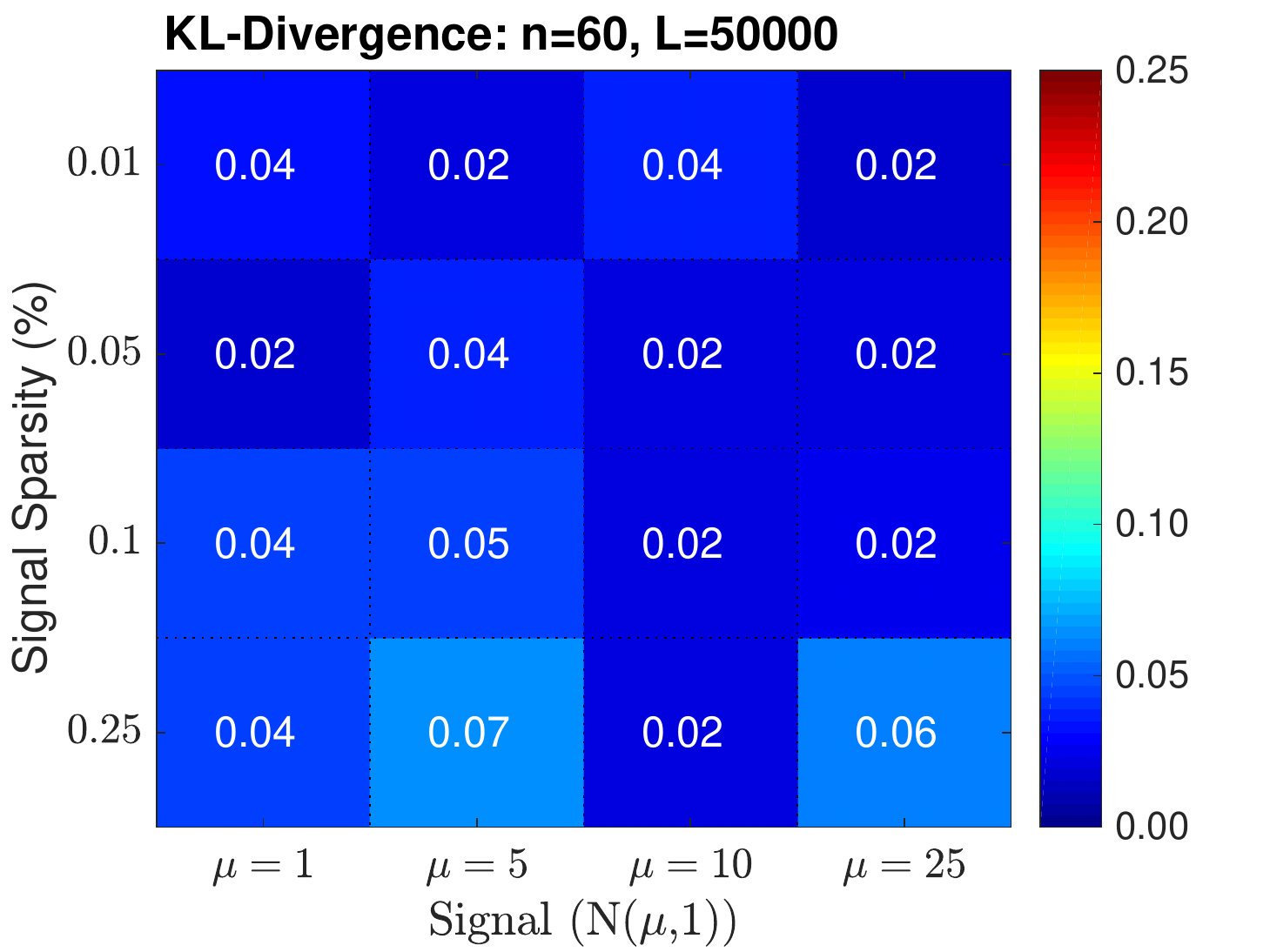}
	\includegraphics[width=0.33\textwidth]{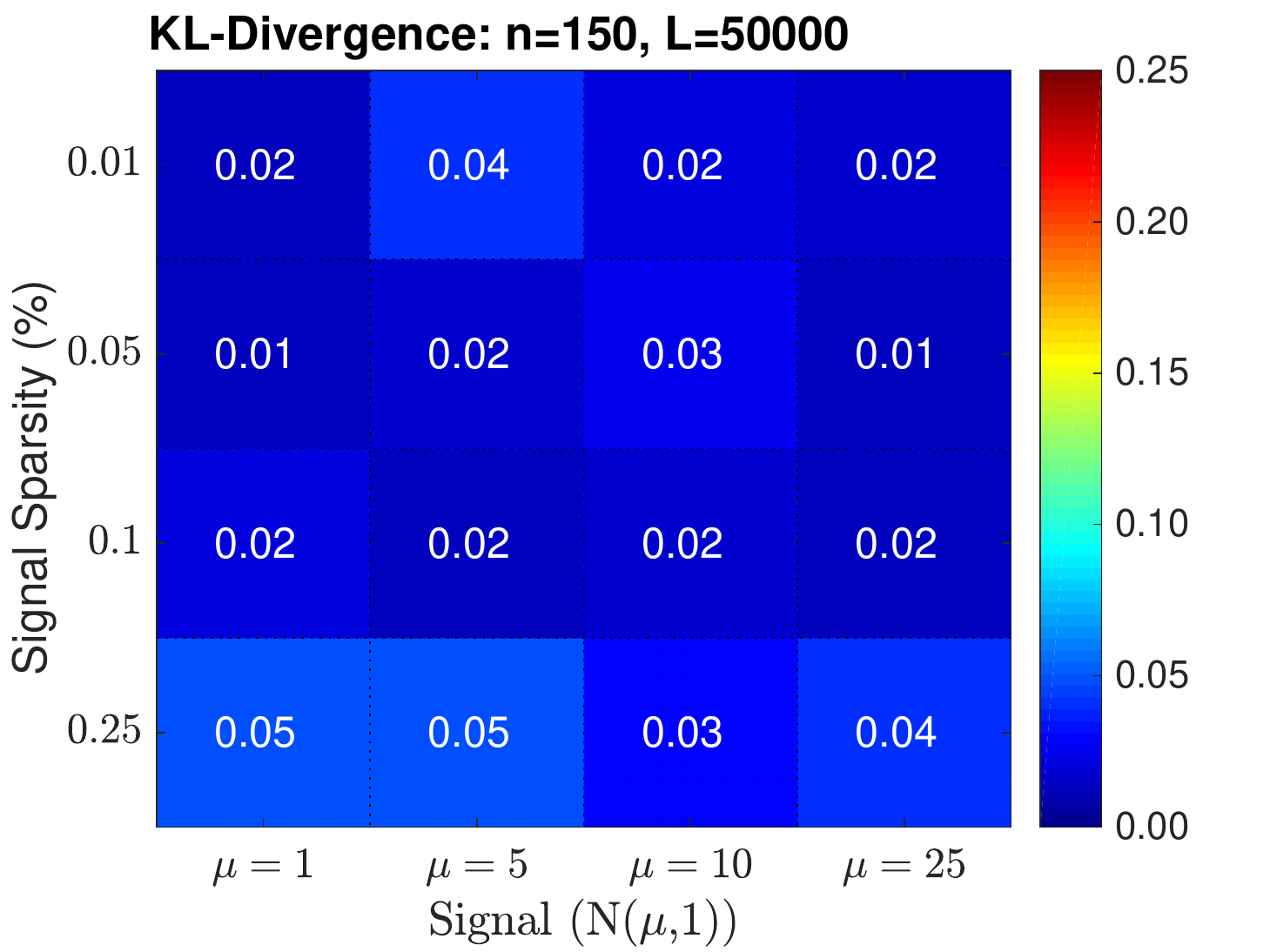}
	\includegraphics[width=0.33\textwidth]{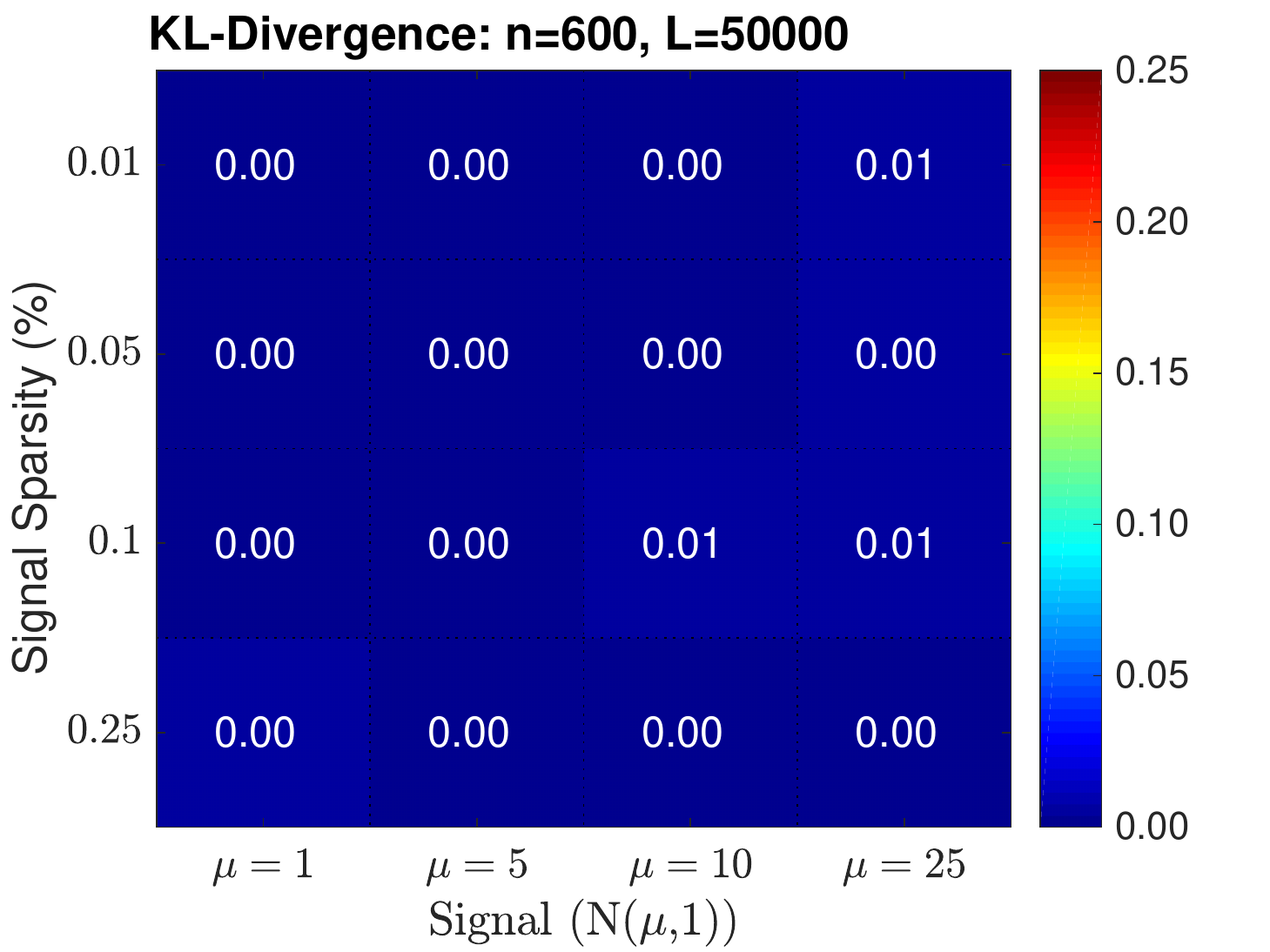}
}%

\caption{KL-Divergence between the true max null and the one recovered by RapidPT on 48 datasets. The sub-sampling rate, $\eta$, used for each run was $2\eta_{min}$. The number of training samples, $l$, used for each run was $n$ (i.e., the same as the number of images in the dataset).}

\label{fig:KLDivBreadthSimData}
\end{figure}

  Figure \ref{fig:KLDivBreadthSimData} shows the KL-Divergence between the max null recovered by regular permutation testing and the max null recovered by RapidPT on 48
  synthetically generated datasets (16 in each column). The input hyperparameters used were fixed to $L=50000$, $\eta=2\eta_{min}$, and $l=n$, where $\eta_{min}$ refers to
  the theoretical minimum sub-sampling rate and $n$ is the dataset size. As expected, the strength or sparsity of the signal does not have an impact on the performance of RapidPT.
  The dataset size, however, does have a slight impact on the accuracy but we still find that the recovered $t$-thresholds for the smaller datasets are within ~2$\%$ of the true threshold.

\noindent

\textbf{Results on the ADNI Dataset}
\newline
\textit{Can we recover the max null distribution?}

\begin{figure}[H]
\centerline{%
	\includegraphics[width=0.4\textwidth]{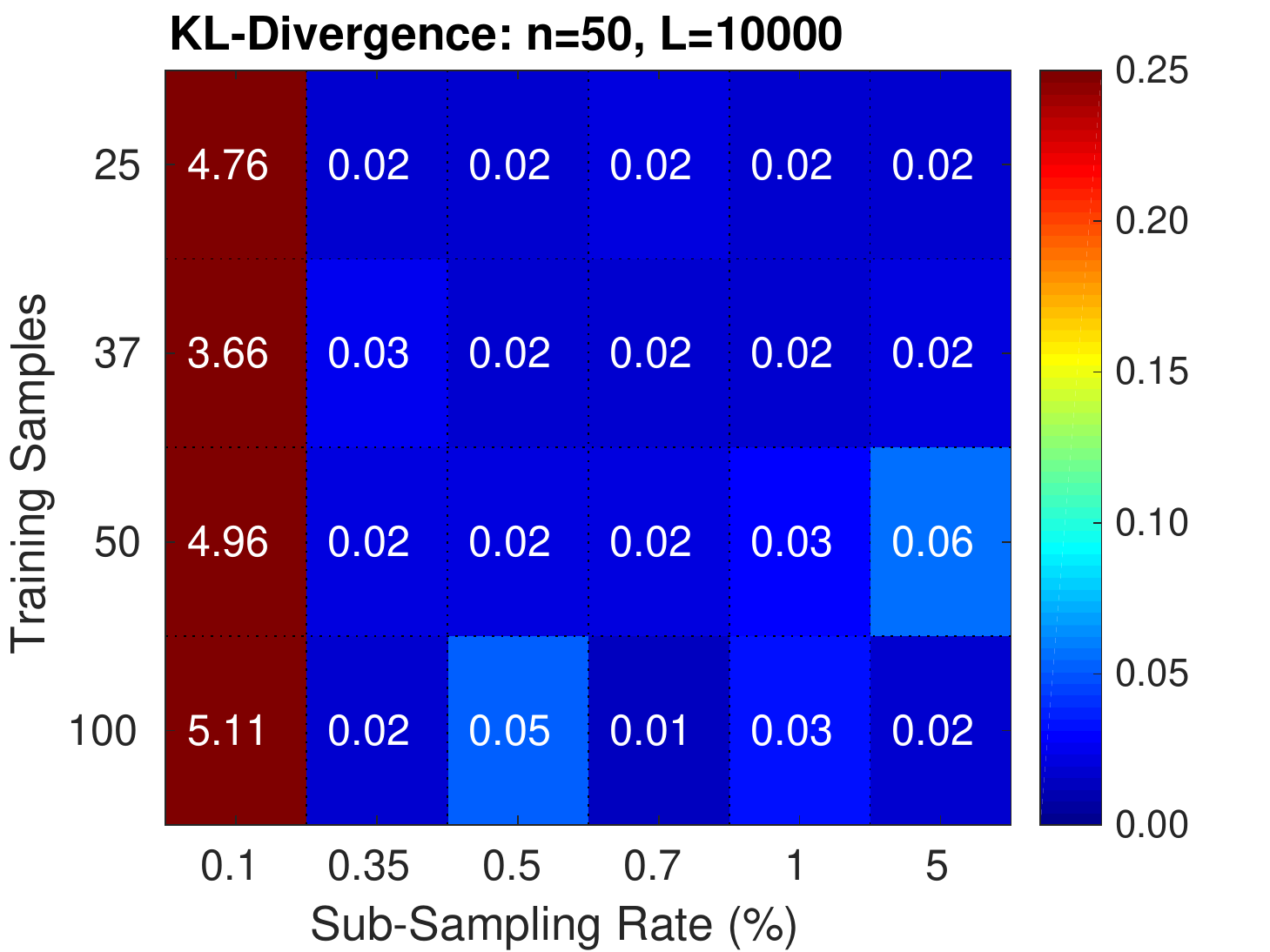}
	\includegraphics[width=0.4\textwidth]{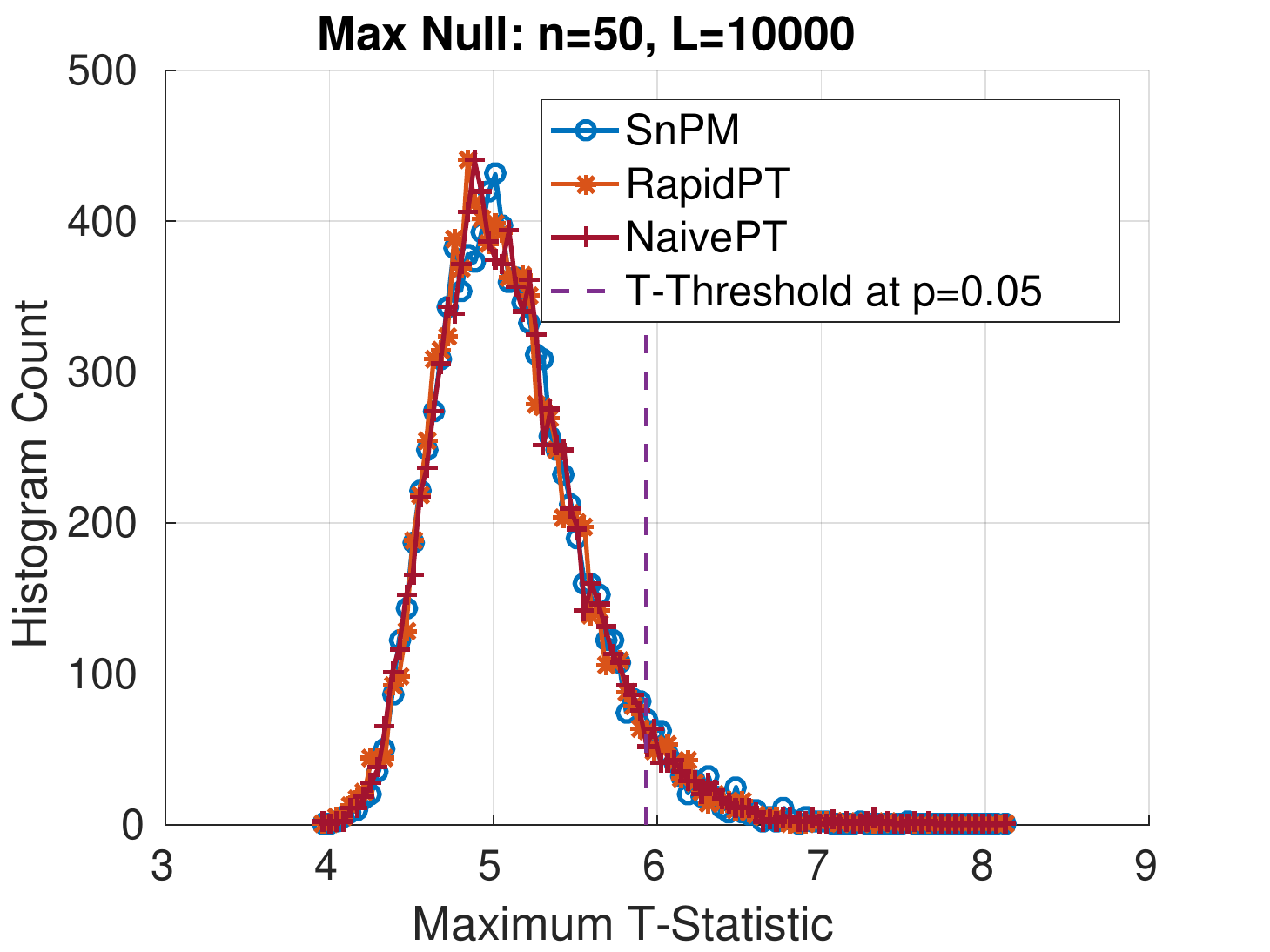}%
}%
\centerline{%
	\includegraphics[width=0.4\textwidth]{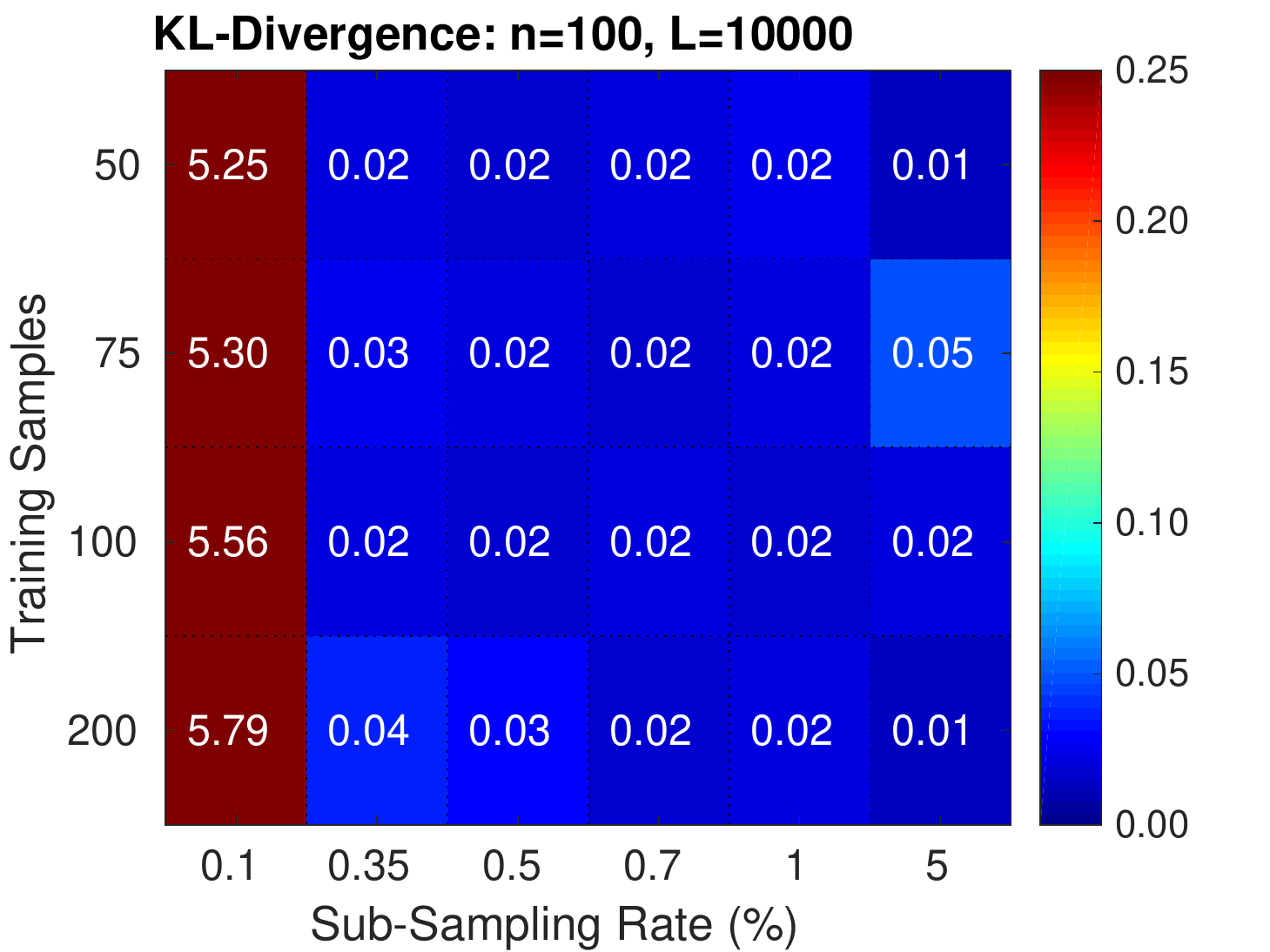}%
	\includegraphics[width=0.4\textwidth]{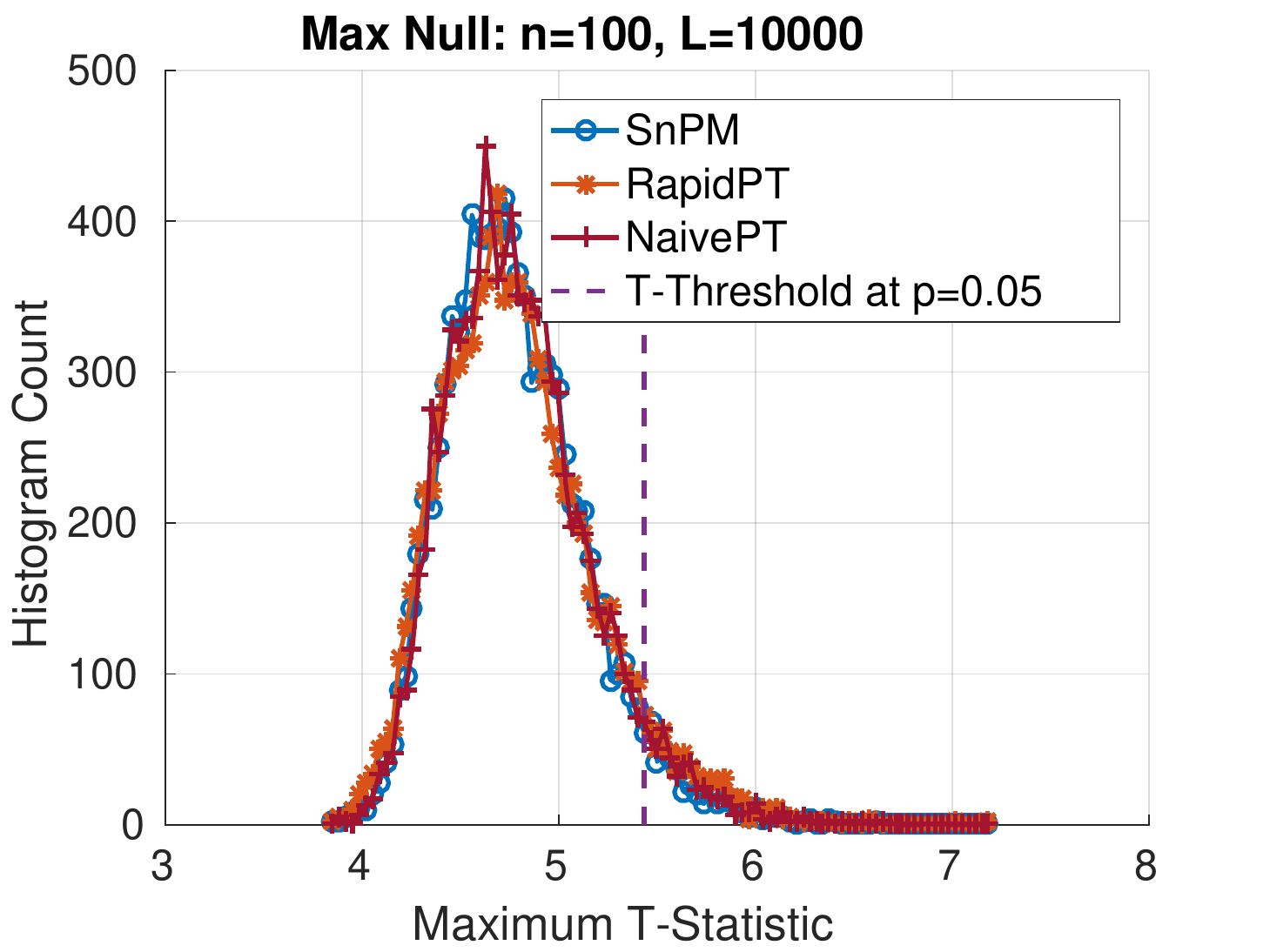}%
}%
\centerline{%
	\includegraphics[width=0.4\textwidth]{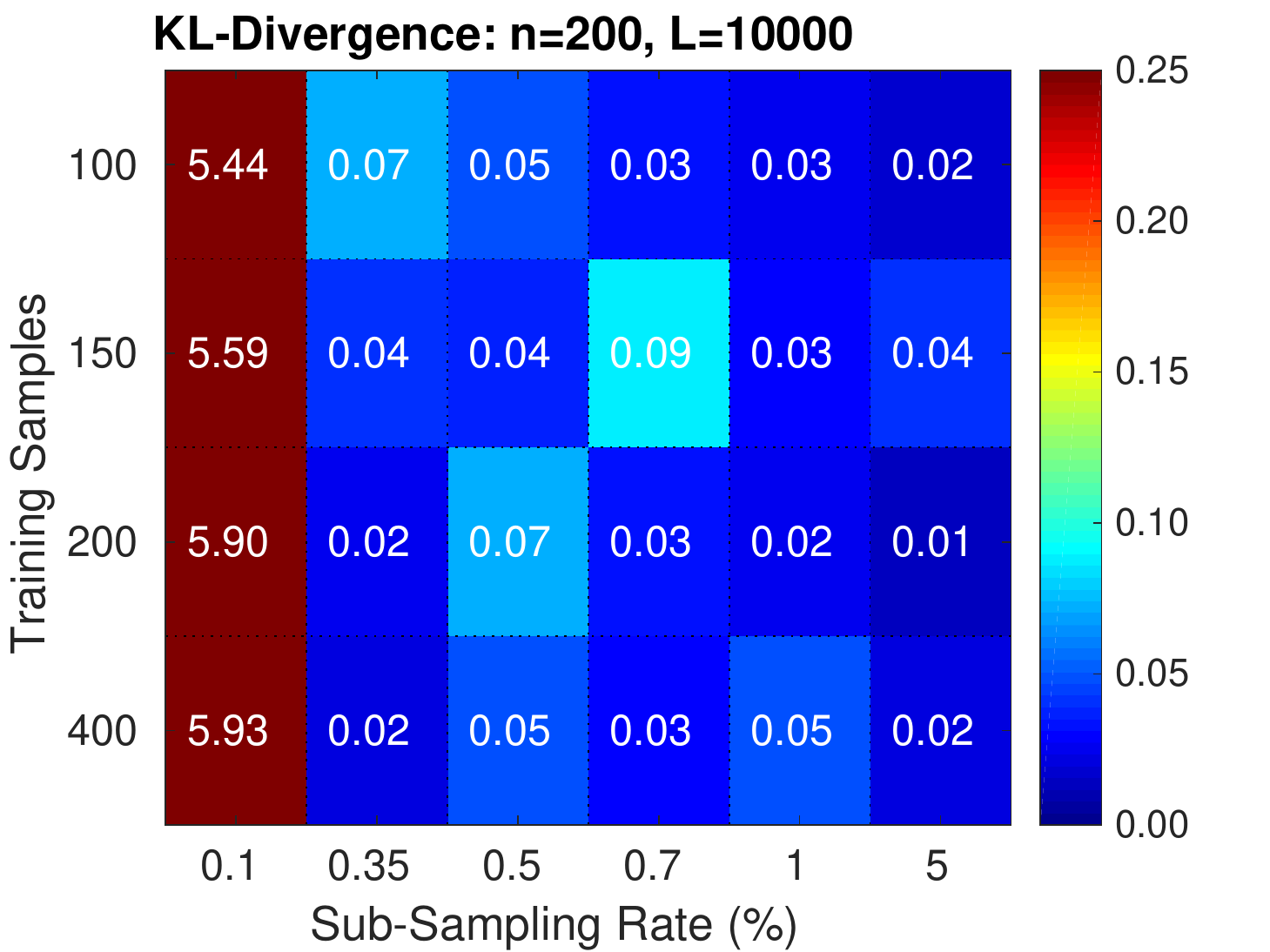}%
	\includegraphics[width=0.4\textwidth]{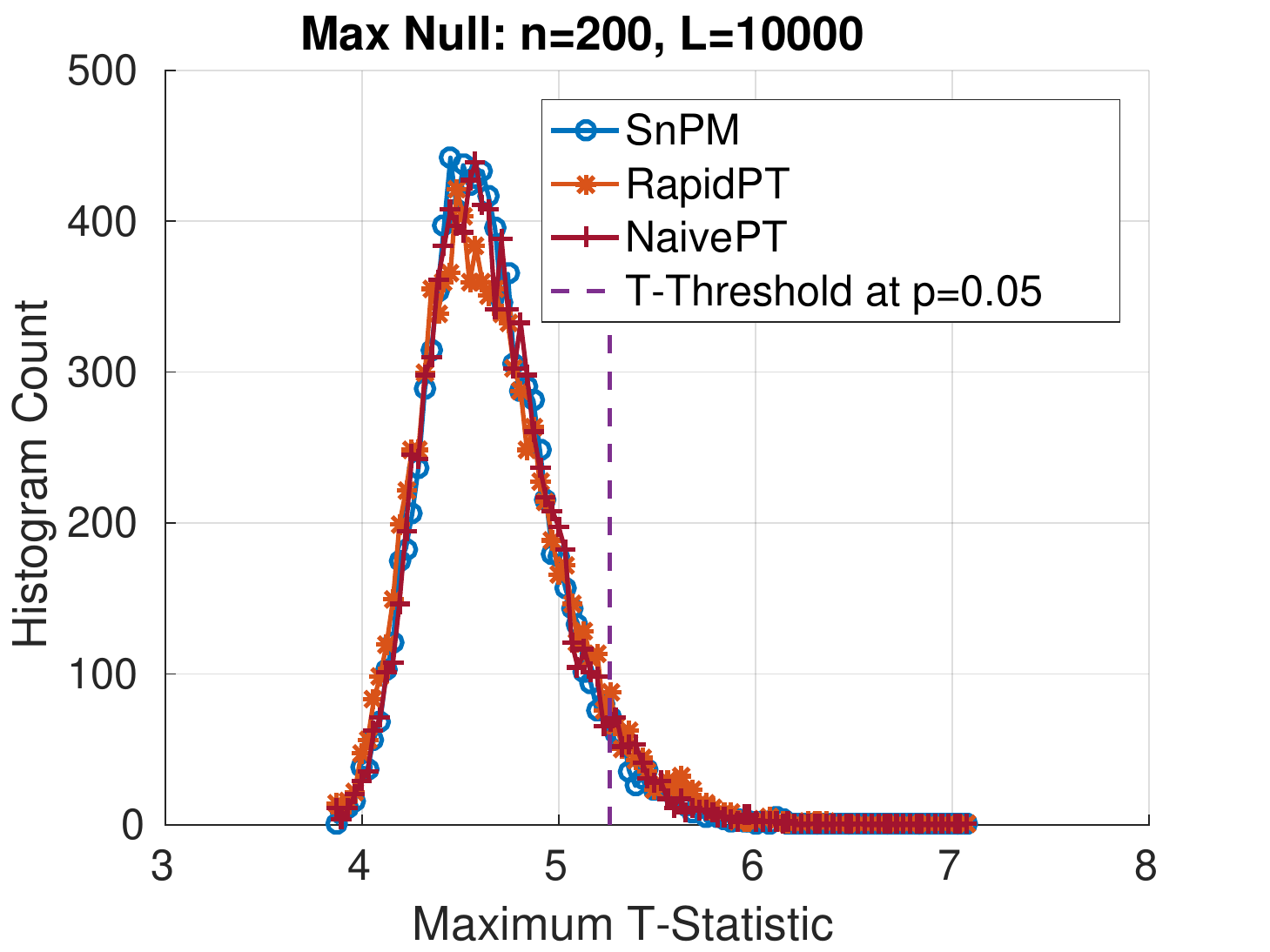}%
}%
\centerline{%
	\includegraphics[width=0.4\textwidth]{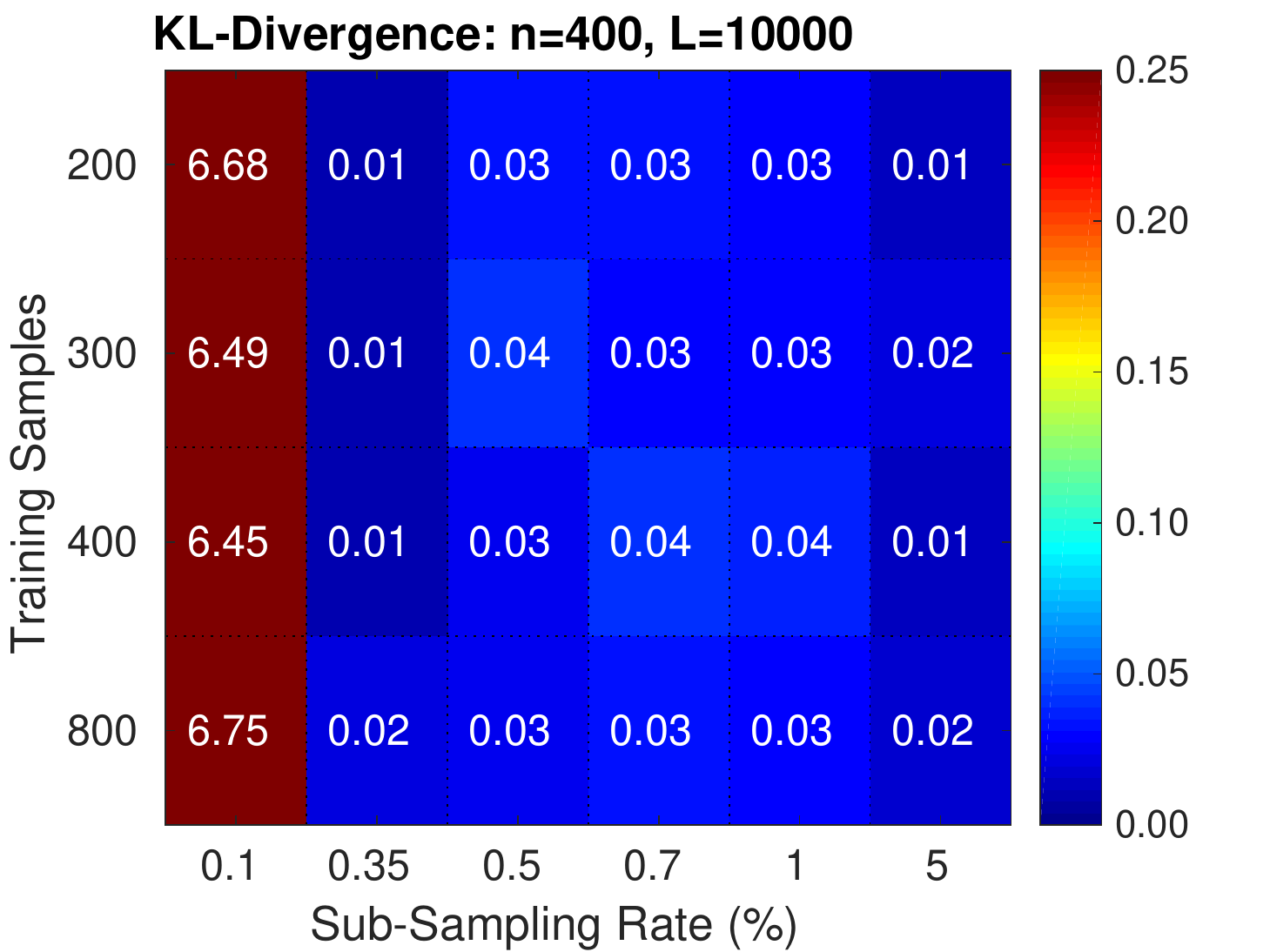}%
	\includegraphics[width=0.4\textwidth]{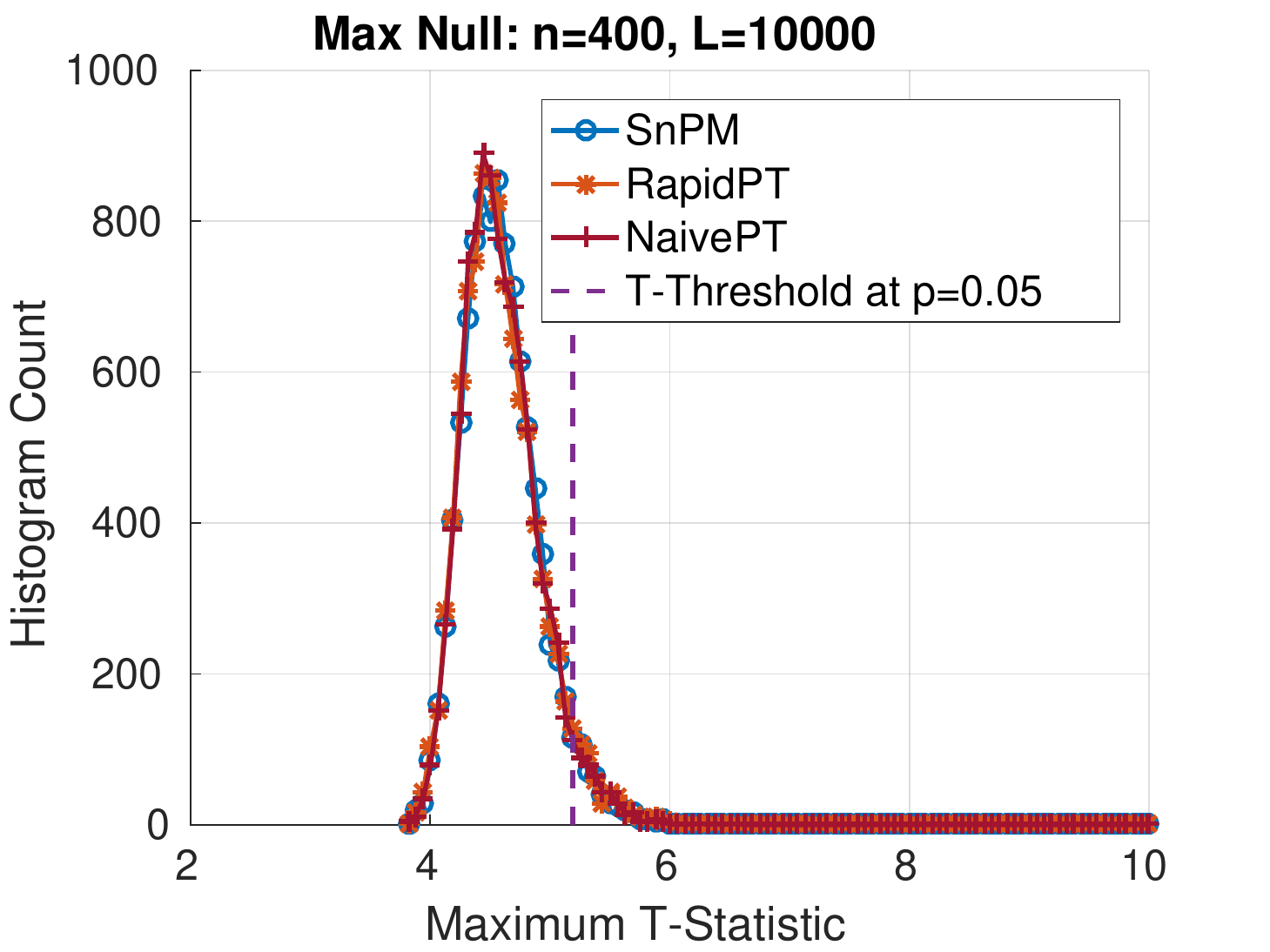}%
}%

\caption{\textit{Left:} Colormap of the KL-Divergence between the max null distributions of RapidPT and SnPM. Each colormap is associated to a run on one of the datasets and a fixed number of permutations.  The resulting KL-Divergence from 24 hyperparameter combinations is displayed on each colormap. Rows 1, 2, 3, and 4 of this figure are associated to the 50, 100, 200, and 400 subject datasets respectively. \textit{Right:} The max null distributions given by RapidPT, SnPM, and NaivePT for the hyperparameters: $L=10000$, $l=n$, and $\eta=0.35\%$.}
\label{fig:KLDivSnPM}
\end{figure}

The left column of Figure \ref{fig:KLDivSnPM} uses a colormap to summarize the KL-Divergence results obtained from comparing the max null distributions of a single run of 
SnPM versus multiple RapidPT runs with various hyperparameters. The right column of Figure \ref{fig:KLDivSnPM} puts the numbers displayed in the colormaps 
into context by showing the actual max null distributions for a single combination of hyperparameters. Each row corresponds to each of the four datasets used in our evaluations. 

The sub-sampling rate was the hyperparameter that had the most significant impact on the KL-Divergence. As shown in Figure \ref{fig:KLDivSnPM}, a sub-sampling rate of 
$0.1\%$ led to high KL-Divergence, i.e., the max null distribution was not recovered in this case. 
For every other combination of hyperparameters RapidPT was able to sample at rates as low as $0.35\%$ and still recover an accurate max null distribution. 
Most KL-Divergence values were in the $0.01-0.05$ range with some occasional values between $0.05-0.15$. However, using the max null distributions derived from only 
$2000$ permutations leads to the resulting KL-Divergence to range mainly between $0.05-0.15$, as shown in the supplementary results.

\noindent
\textit{Are we rejecting the correct null hypotheses?}

\begin{figure}[H]
\centerline{%
\includegraphics[width=0.4\textwidth]{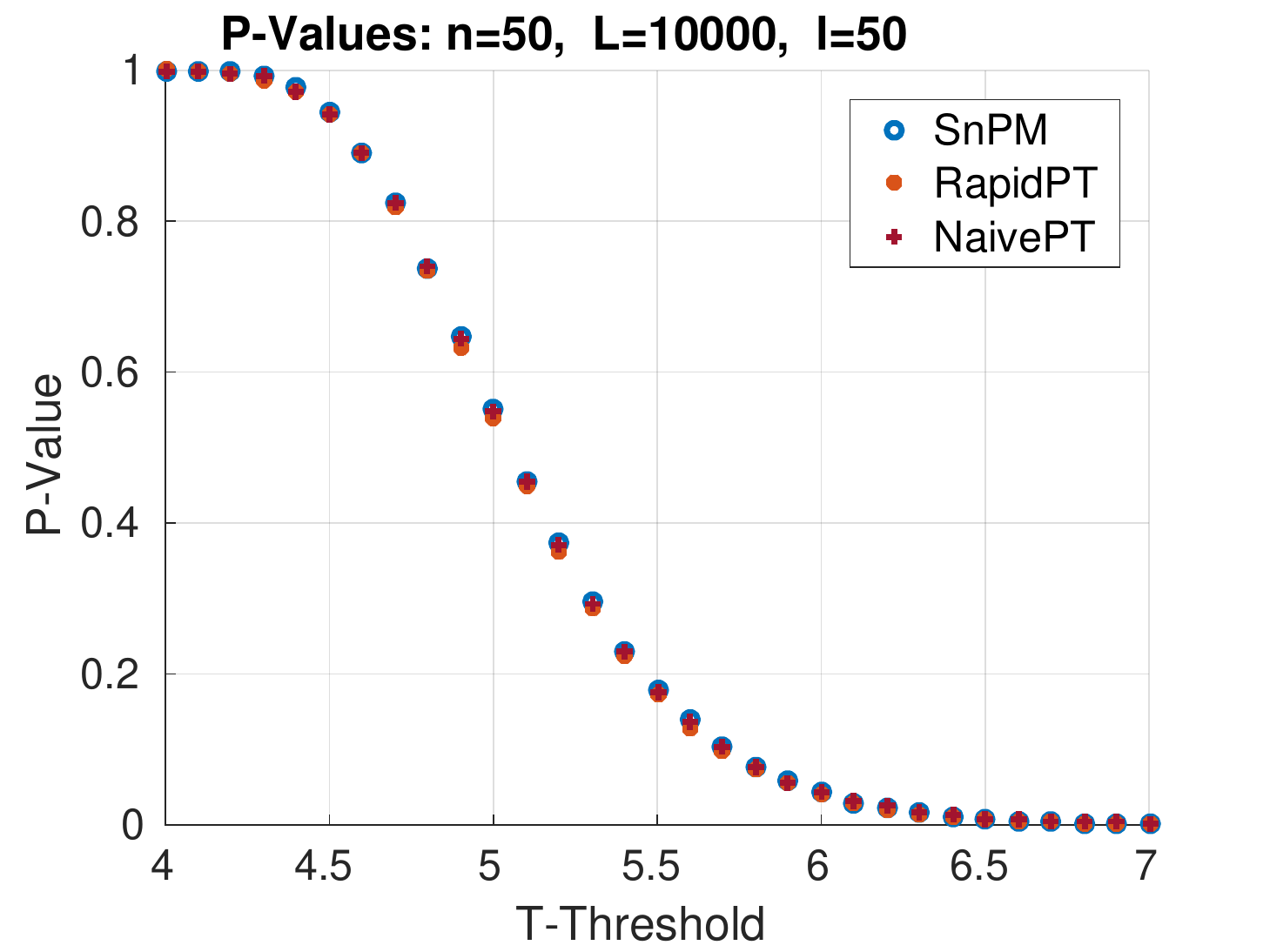}%
\includegraphics[width=0.4\textwidth]{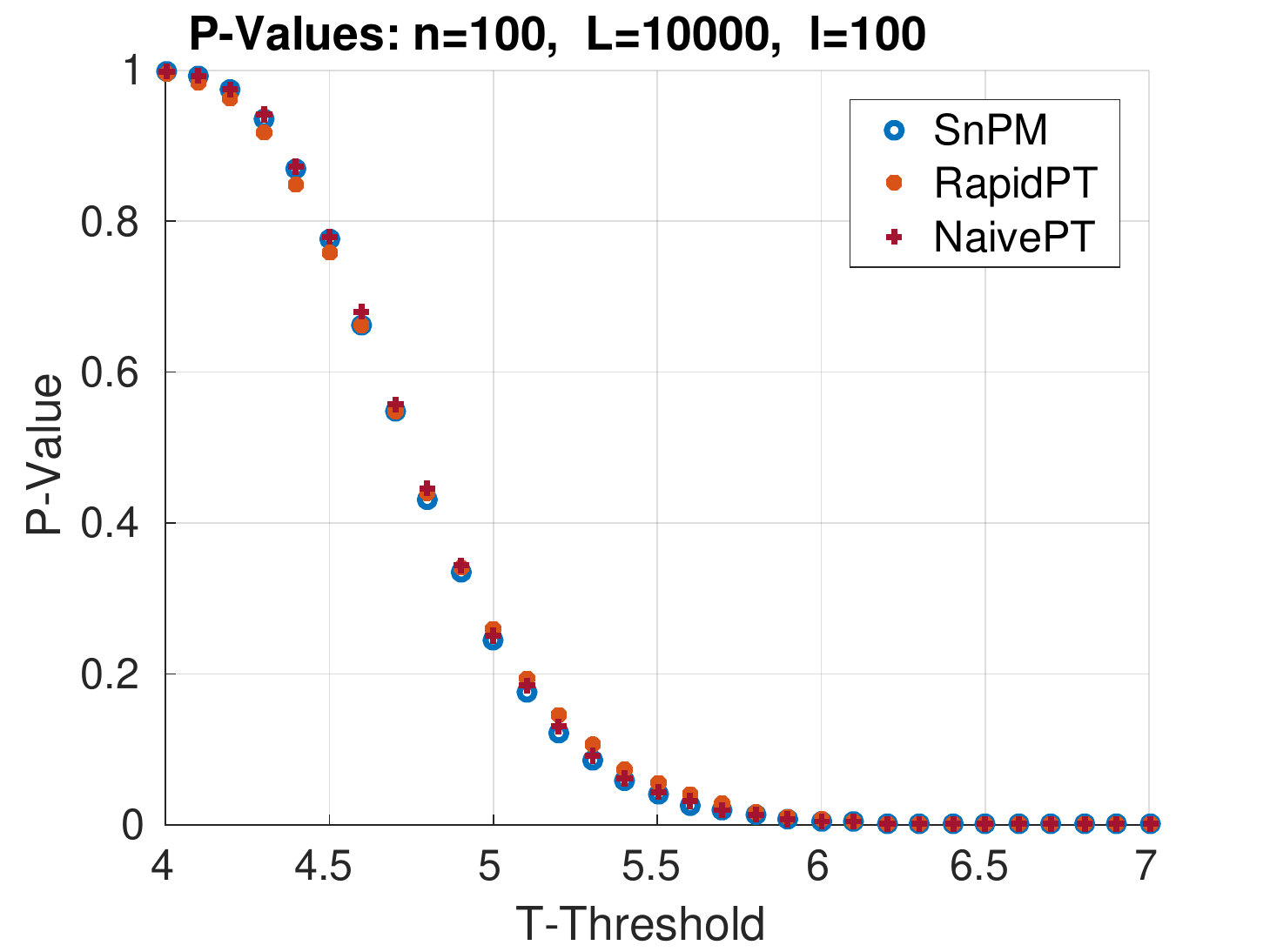}%
}%
\centerline{%
\includegraphics[width=0.4\textwidth]{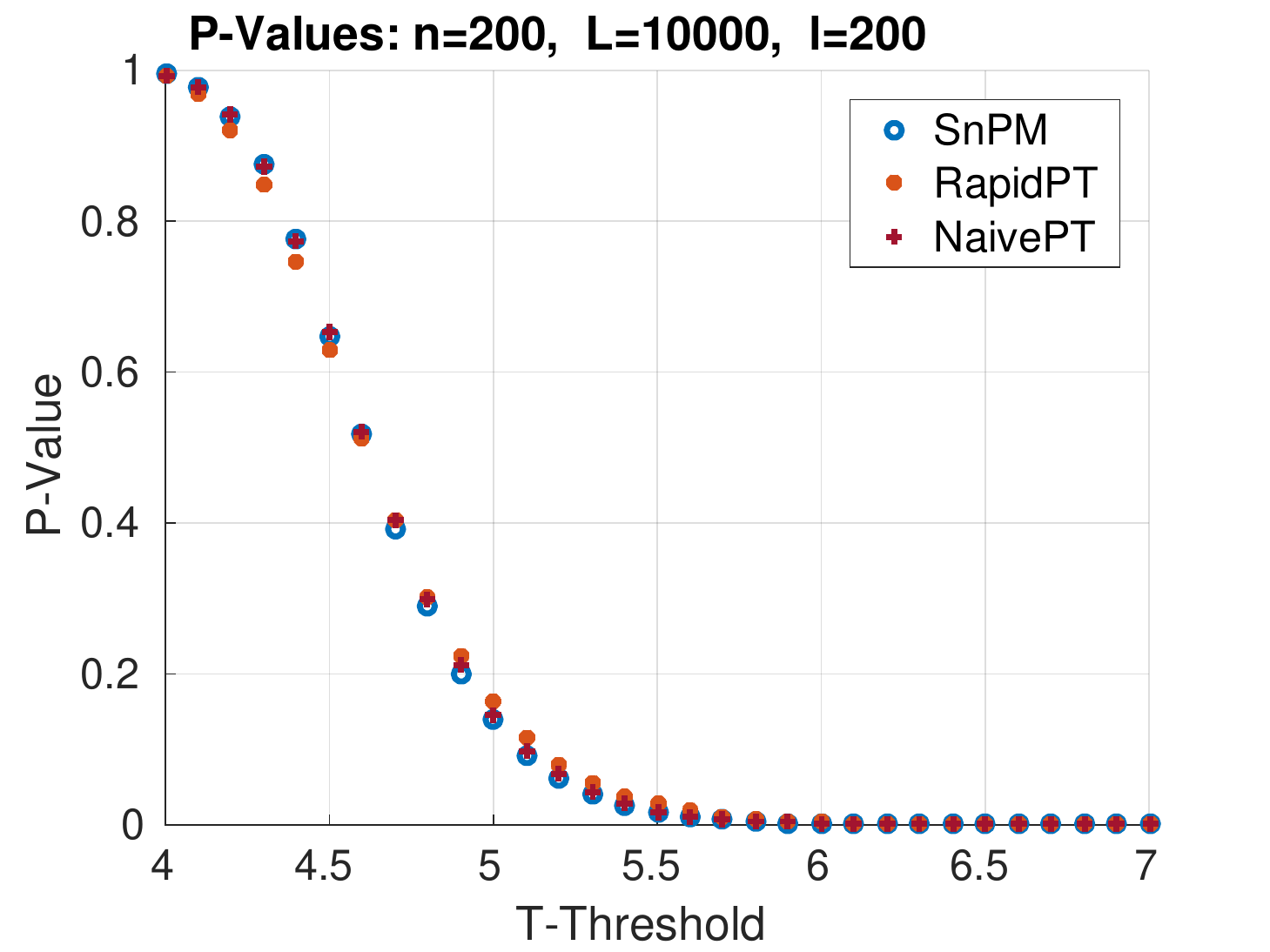}%
\includegraphics[width=0.4\textwidth]{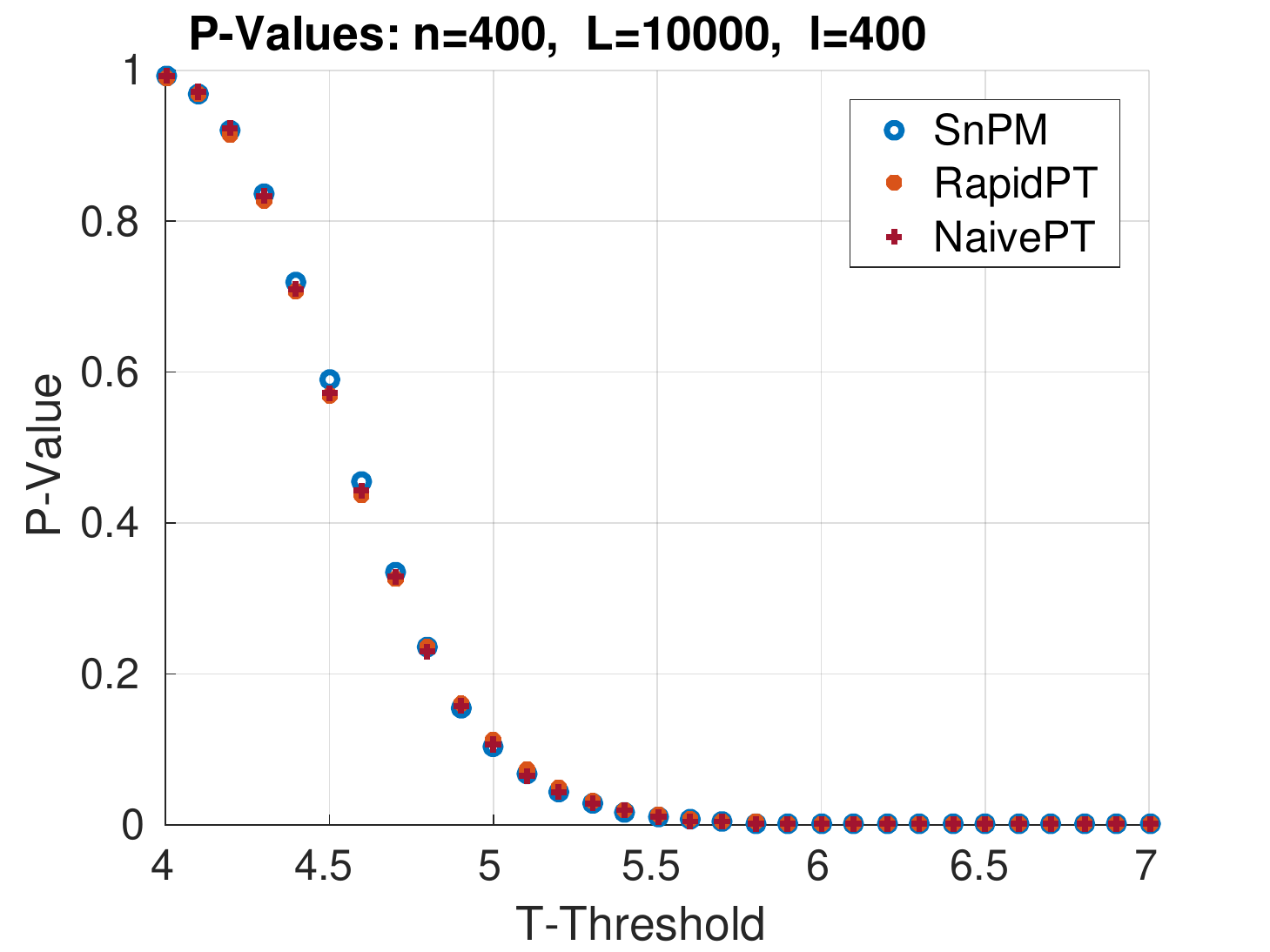}%
}%
\caption{$p$-values for SnPM, RapidPT, and NaivePT. The hyperparameters used were: $\eta = 0.35\%$, $L = 10000$, and $l = n$. The results
  in this plot were obtained from the max null distributions shown in the right hand side of Figure \ref{fig:KLDivSnPM}.}
\label{fig:PVals}
\end{figure}

\begin{figure}[H]
\centerline{%
\includegraphics[width=0.4\textwidth]{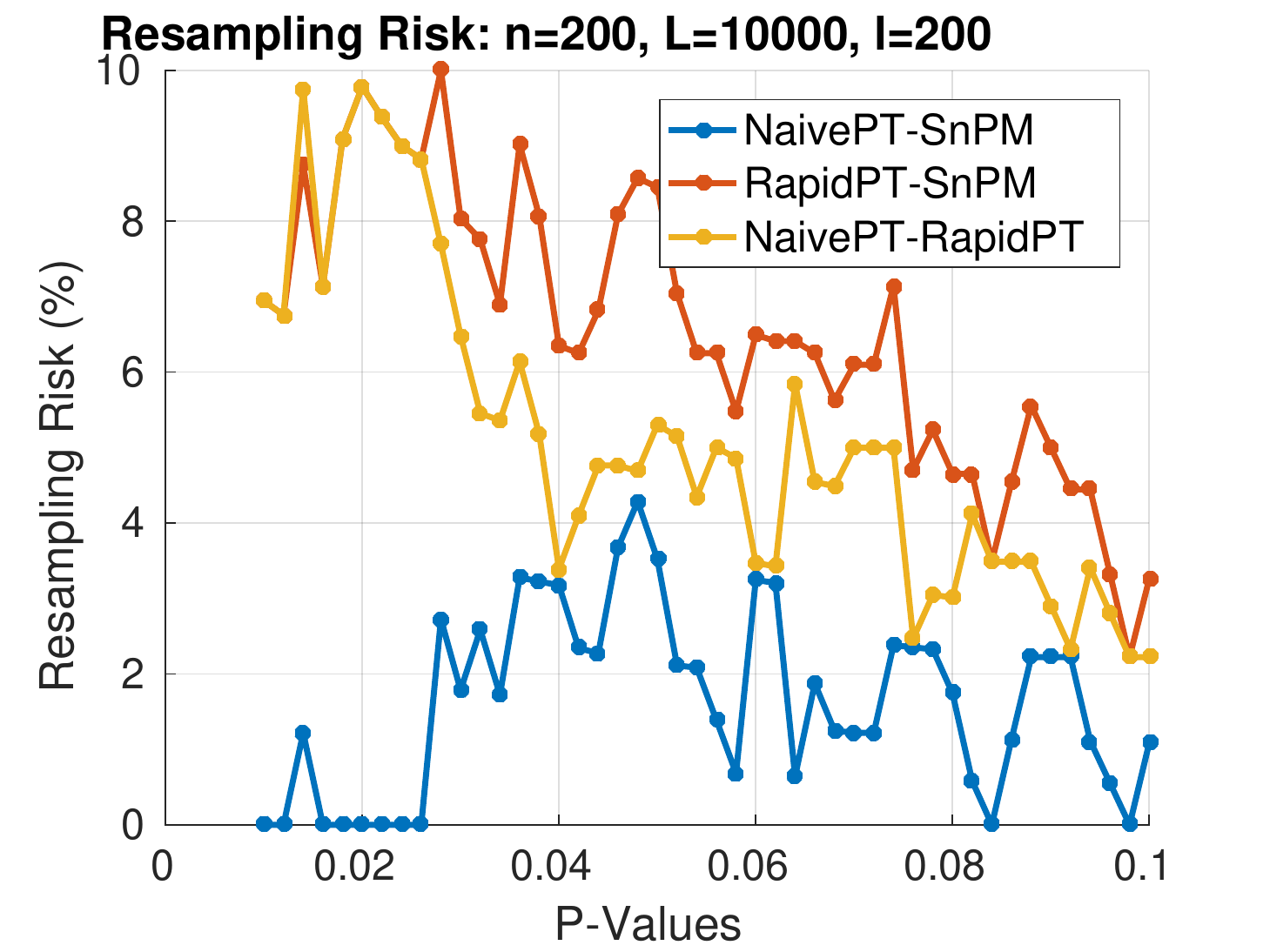}%
\includegraphics[width=0.4\textwidth]{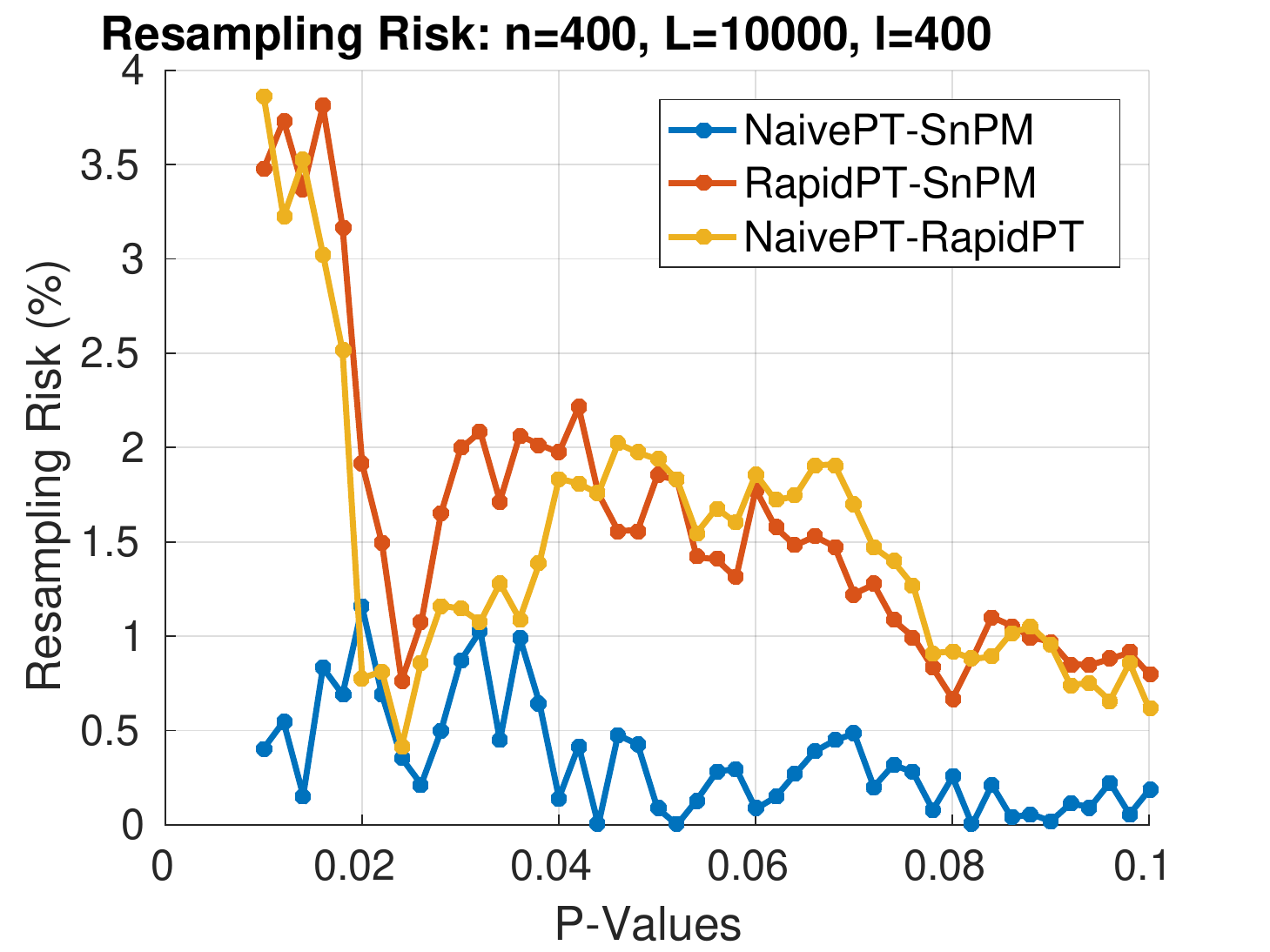}%
}%
\caption{Resampling risk of NaivePT-SnPM, RapidPT-SnPM, and NaivePT-RapidPT. The hyperparameters used were: $\eta = 0.35\%$, $L = 10000$, and $l = n$.}
\label{fig:ResamplingRisk}
\end{figure}

The test statistics obtained using the original data labels whose value exceed the $t$-threshold associated to a given $p$-value will 
correspond to the null hypothesis rejected. Figure \ref{fig:PVals} shows the resultant mapping between $t$-threshold and $p$-values for the max null 
distribution for a given set of hyperparameters. It is evident that the difference across methods is minimal. Moreover,  Figure \ref{fig:PVals} 
shows that low $p$-values ($p < 0.1$), which are the main object of interest, show the lowest differences. However, despite the low percent differences 
between the $p$-values, in the larger datasets ($100$, $200$, and $400$ subjects) RapidPT consistently yields slightly more conservative $p$-values near the 
tails of the distribution. Nonetheless, Figure \ref{fig:ResamplingRisk} shows that the resampling risk between RapidPT and the two baselines remains very 
close to the resampling risk between both baselines. In practice, these plots show that RapidPT will reject the null hypothesis for a slightly lower number 
of voxels than SnPM or NaivePT. 

\begin{figure}[H]
\centering
    \textbf{ADNI Statistic Maps}
	\\[-1pt]
	\subfloat[SnPM Statistic Maps. Slice -44 to 58 from left to right top to bottom.]{\label{fig:sidesnpm1}{\includegraphics[width=0.9\textwidth]{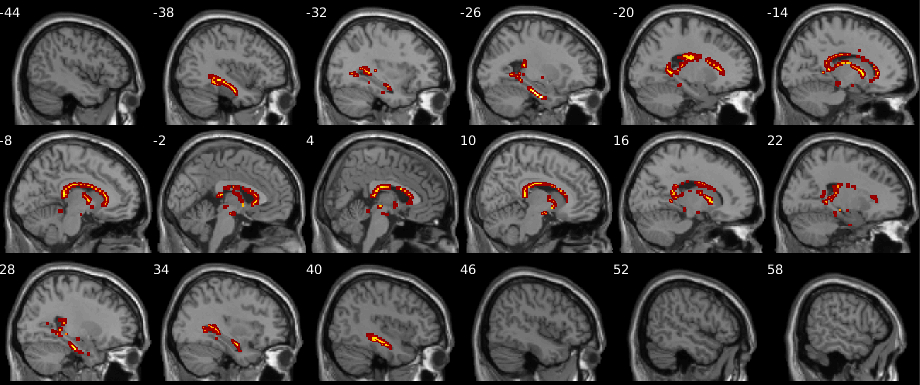}}}
	\\[-1pt]
	\subfloat[RapidPT Statistic Maps. Slice -44 to 58 from left to right top to bottom. $\eta=0.5\%$,$l=200$.]{\label{fig:siderpt1}{\includegraphics[width=0.9\textwidth]{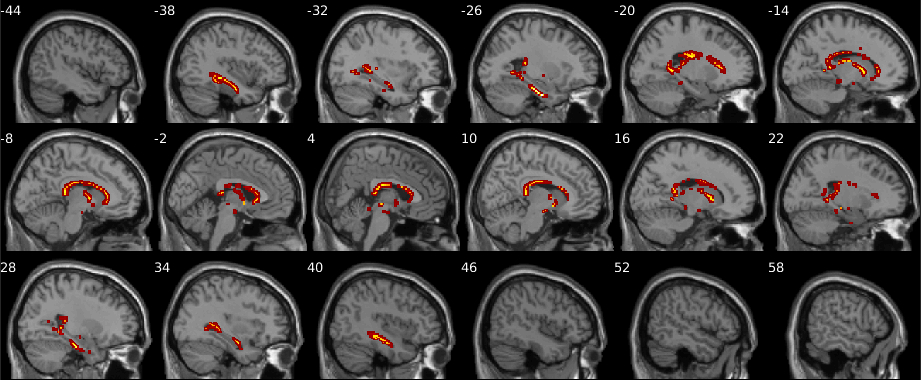}}}
	\\[-2ex]
	\subfloat[SnPM Report]{\label{fig:reportsnpm}{\includegraphics[width=0.5\textwidth]{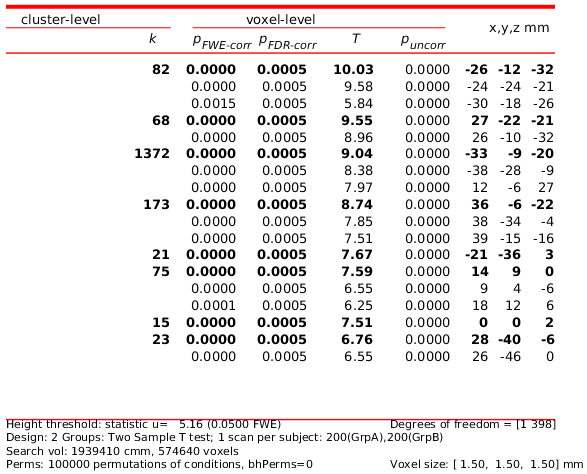}}}
	\subfloat[RapidPT Report]{\label{fig:reportrpm}{\includegraphics[width=0.5\textwidth]{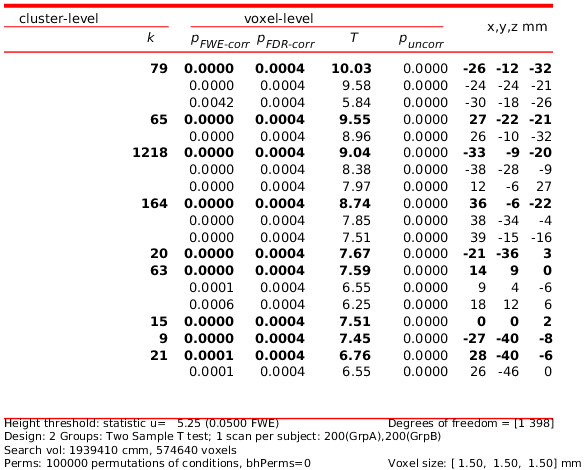}}}


\caption{Thresholded FWER corrected statistical maps at ($\alpha = 0.05$) with the $n=400$ dataset. The hyperparameters used were: $\eta=0.5\%$, $l=n$, and $L=100000$. The images show the test statistics for which the null was rejected in SnPM (top) and RapidPT (bottom). The tables show a numerical summary of the images. The columns refer to: $k$ - cluster size, $p_{FWE-corr}$ - corrected $p$-values, $T$ - max cluster t-statistic. $p_{FWE-corr}$ that appears as $0.0000$ are $1e^{-5}$. These results were obtained from a run with the hyperparameters specified in the title.}
\label{fig:pmaps400}
\end{figure}

\begin{figure}[H]
\centering
    \textbf{ADNI Statistic Maps}
	\\[-1pt]
	\subfloat[SnPM P-Map. Slice -44 to 58 from left to right top to bottom.]{\label{fig:sidesnpm1}{\includegraphics[width=0.9\textwidth]{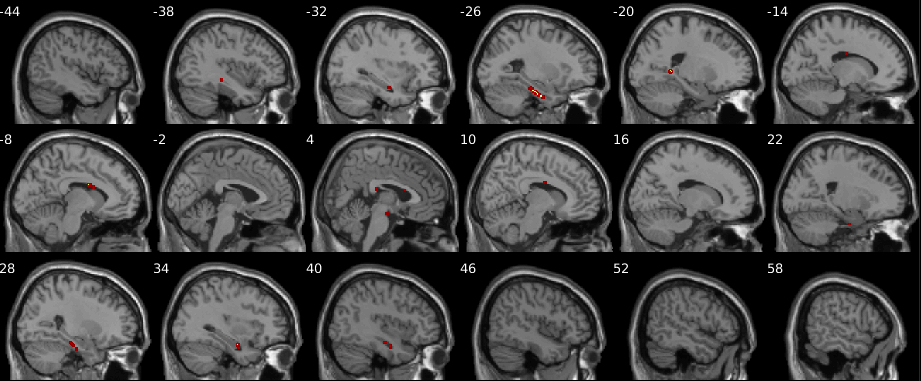}}}
	\\[-1pt]
	\subfloat[RapidPT P-Map. Slice -44 to 58 from left to right top to bottom. $\eta=0.5\%$,$l=200$.]{\label{fig:siderpt1}{\includegraphics[width=0.9\textwidth]{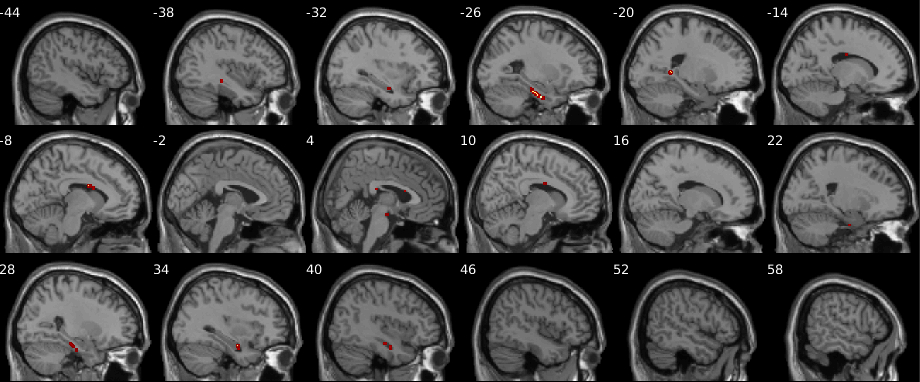}}}
	\\[-2ex]
	\subfloat[SnPM Report]{\label{fig:reportsnpm}{\includegraphics[width=0.5\textwidth]{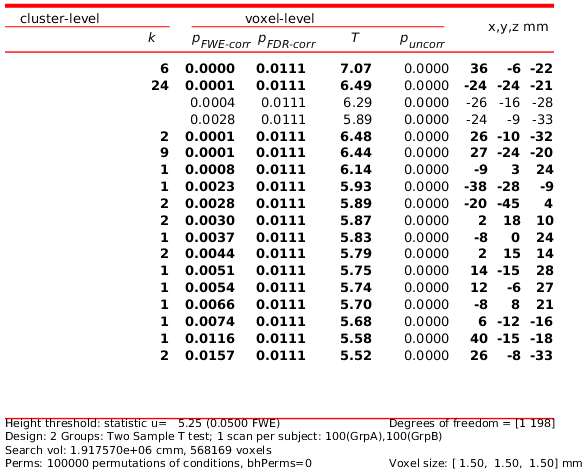}}}
	\subfloat[RapidPT Report]{\label{fig:reportrpm}{\includegraphics[width=0.5\textwidth]{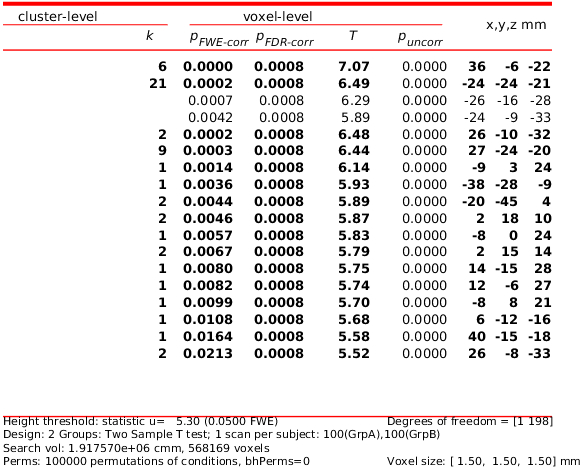}}}

\caption{Thresholded FWER corrected statistical maps at ($\alpha = 0.05$) with the $n=200$ dataset. The hyperparameters used were: $\eta=0.5\%$, $l=n$, and $L=100000$.  The images show the test statistics for which the null was rejected in SnPM (top) and RapidPT (bottom). The tables show a numerical summary of the images. The columns refer to: $k$ - cluster size, $p_{FWE-corr}$ - corrected $p$-values, $T$ - max cluster t-statistic. $p_{FWE-corr}$ that appears as $0.0000$ are $1e^{-5}$. These results were obtained from a run with the hyperparameters specified in the title.}
\label{fig:pmaps200}
\end{figure}

Despite the slight difference in thresholds, the actual brain regions whose null hypotheses were rejected consistently match between both methods as shown in Figures \ref{fig:pmaps400} and \ref{fig:pmaps200}. Additionally, the regions picked up by both RapidPT and SnPM in Figure \ref{fig:pmaps400} correspond to the {\it Hippocampus} -- which is one of the primary structural brain imaging region that corresponds to the signature of cognitive decay at the onset of Alzheimer's disease. The regions in Figure \ref{fig:pmaps200} contain a subset of the brain regions in Figure \ref{fig:pmaps400} which is expected from the thresholds shown in the right column of Figure \ref{fig:KLDivSnPM}.

\subsection{Runtime Performance} \label{sec:performance}

\subsubsection{Effect of hyperparameters on the speed of RapidPT}

\begin{figure}[h]

 \centerline{
 	\includegraphics[width=0.25\textwidth]{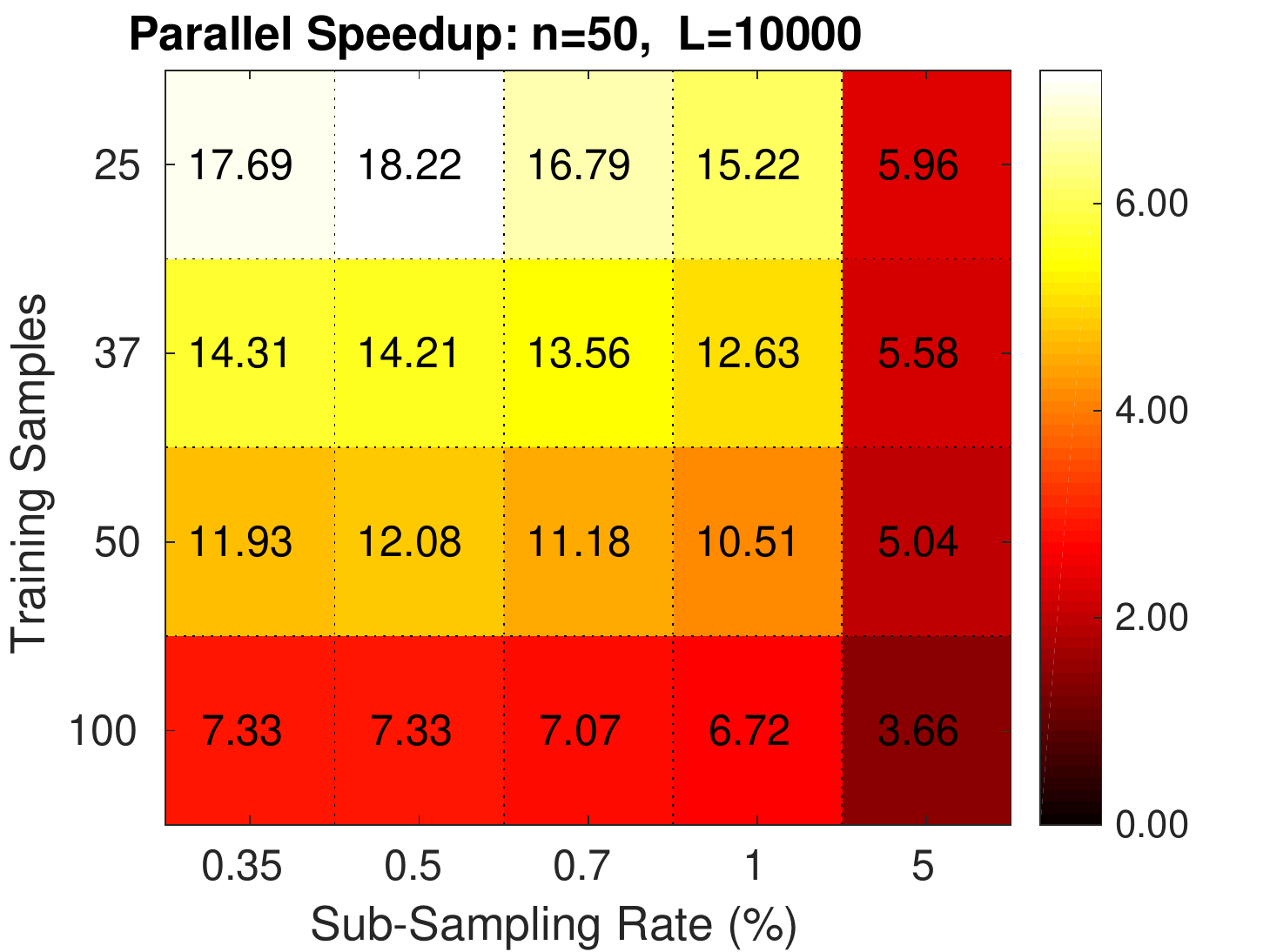}
 	\includegraphics[width=0.25\textwidth]{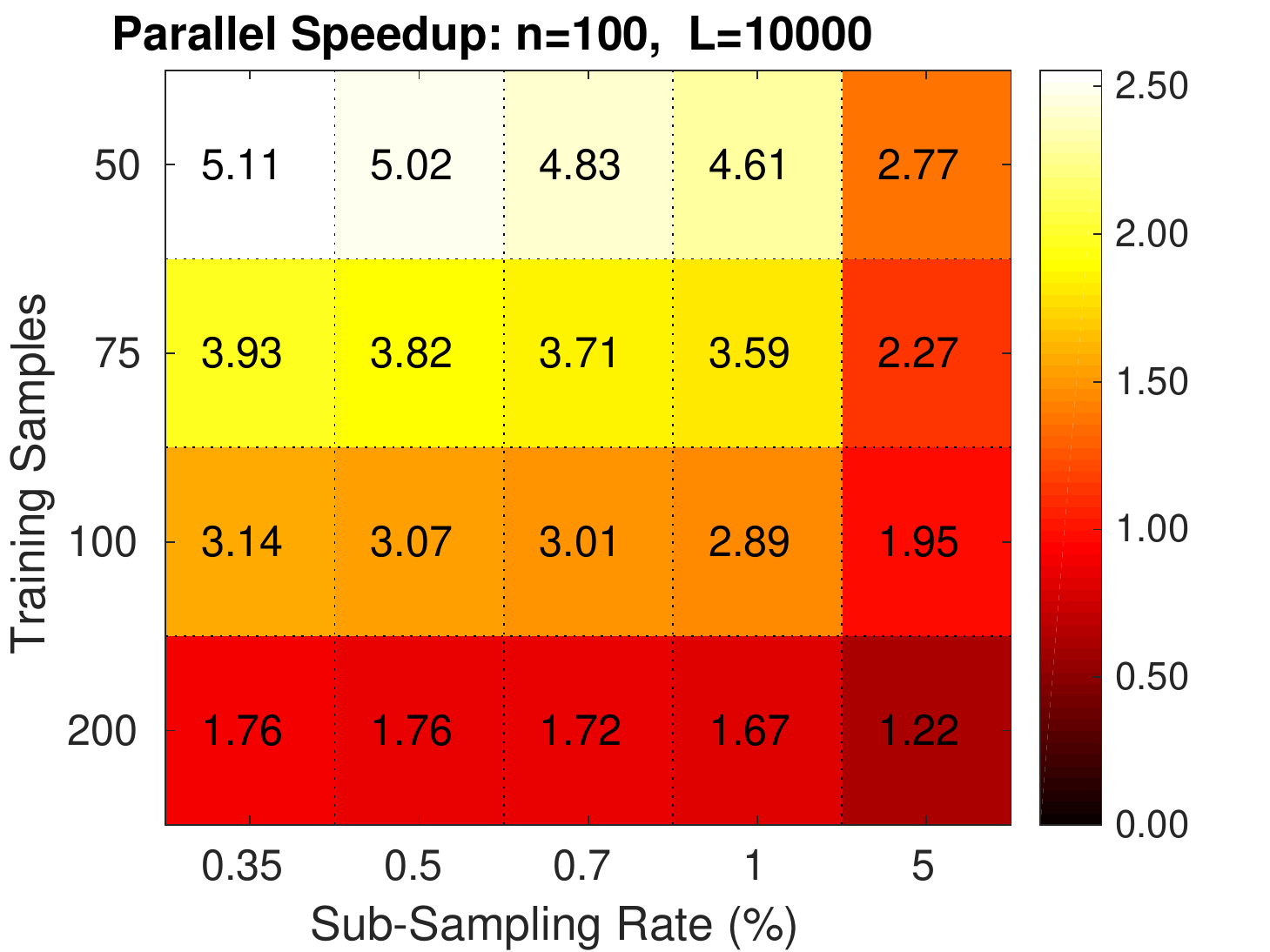}
 	\includegraphics[width=0.25\textwidth]{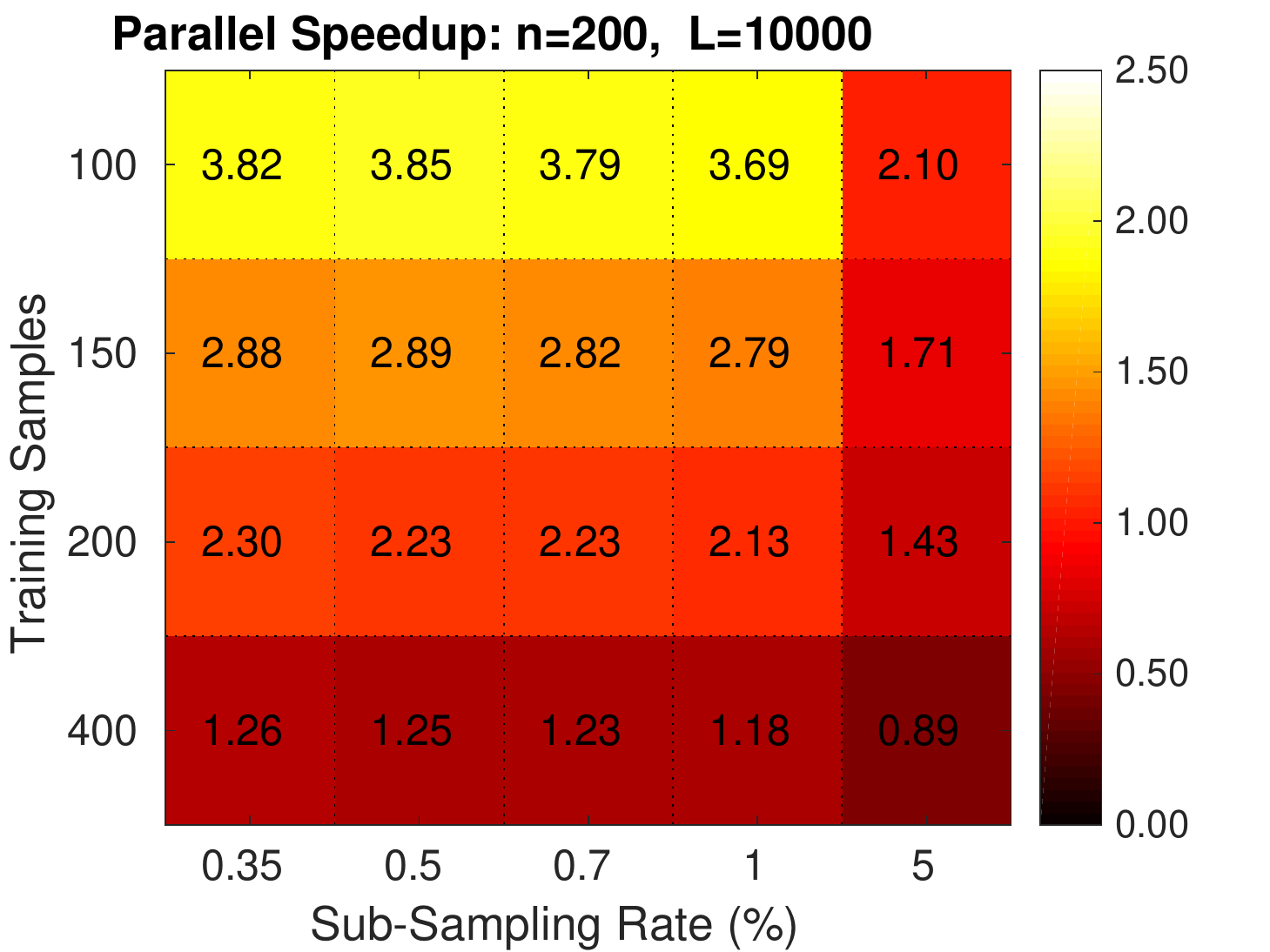}
 	\includegraphics[width=0.25\textwidth]{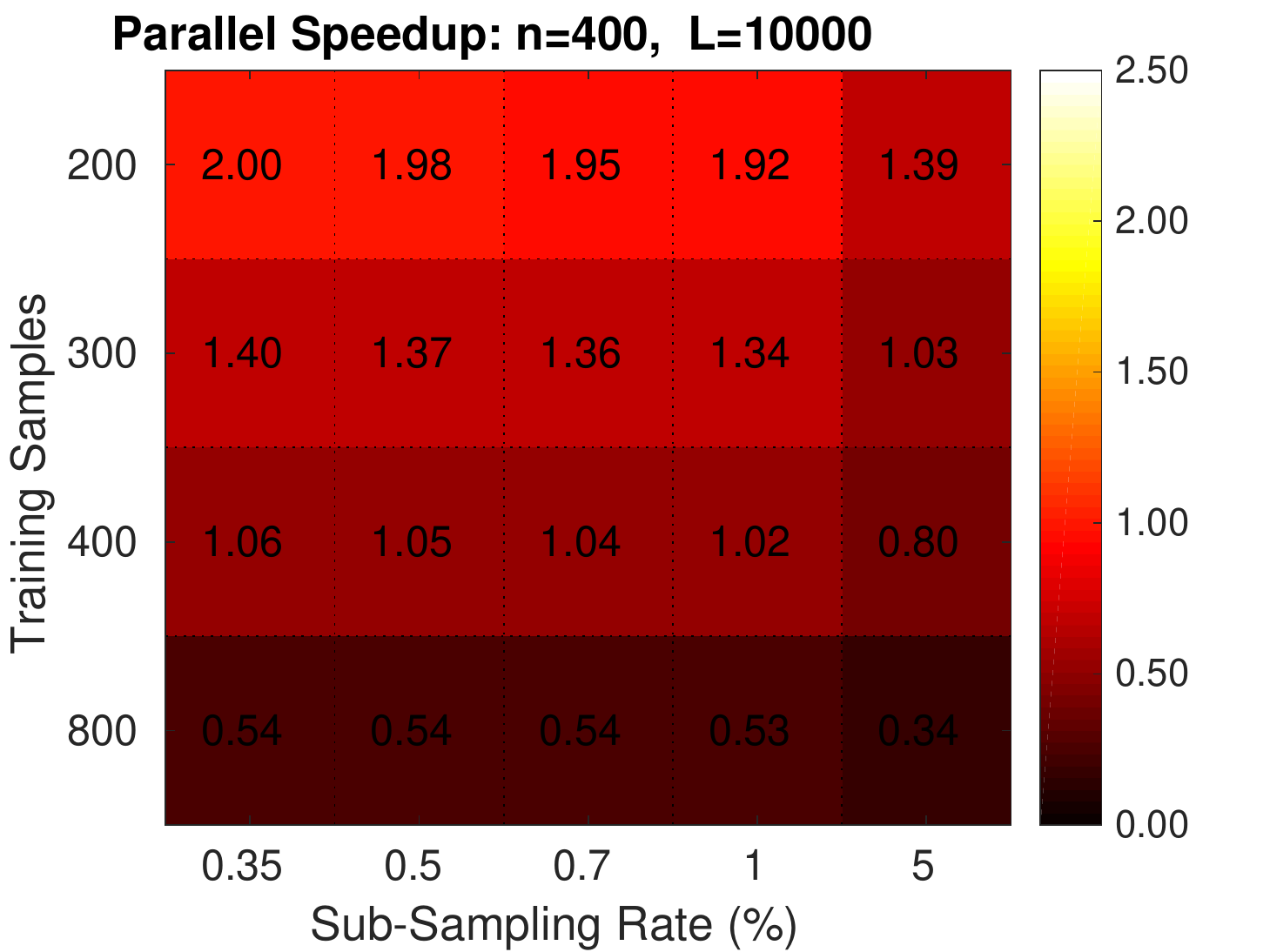}
 }
 \centerline{
 	\includegraphics[width=0.25\textwidth]{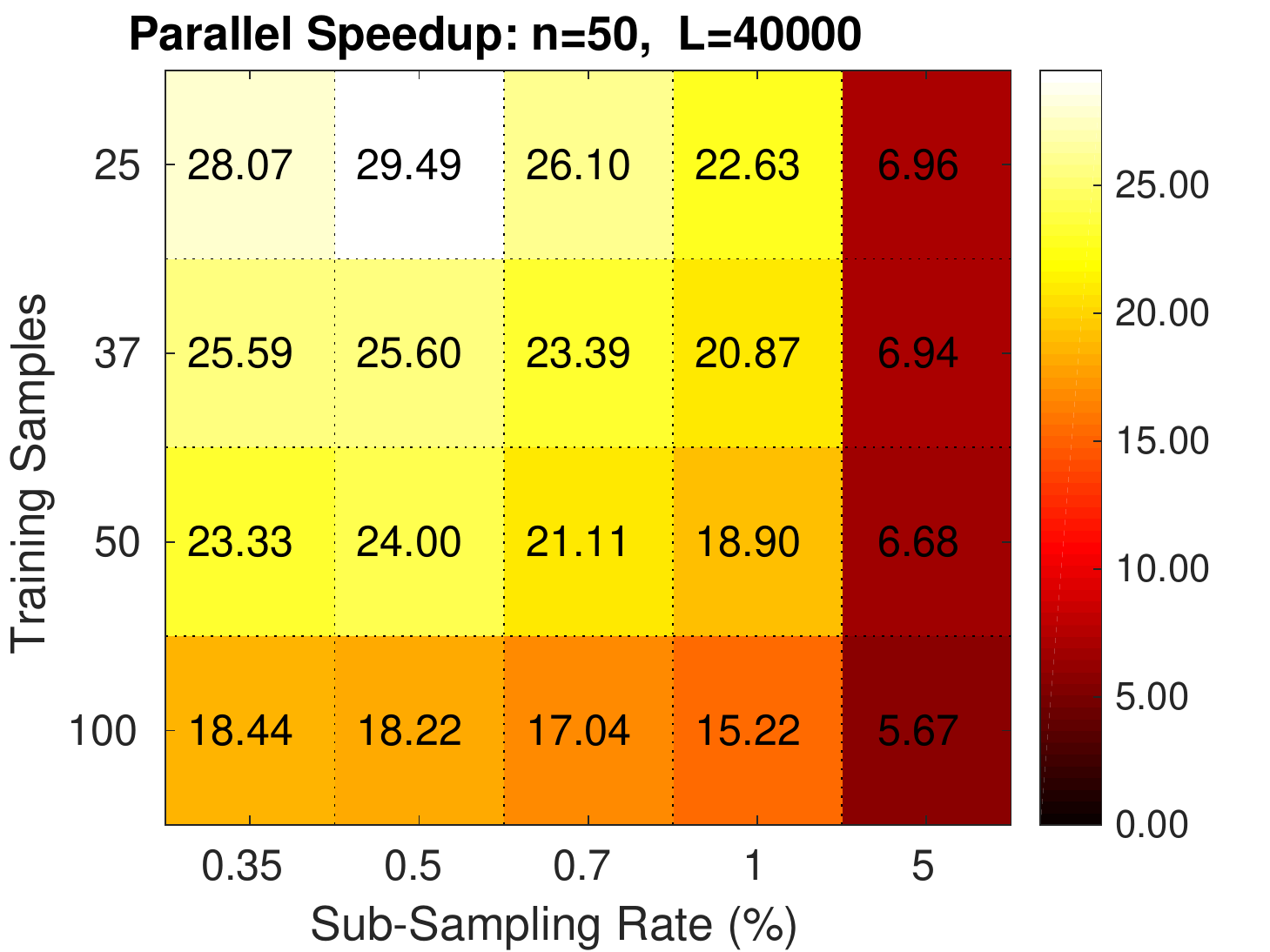}
 	\includegraphics[width=0.25\textwidth]{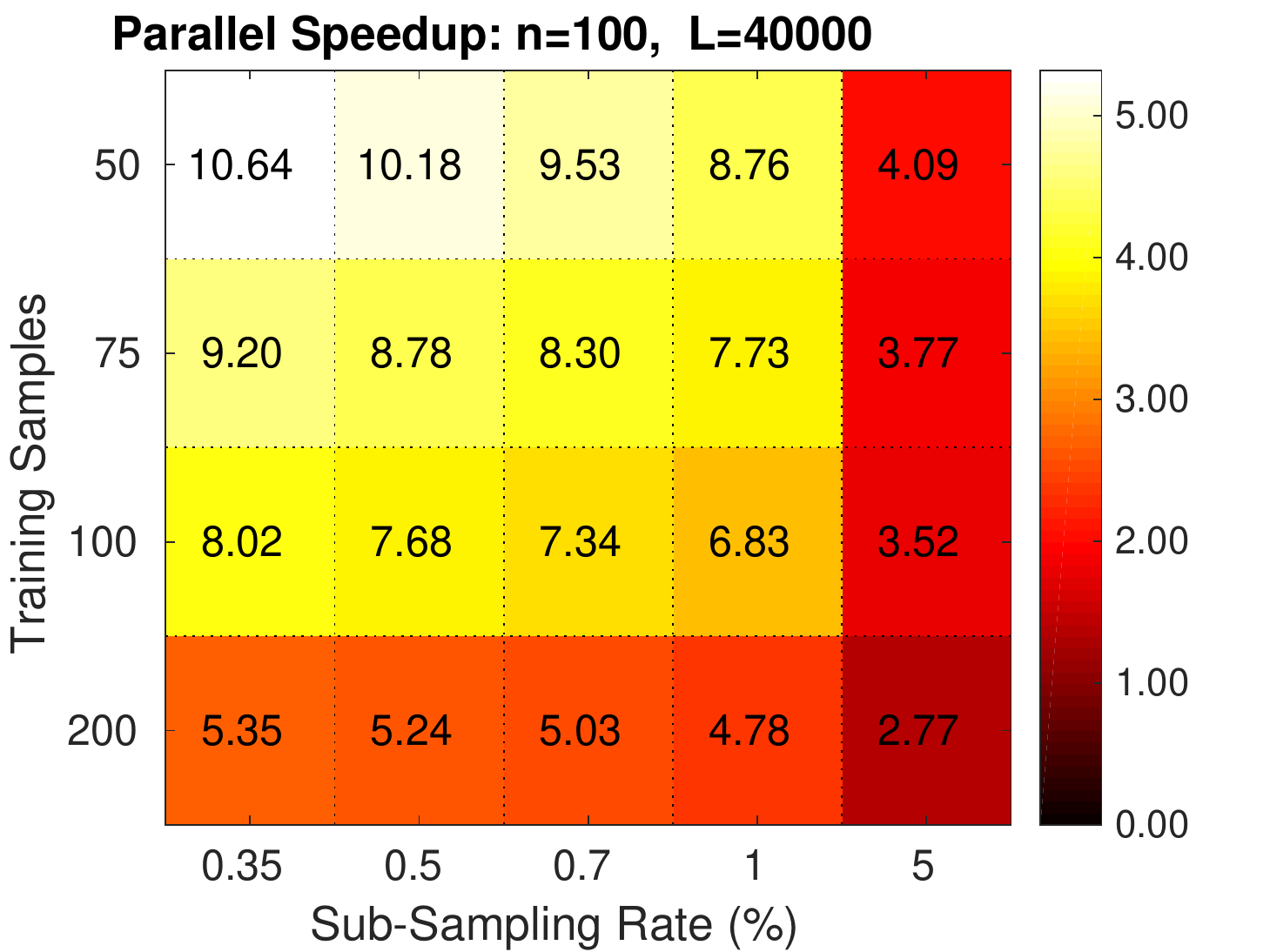}
 	\includegraphics[width=0.25\textwidth]{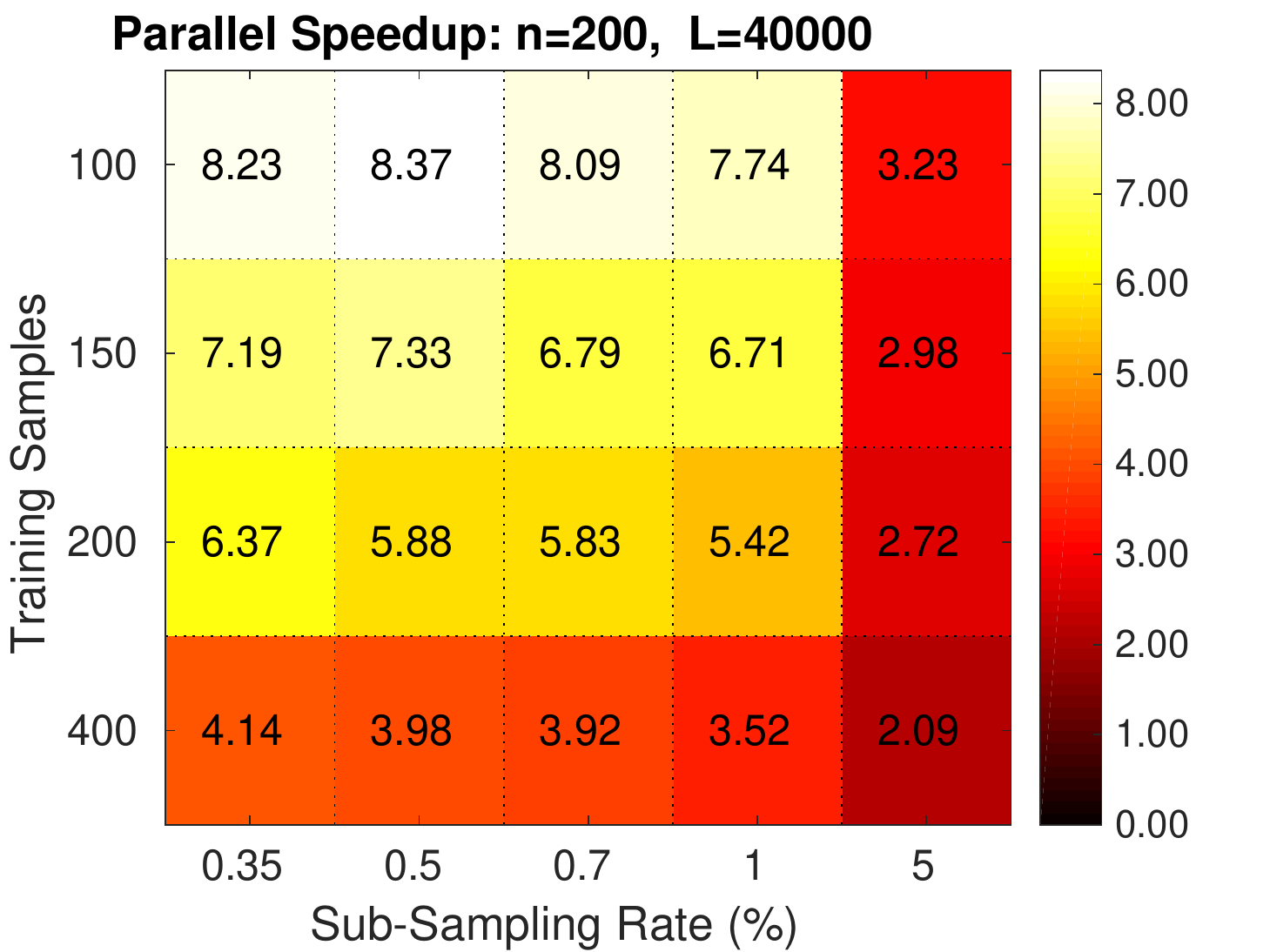}
 	\includegraphics[width=0.25\textwidth]{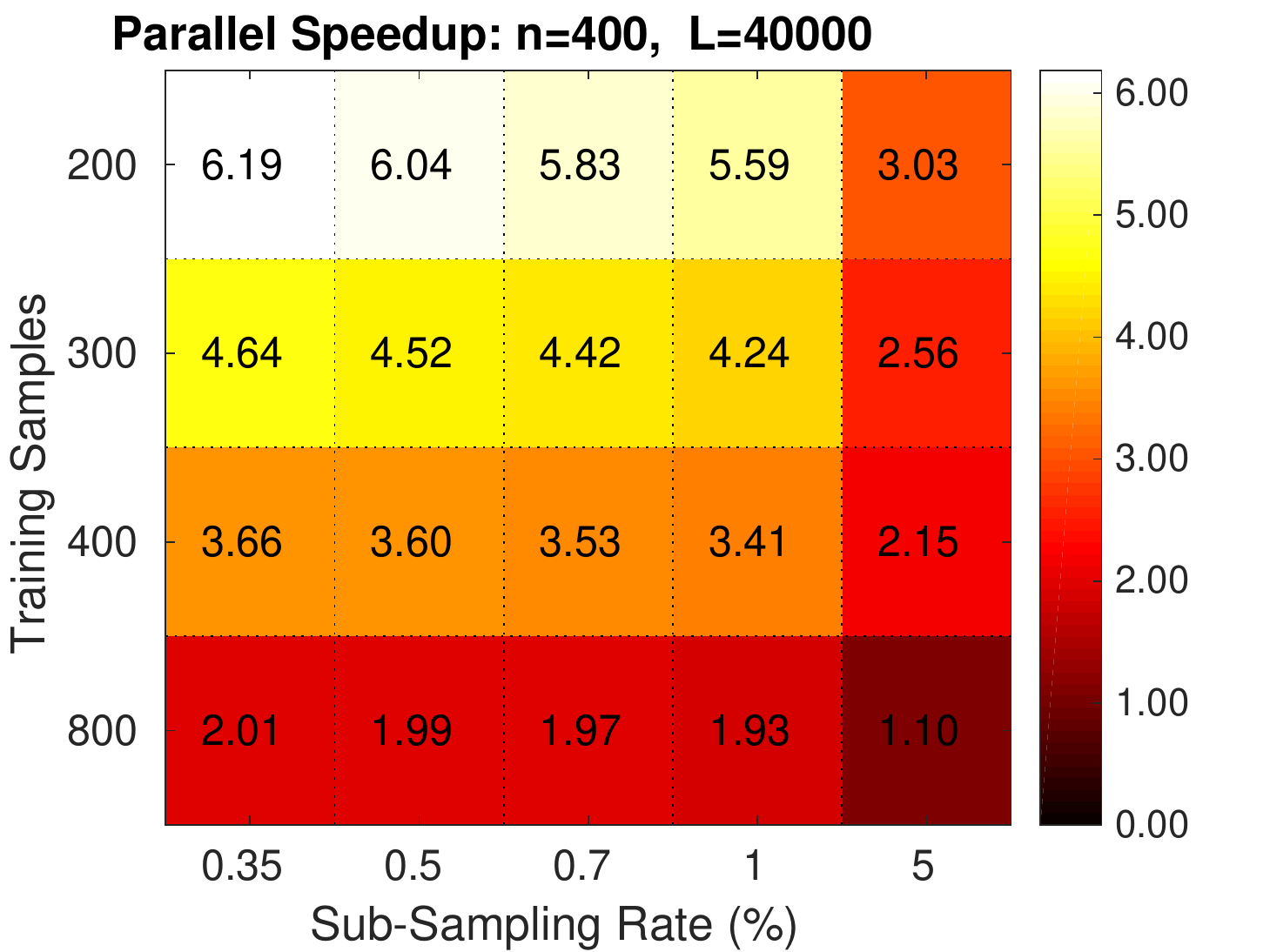}
 }
 \centerline{
 	\includegraphics[width=0.25\textwidth]{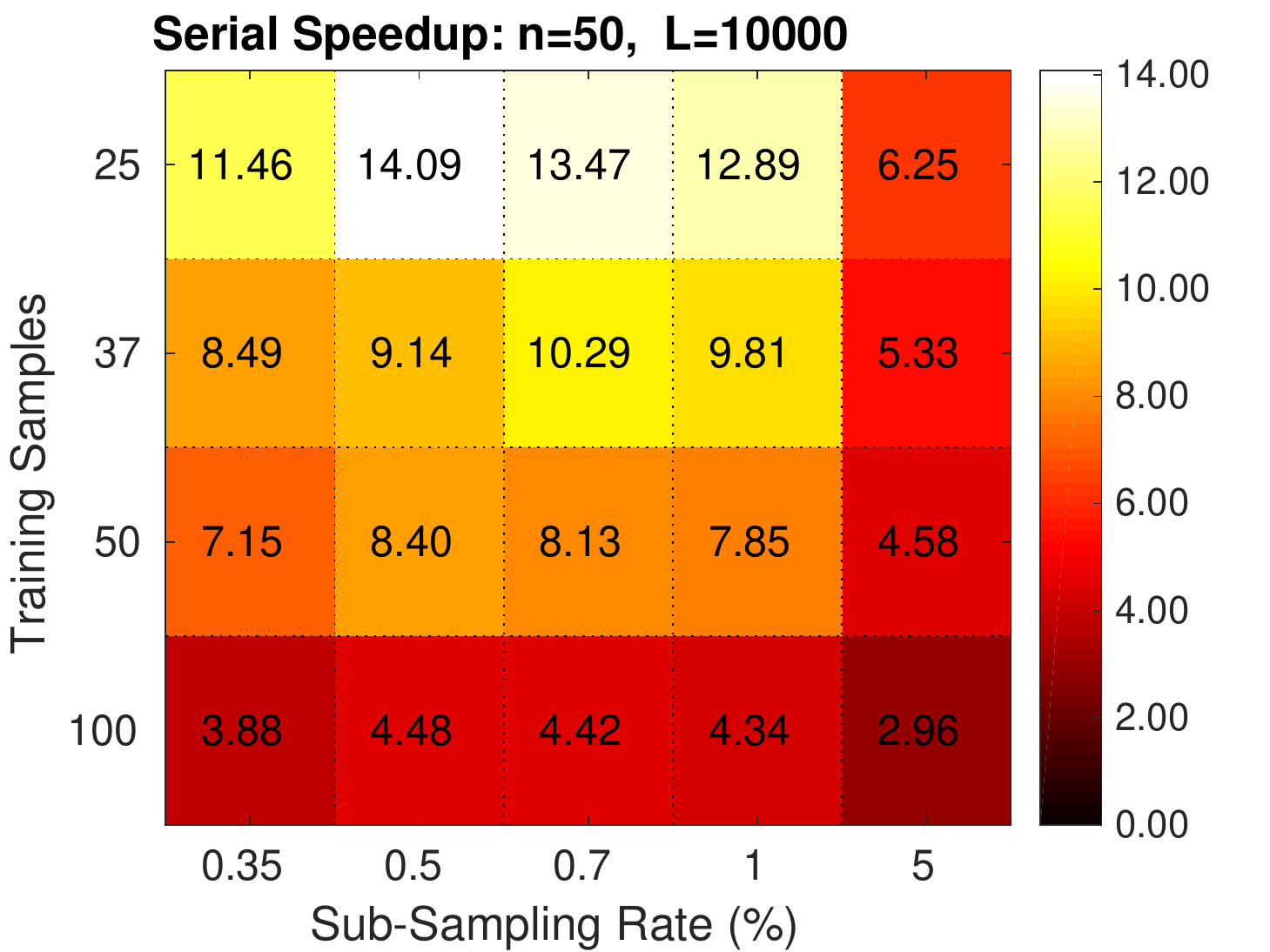}
 	\includegraphics[width=0.25\textwidth]{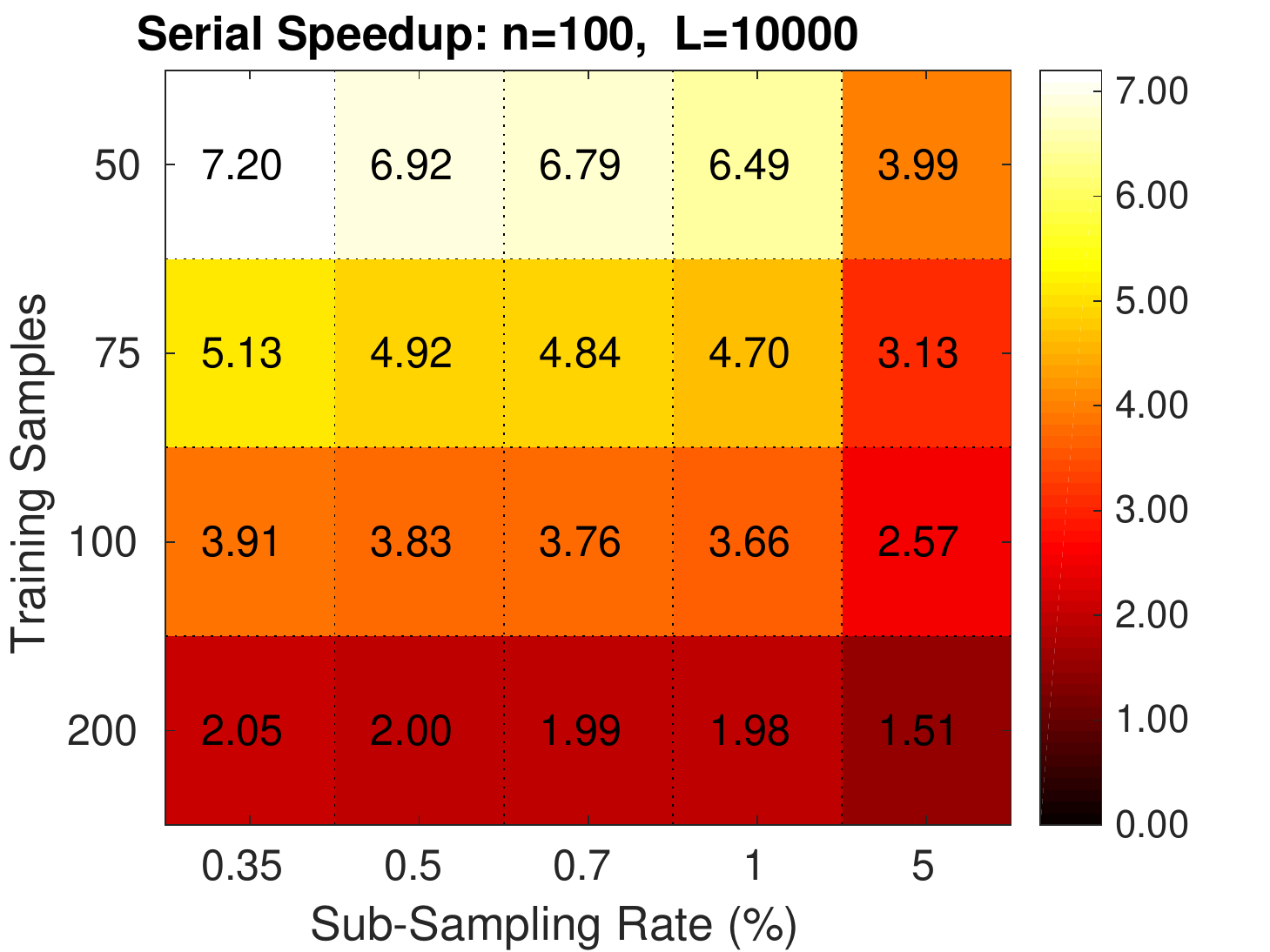}
 	\includegraphics[width=0.25\textwidth]{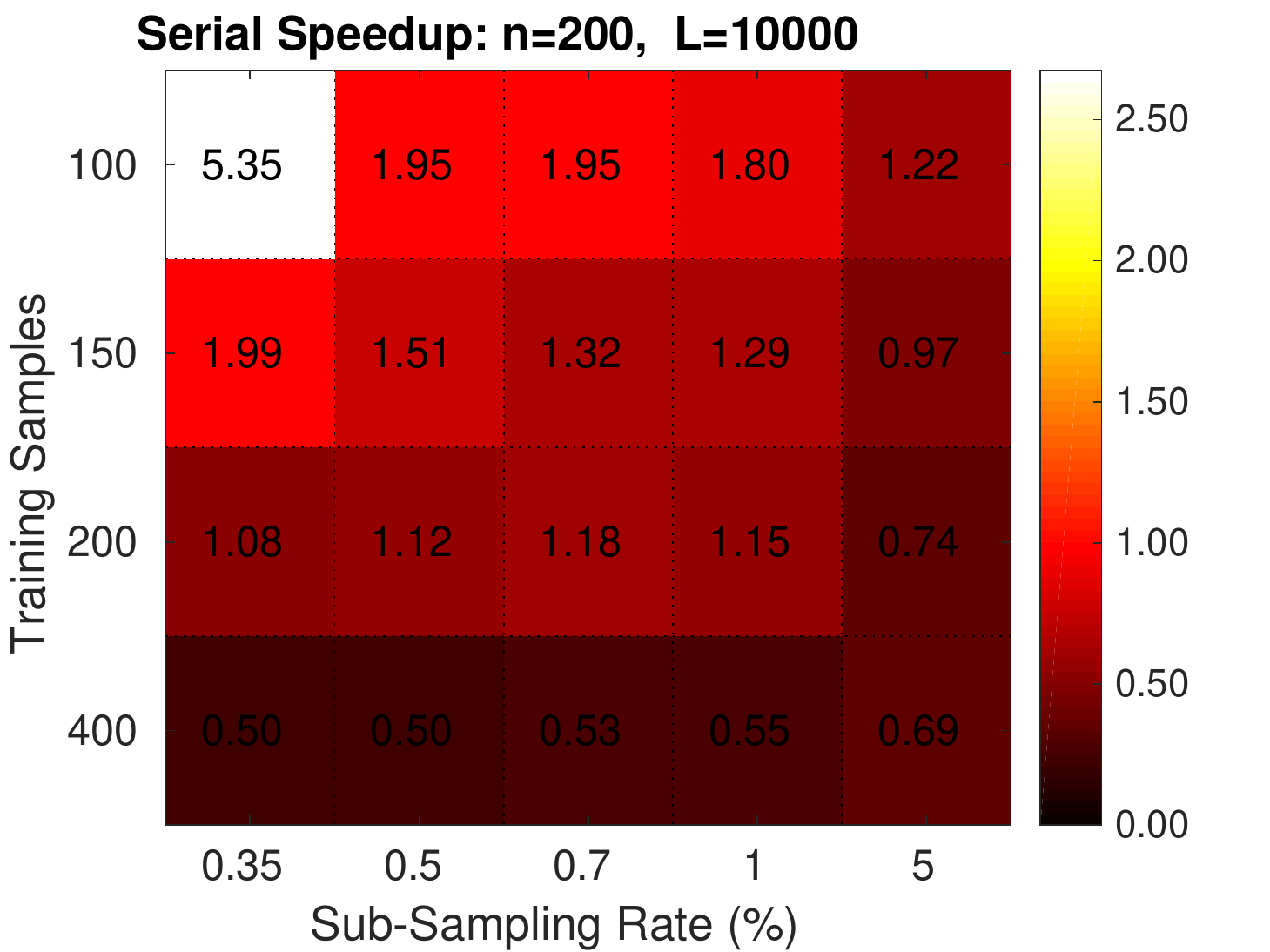}
 	\includegraphics[width=0.25\textwidth]{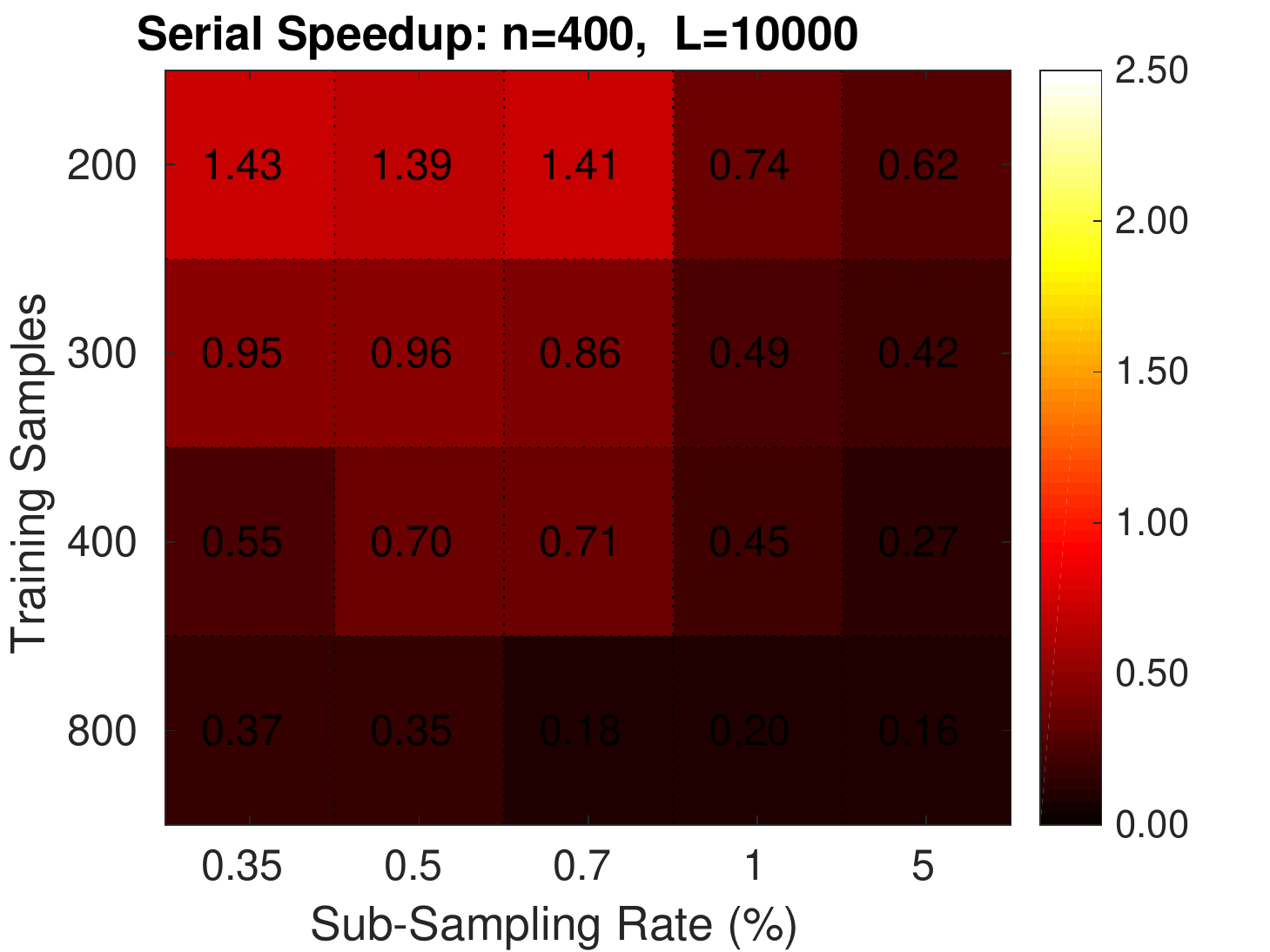}
 }
 \centerline{
 	\includegraphics[width=0.25\textwidth]{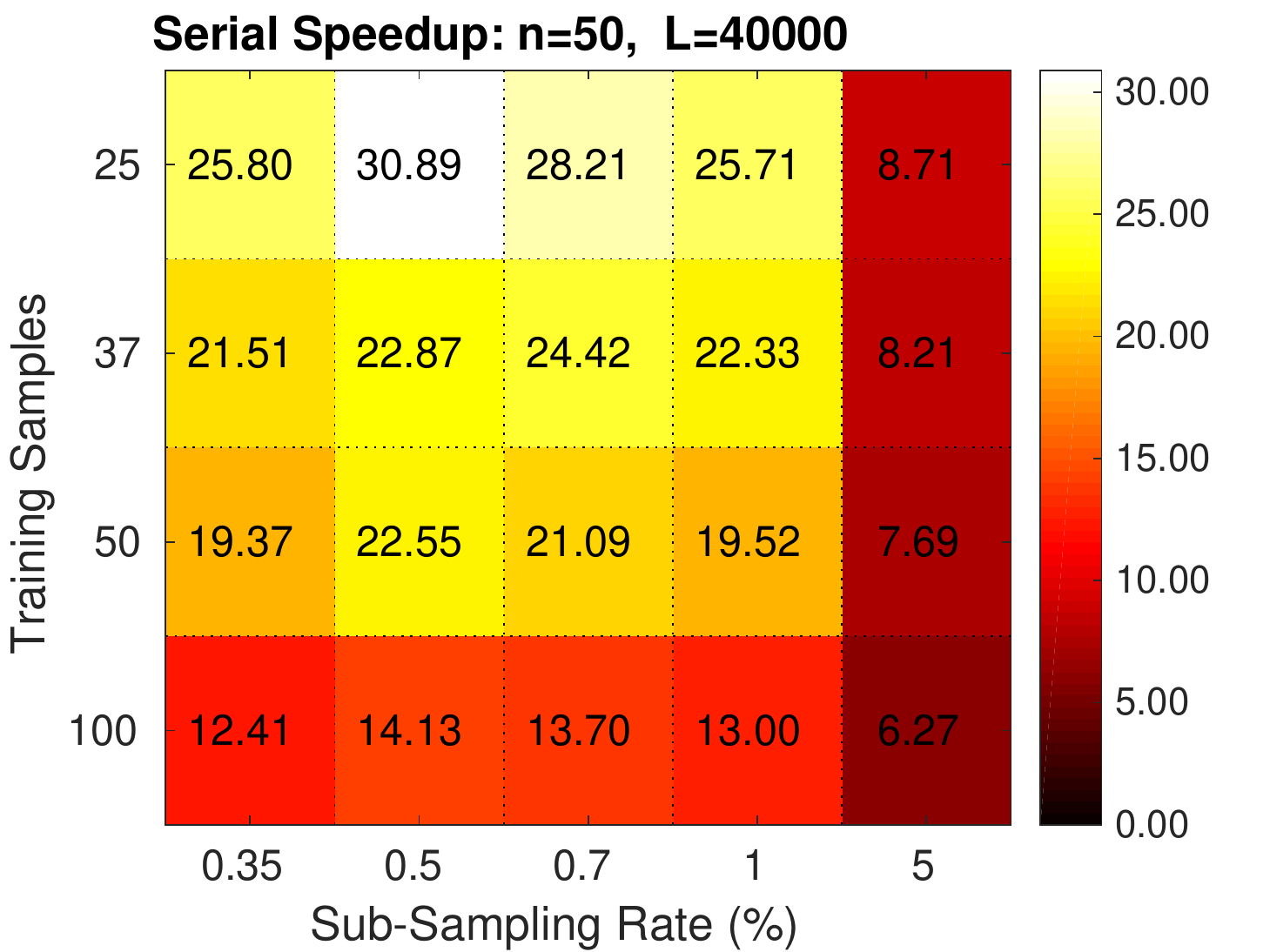}
 	\includegraphics[width=0.25\textwidth]{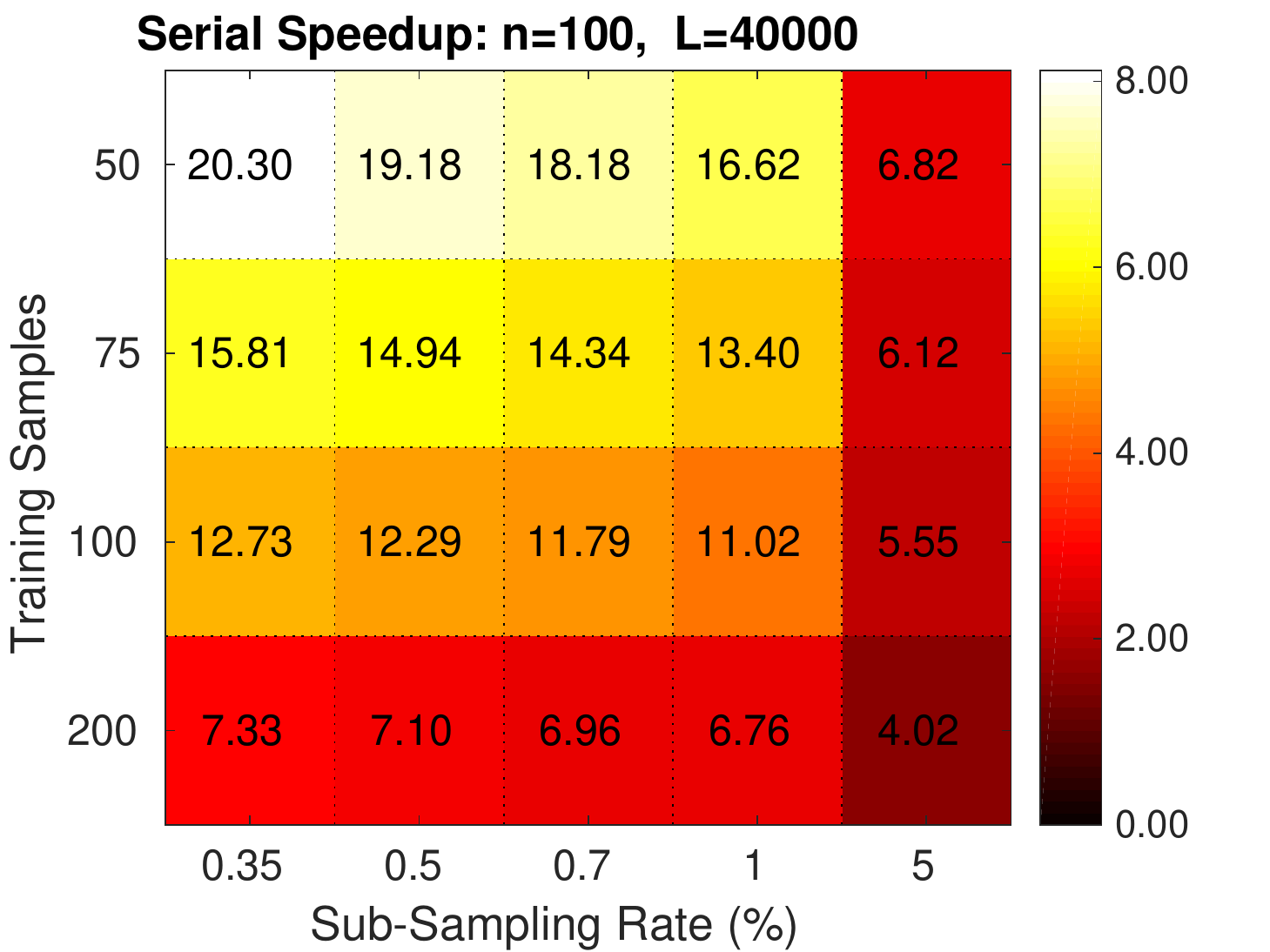}
 	\includegraphics[width=0.25\textwidth]{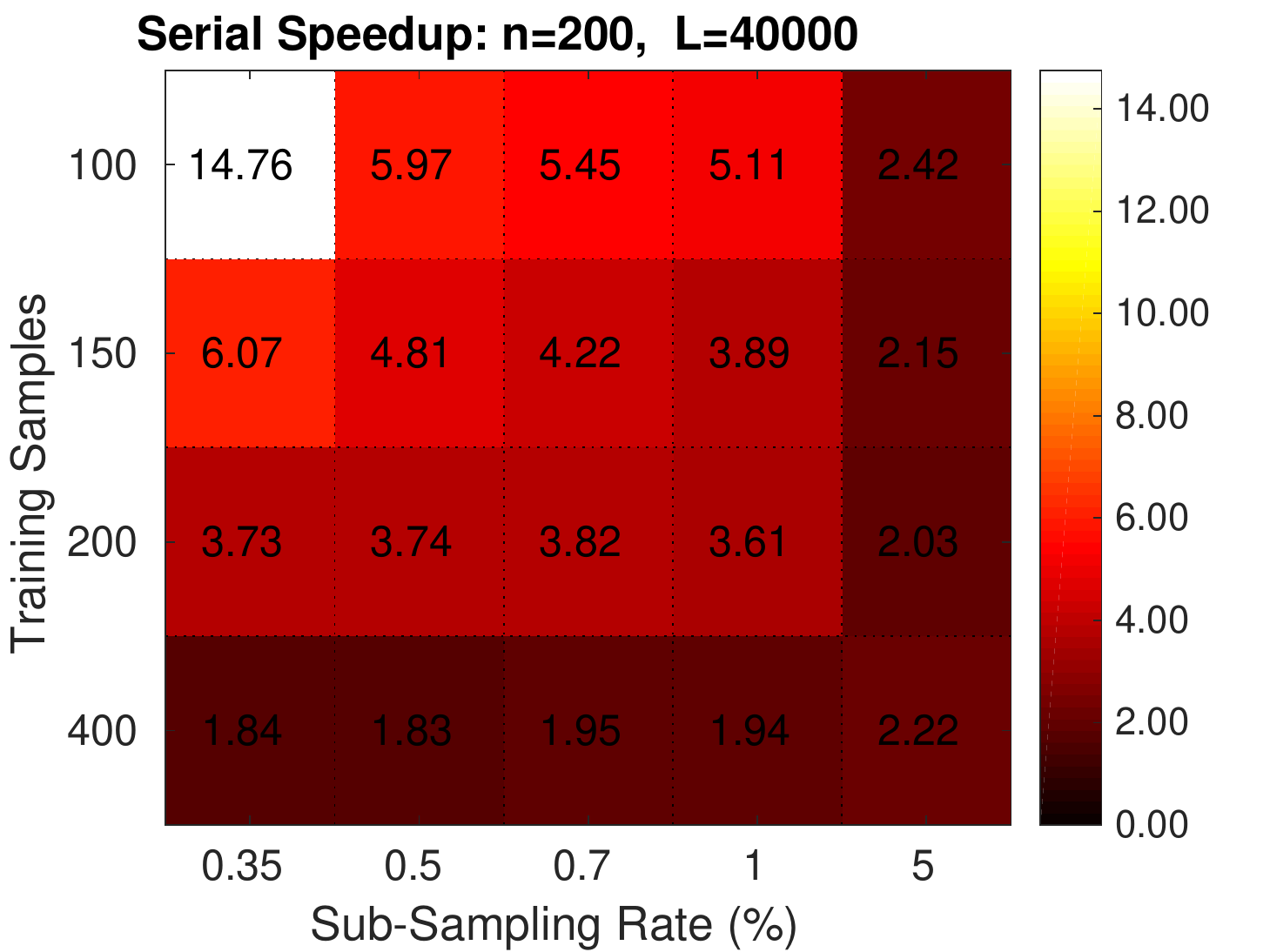}
 	\includegraphics[width=0.25\textwidth]{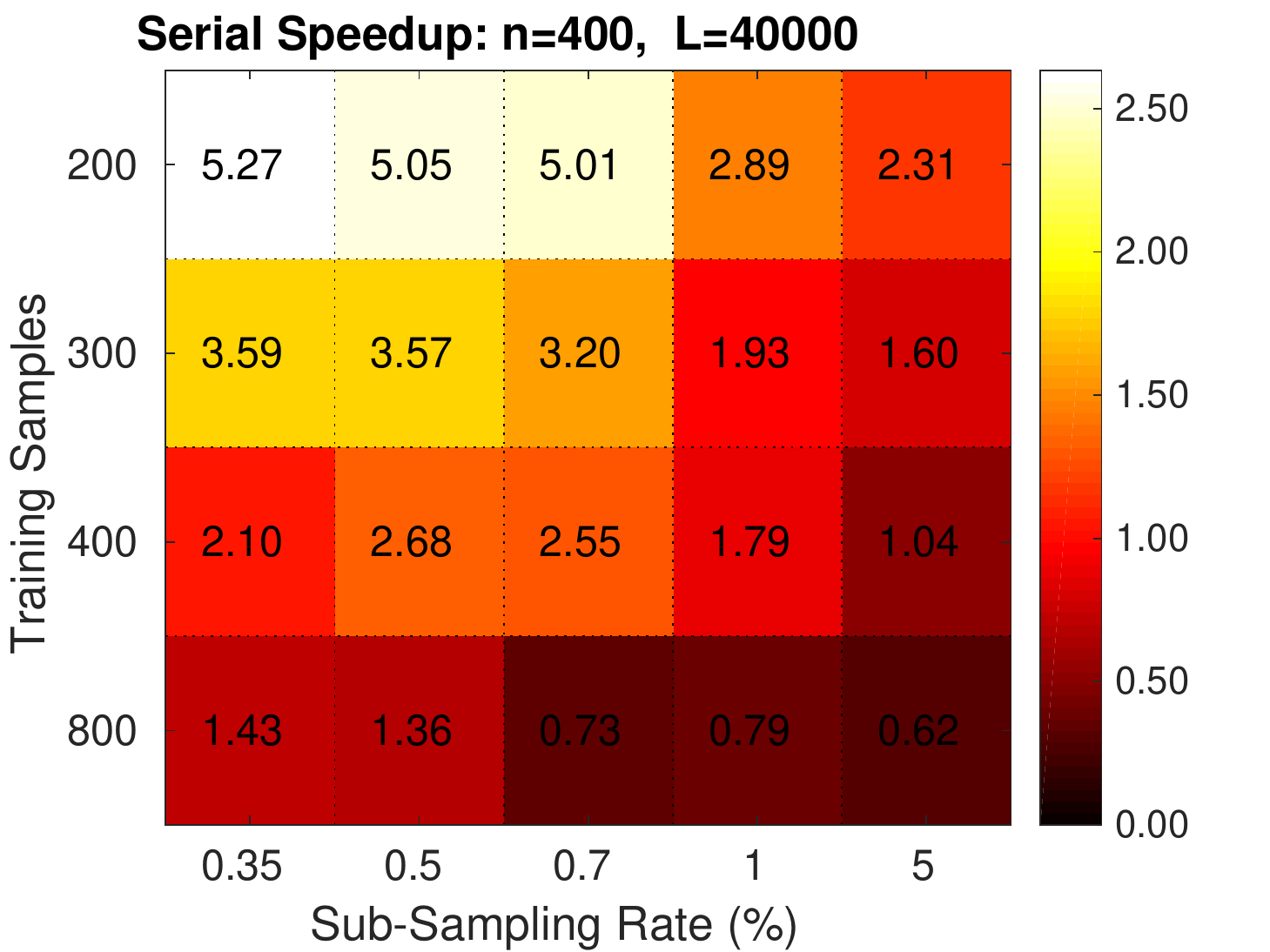}
 }

 \caption{Colormaps of the speedup gains of RapidPT over SnPM in a serial and parallel computing environment. Each colormap corresponds to 
   a run with a given dataset and a fixed number of permutations, and displays 20 different speedups resulting from different hyperparameter combinations. 
   The first two rows correspond to the speedups obtained from the runs on 16 cores, and the last two columns from runs on a single core. 
   Columns 1, 2, 3, and 4 of this figure are correspond to the 50, 100, 200, and 400 subject datasets, respectively.}
   
\label{fig:SpeedupsSnPM}
\end{figure}

\begin{figure}[H]
\centerline{
	\includegraphics[width=0.25\textwidth]{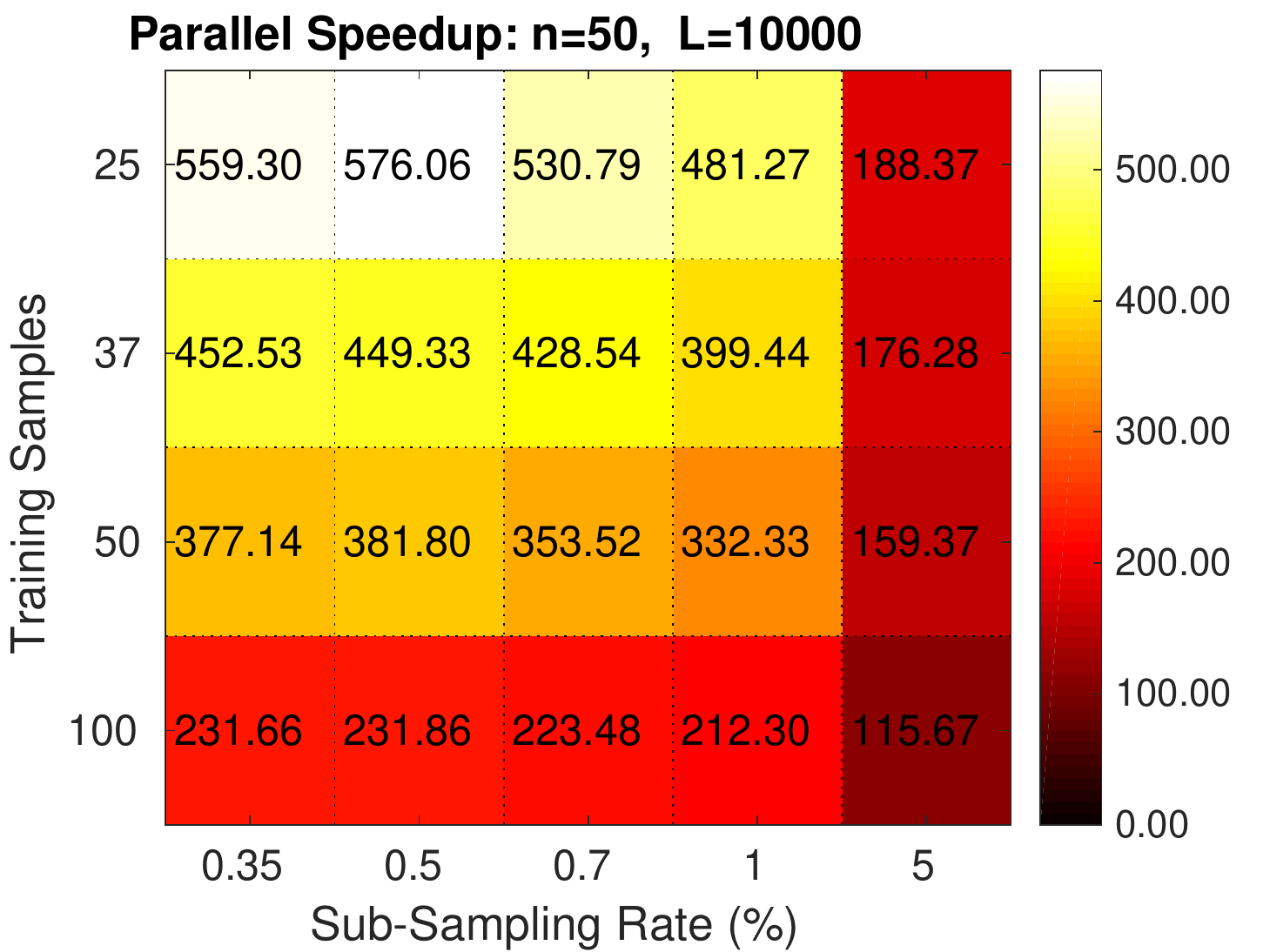}
	\includegraphics[width=0.25\textwidth]{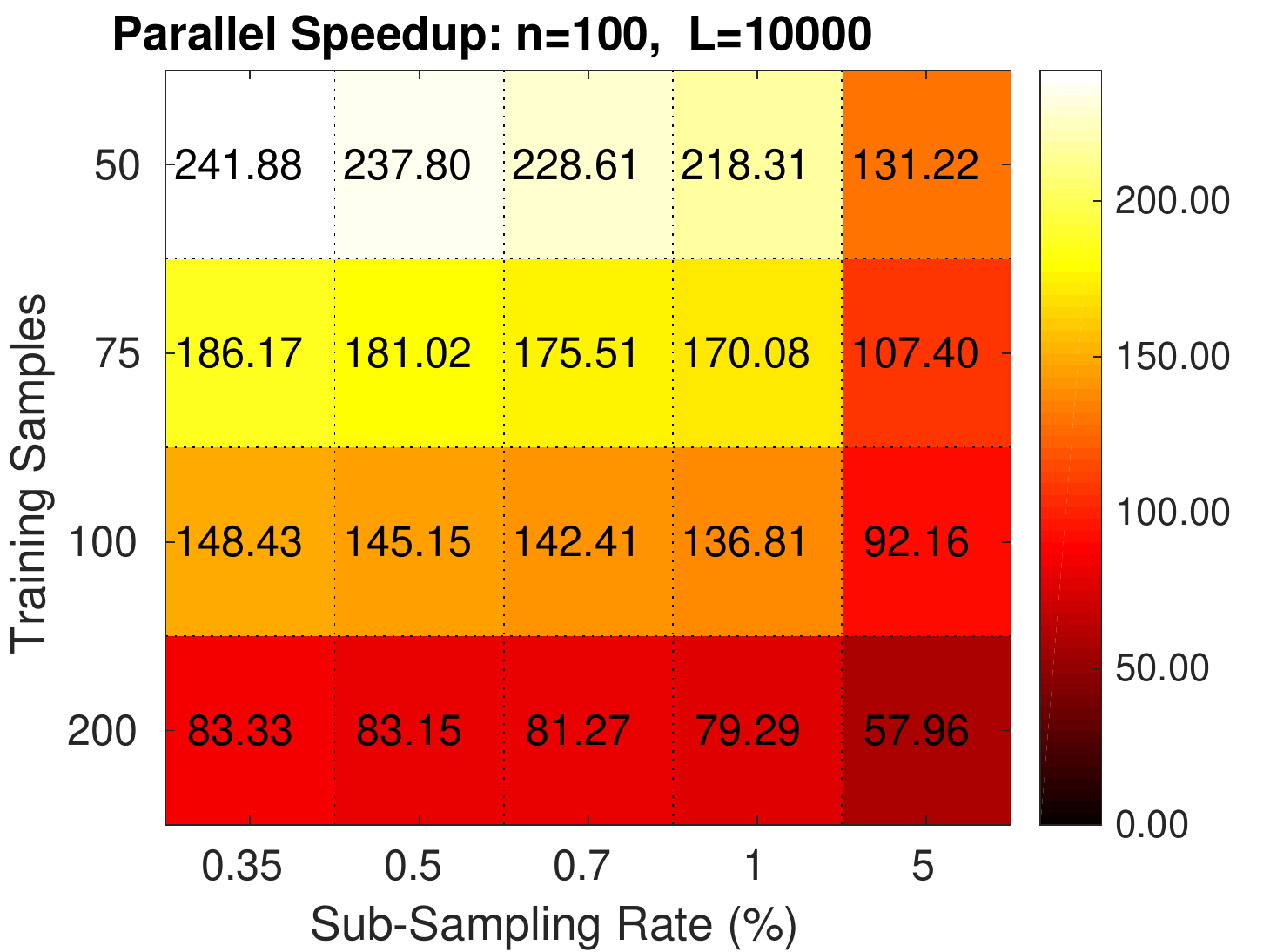}
	\includegraphics[width=0.25\textwidth]{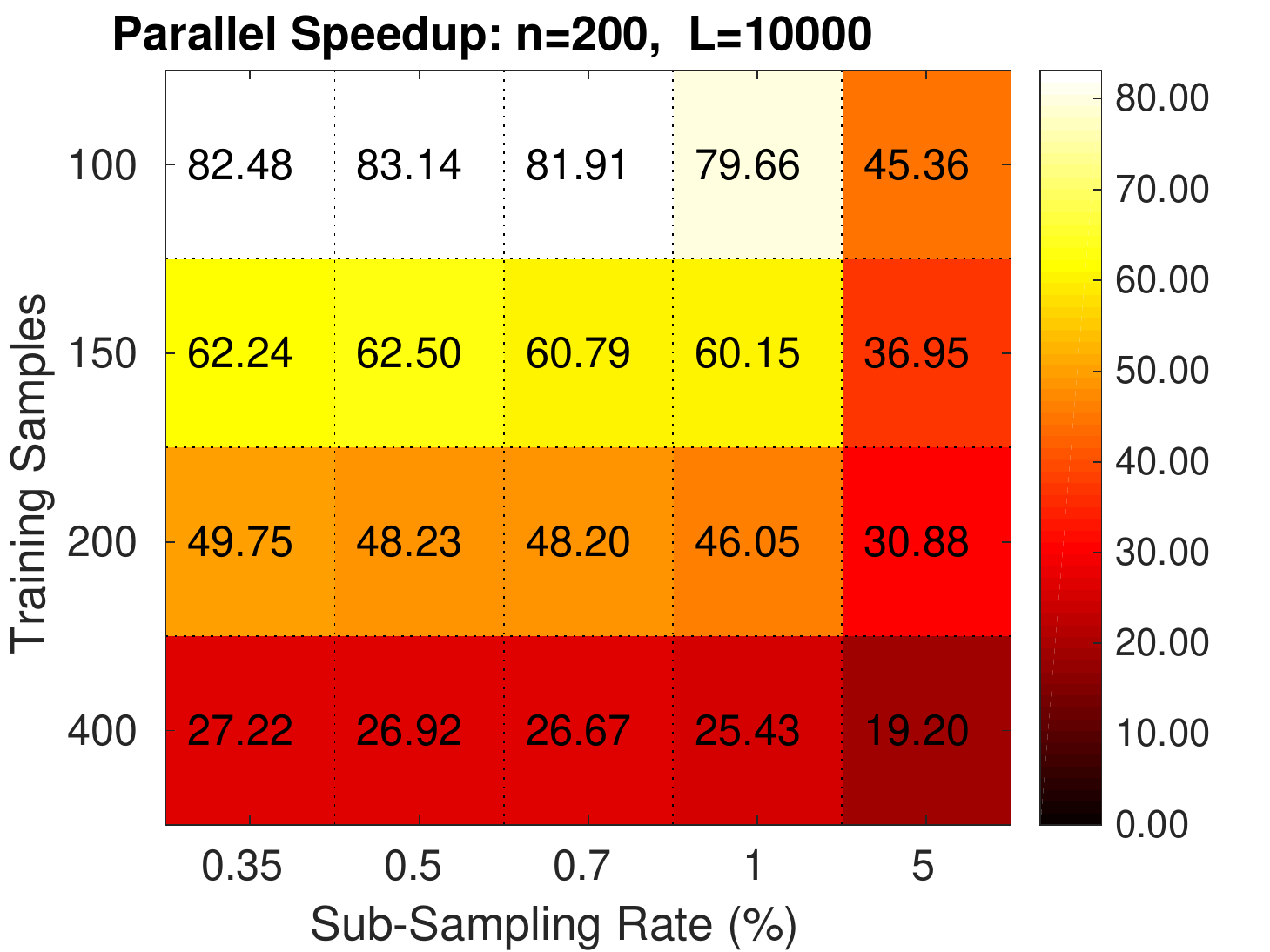}
	\includegraphics[width=0.25\textwidth]{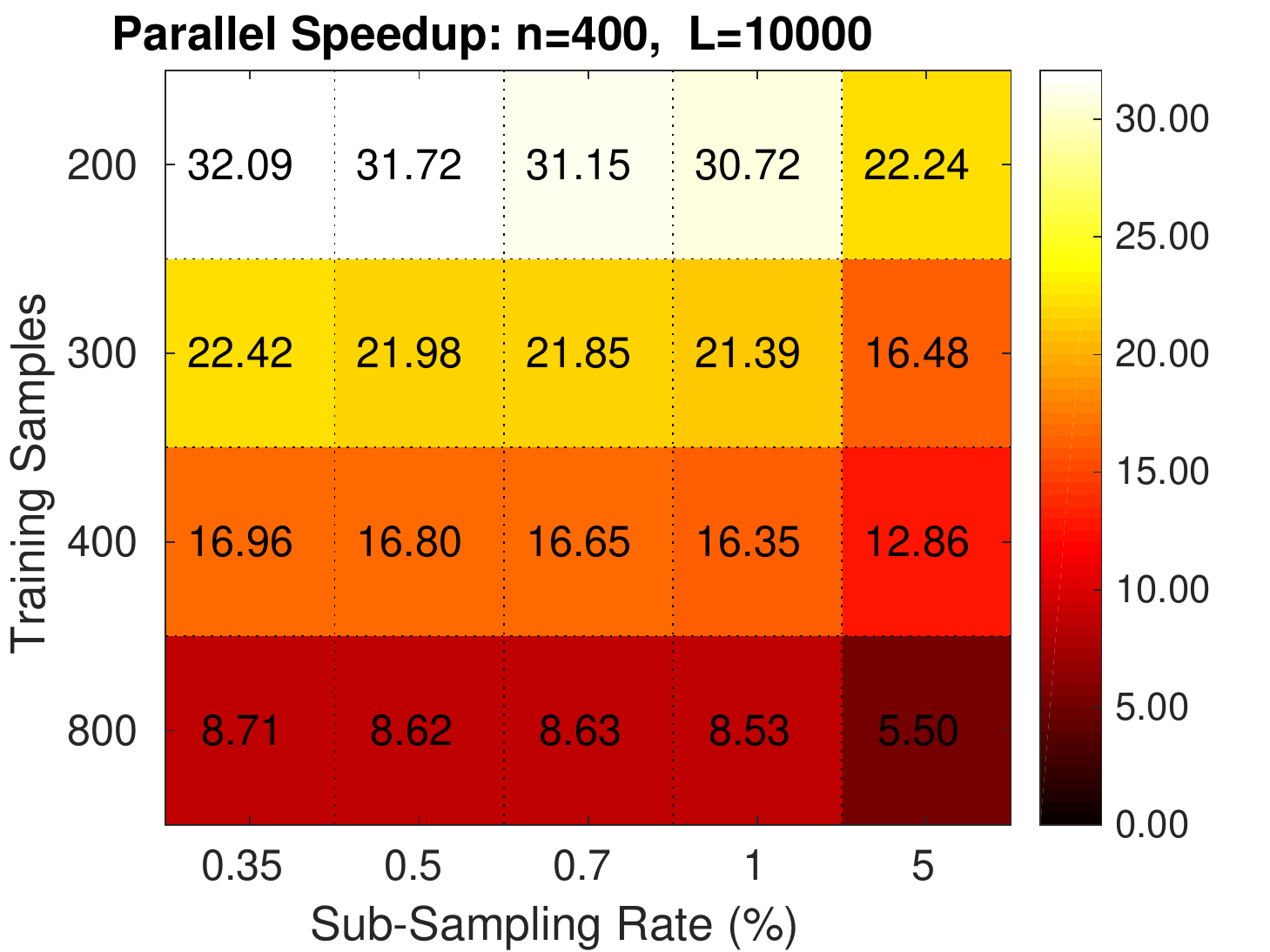}
}
\centerline{
	\includegraphics[width=0.25\textwidth]{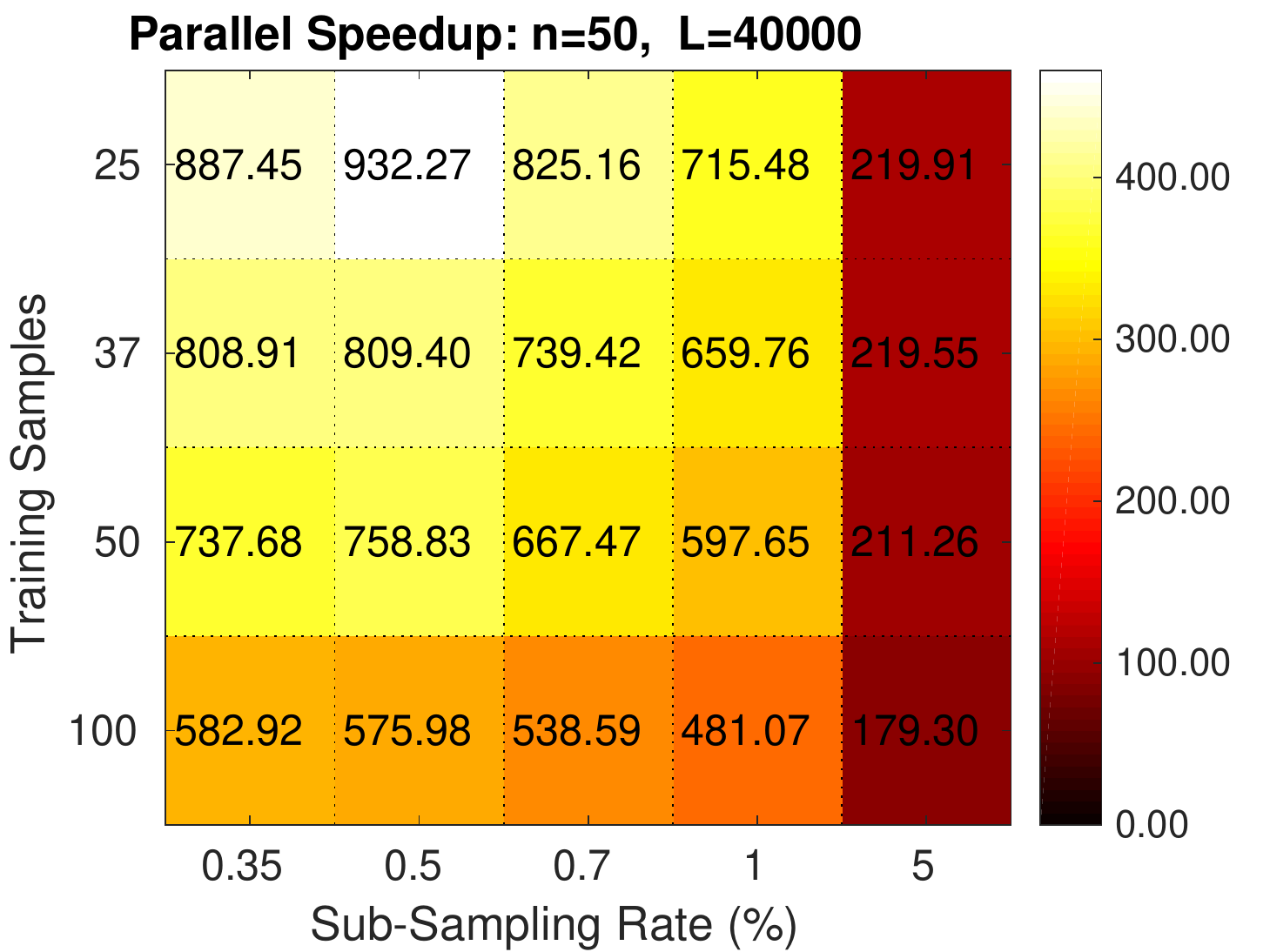}
	\includegraphics[width=0.25\textwidth]{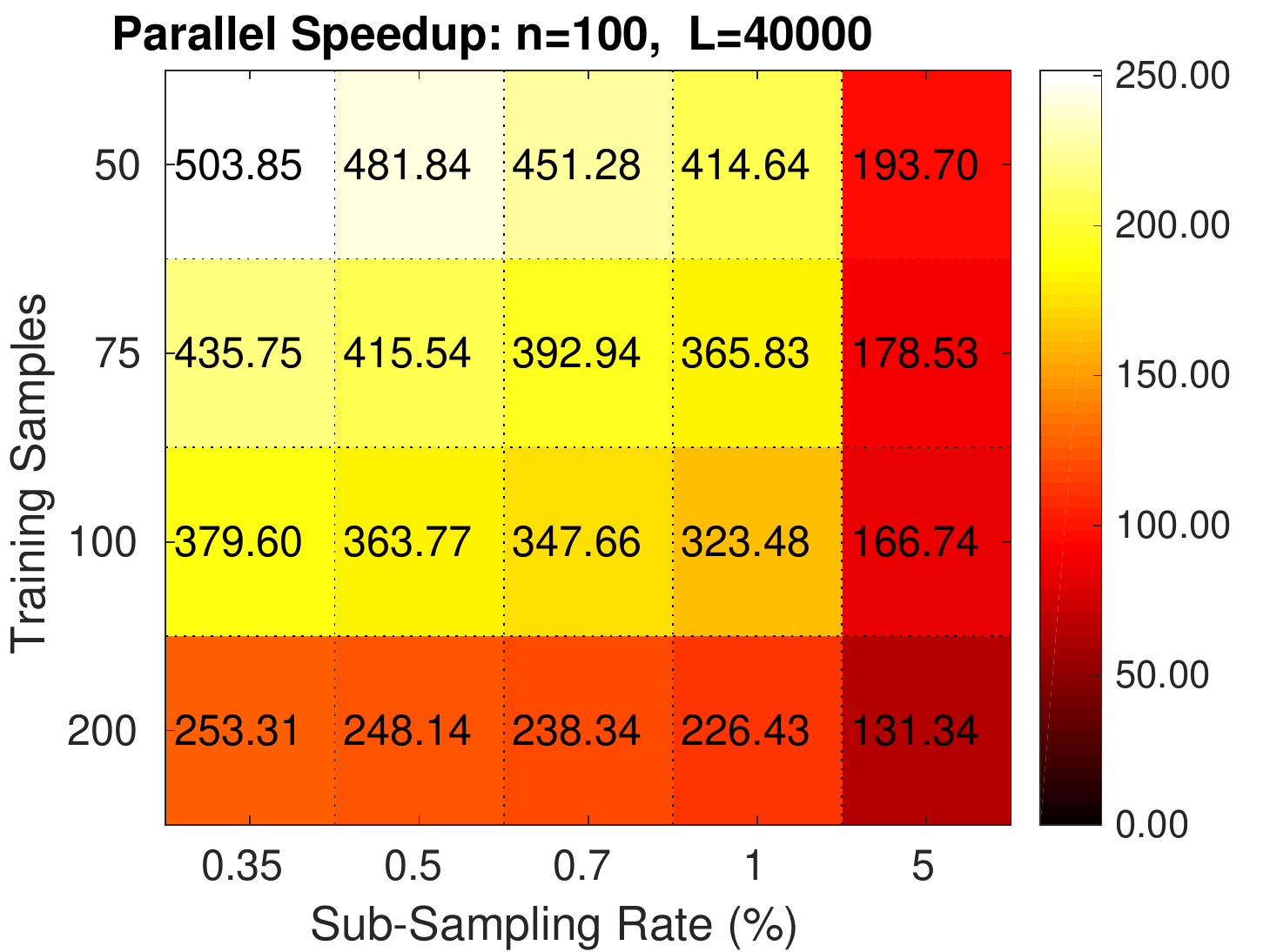}
	\includegraphics[width=0.25\textwidth]{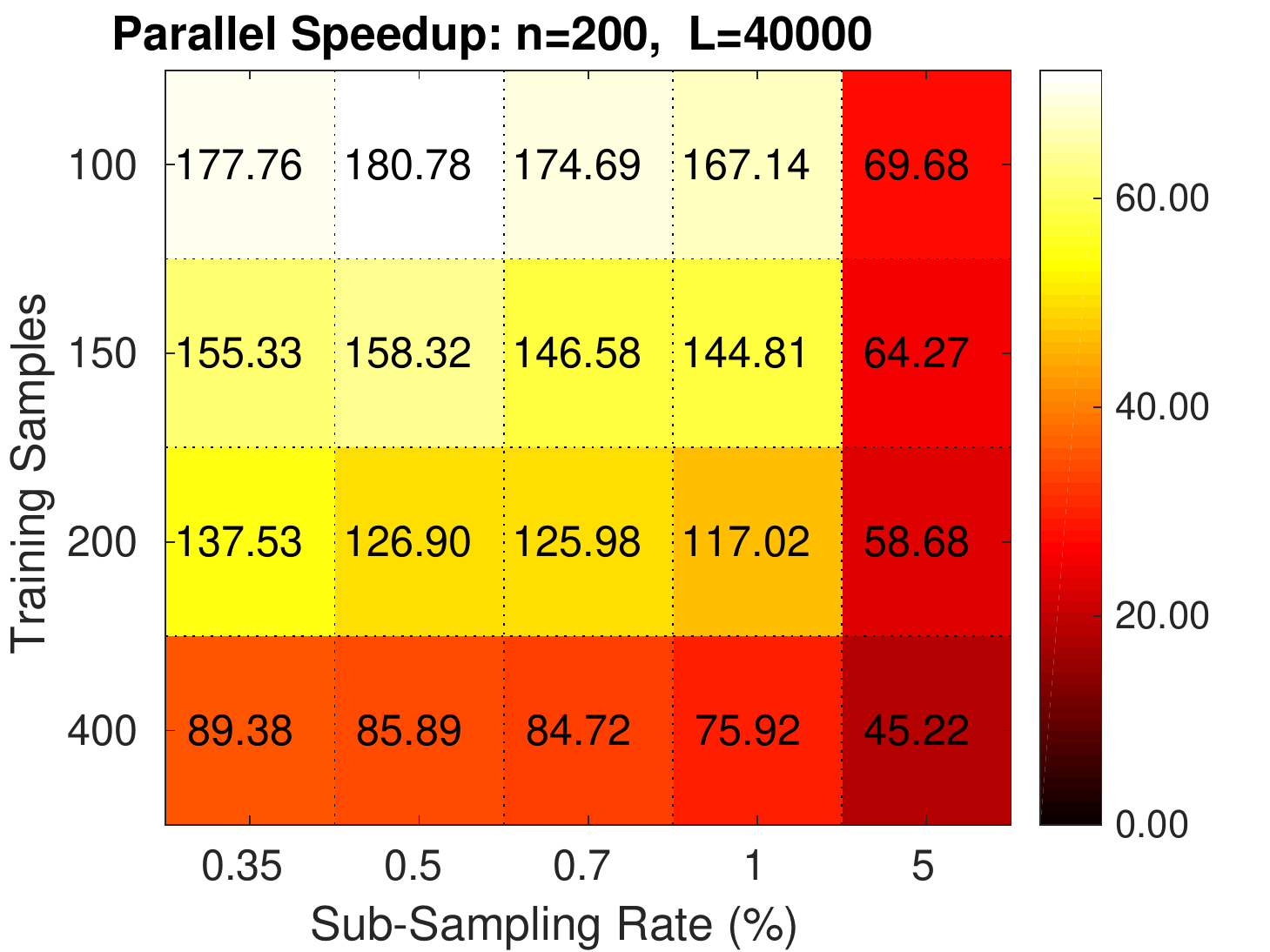}
	\includegraphics[width=0.25\textwidth]{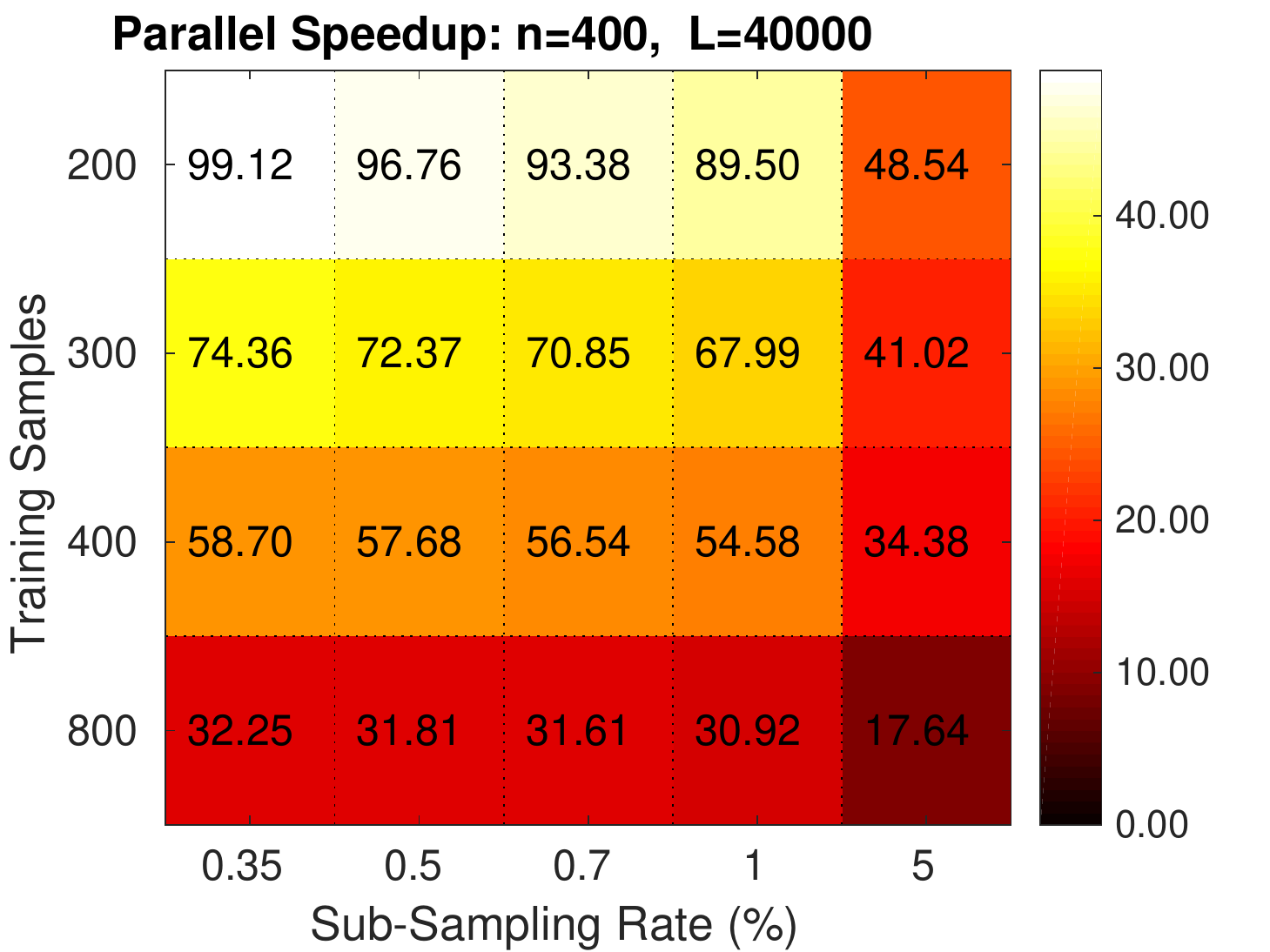}
}
\caption{Colormap of the speedup gains of RapidPT over NaivePT. The organization of the colormaps and the information they display is the same as Figure \ref{fig:SpeedupsSnPM}. These speedups correspond to both programs running on a MATLAB instance that could use all 16 available cores.}
\label{fig:SpeedupNaivePT}
\end{figure}

Figures \ref{fig:SpeedupsSnPM} and \ref{fig:SpeedupNaivePT} show the speedup gains of RapidPT over SnPM and NaivePT, respectively. 
Each column corresponds to a single dataset, and each row corresponds to a different number of permutations. 
The supplementary results include an exhaustive version of these results that show the speedup gains of RapidPT for many additional number of permutations. 

As shown in Figure \ref{fig:SpeedupsSnPM}, 
RapidPT outperforms SnPM in most scenarios. With the exception of the $L = 2000$ and $L = 5000$ runs on the larger datasets ($n = 200$ and $n = 400$), 
the colormaps show that RapidPT is 1.5-30x faster than SnPM. As expected, a low $\eta$ ($0.35\%$,$0.5\%$) and $l$ ($\frac{n}{2}$,$\frac{3n}{4}$) leads to the 
best runtime performance without a noticeable accuracy tradeoff, as can be seen also in Fig. \ref{fig:KLDivSnPM} earlier. 

Figure \ref{fig:SpeedupNaivePT} shows how RapidPT performs against a non-optimized permutation testing implementation. In this setup, RapidPT outperforms NaivePT in 
every single combination of the hyperparameters. The same speedup trends that were seen when comparing RapidPT and SnPM are seen between RapidPT and NaivePT but with a 
much larger magnitude. For the remainder of this section, the runtime results of NaivePT are no longer compared because the difference is too large.


\subsubsection{Scaling of RapidPT vs. SnPM}

\begin{figure}[h!]
\centerline{%
	\includegraphics[width=0.25\textwidth]{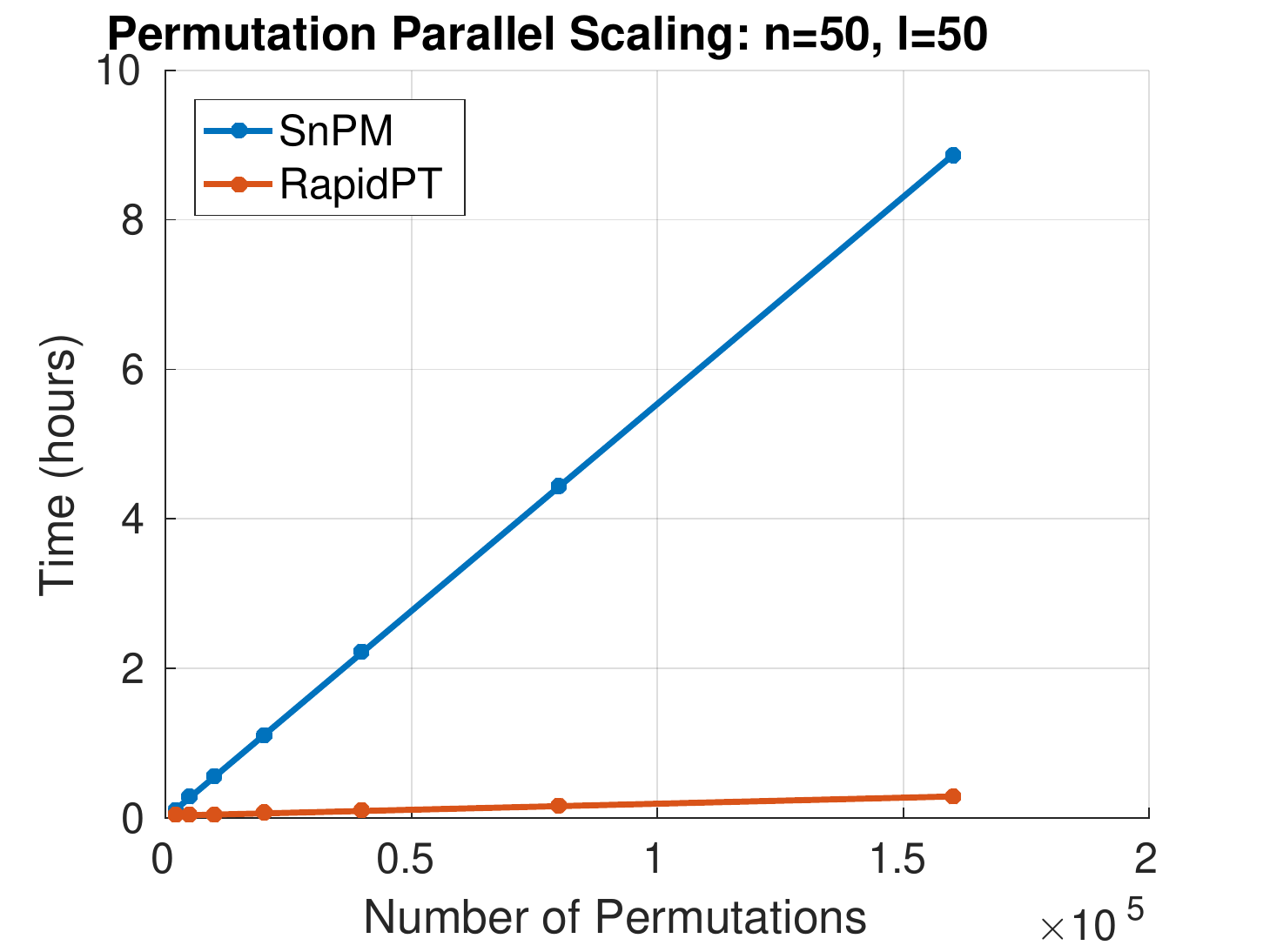}
	\includegraphics[width=0.25\textwidth]{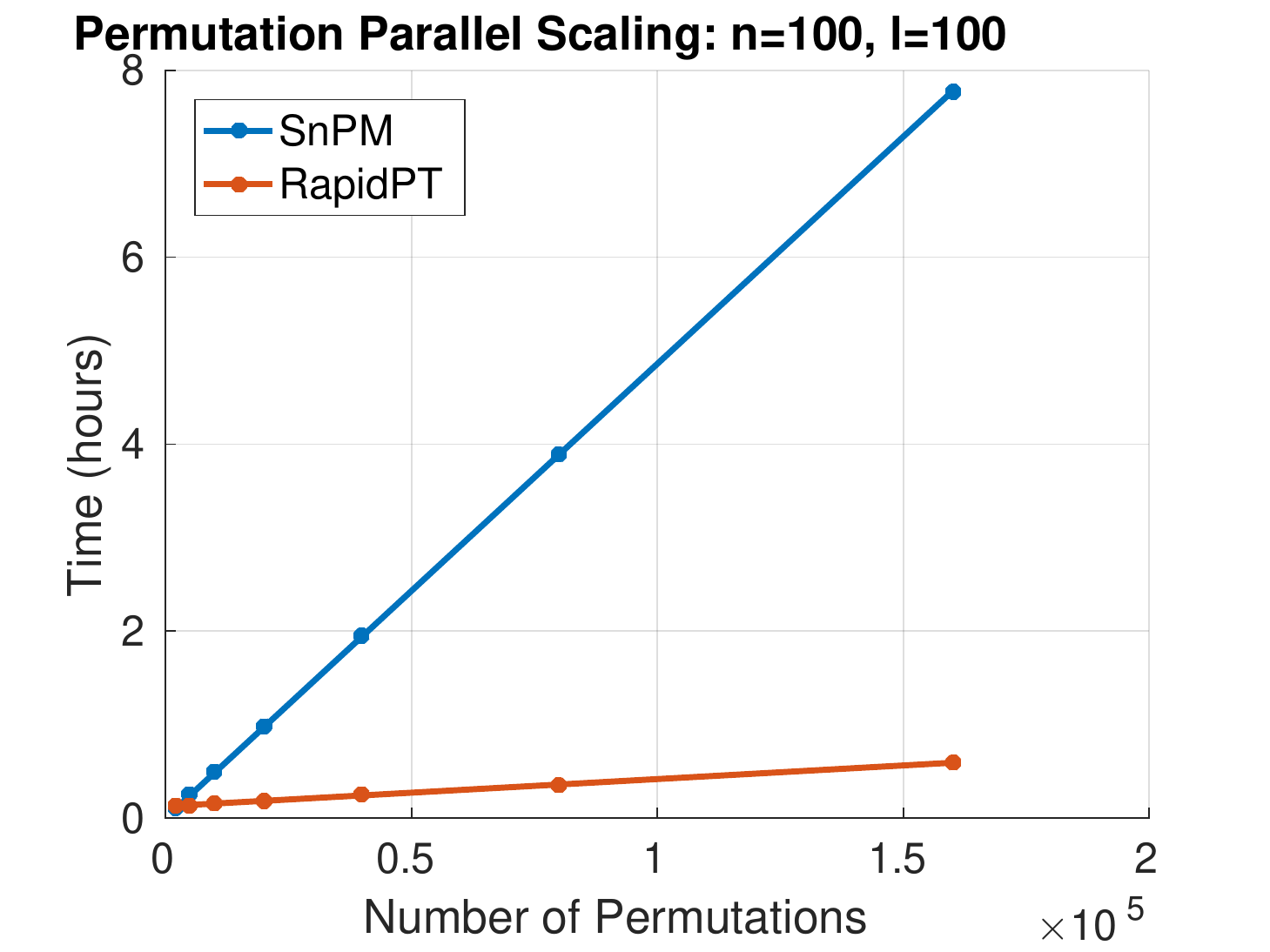}
	\includegraphics[width=0.25\textwidth]{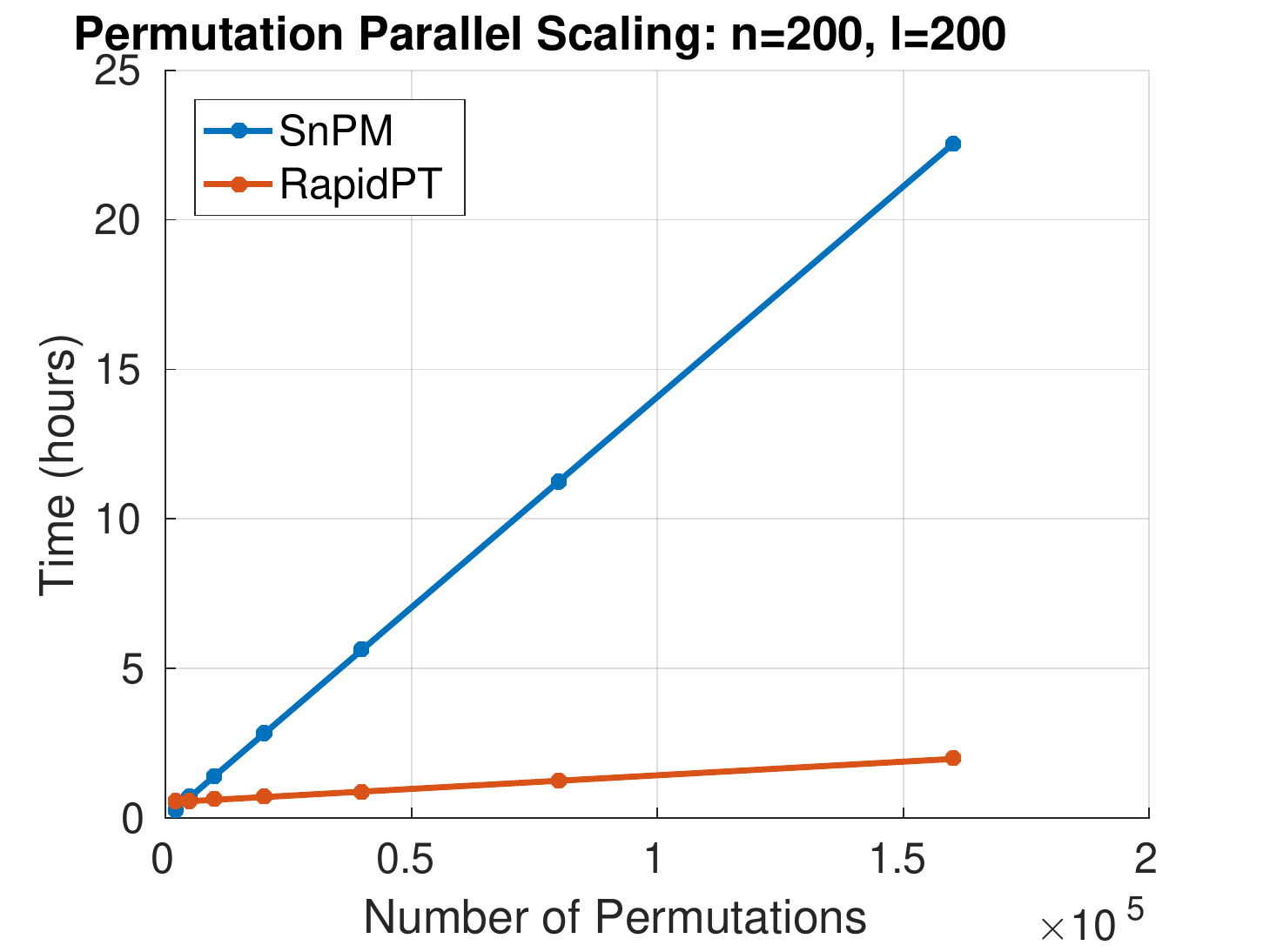}
	\includegraphics[width=0.25\textwidth]{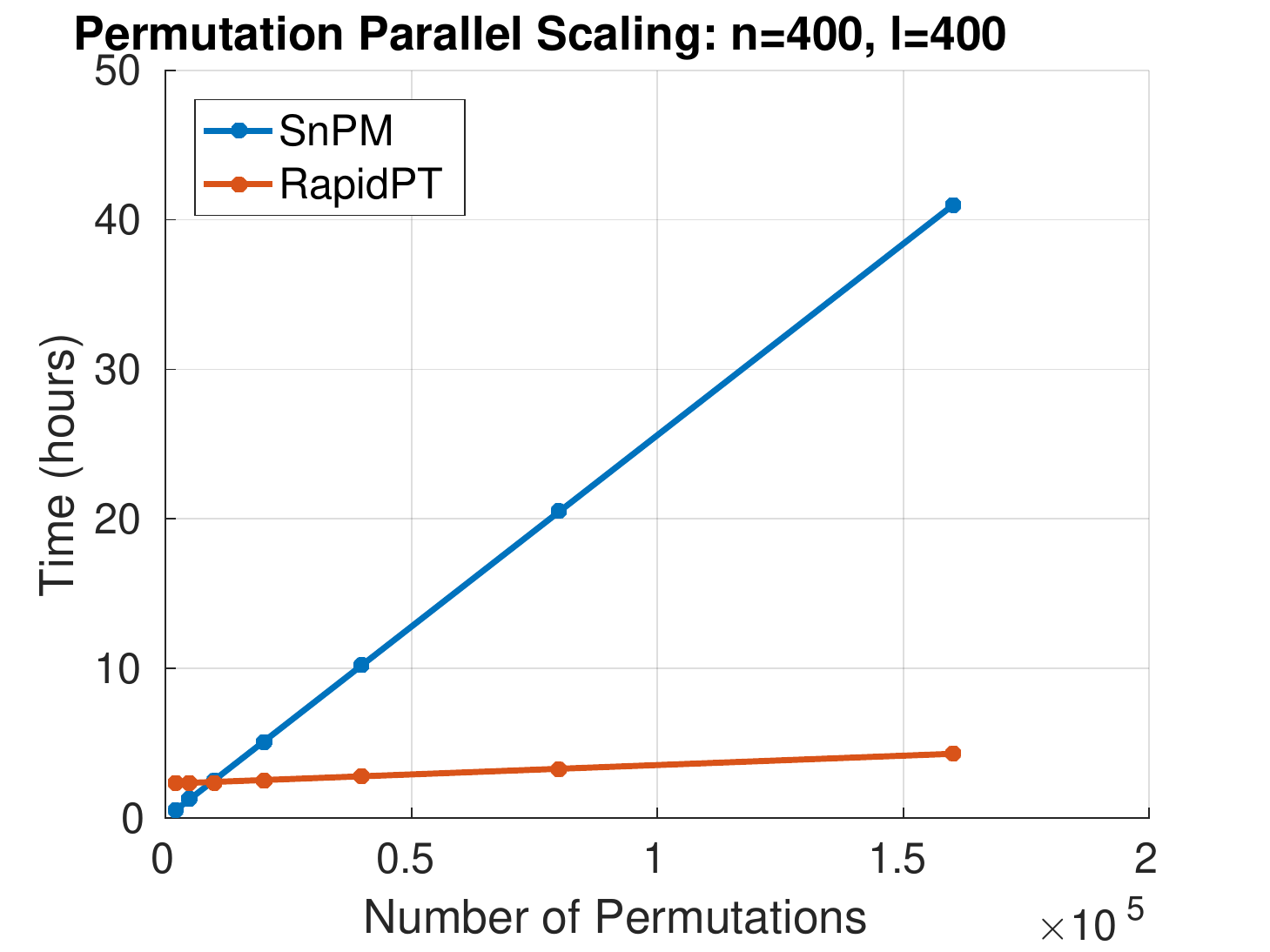}
}
\centerline{%
	\includegraphics[width=0.25\textwidth]{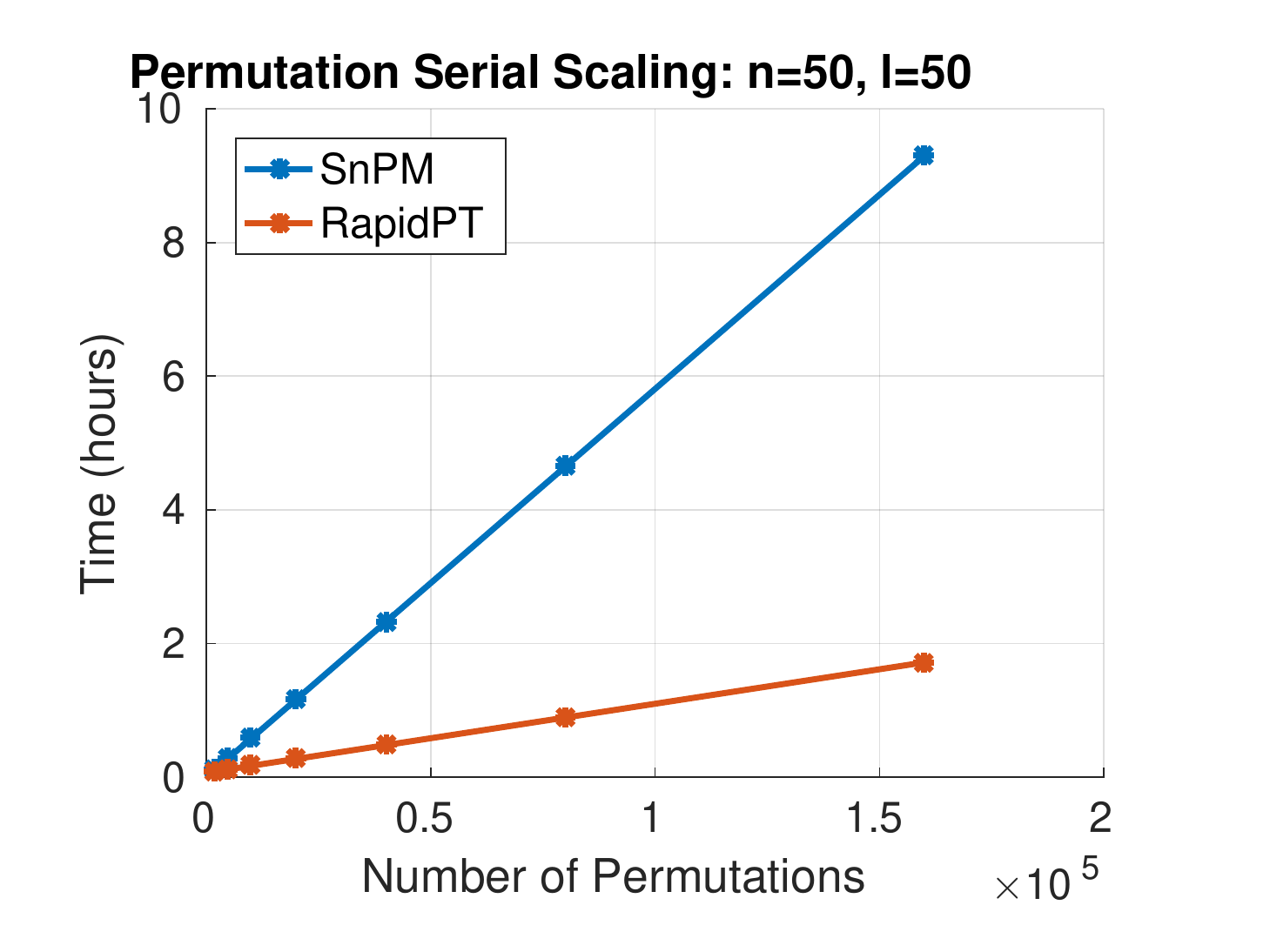}
	\includegraphics[width=0.25\textwidth]{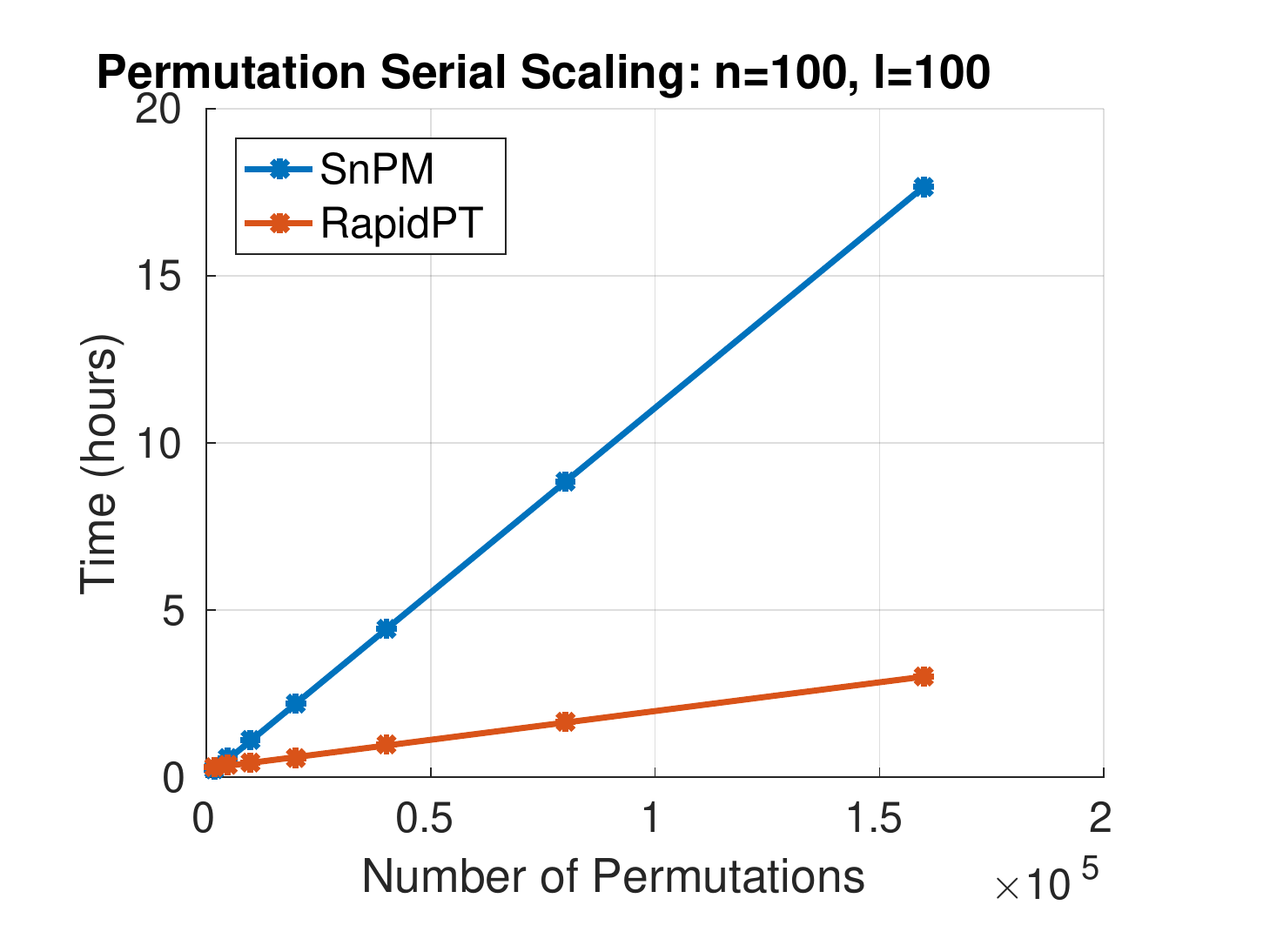}
	\includegraphics[width=0.25\textwidth]{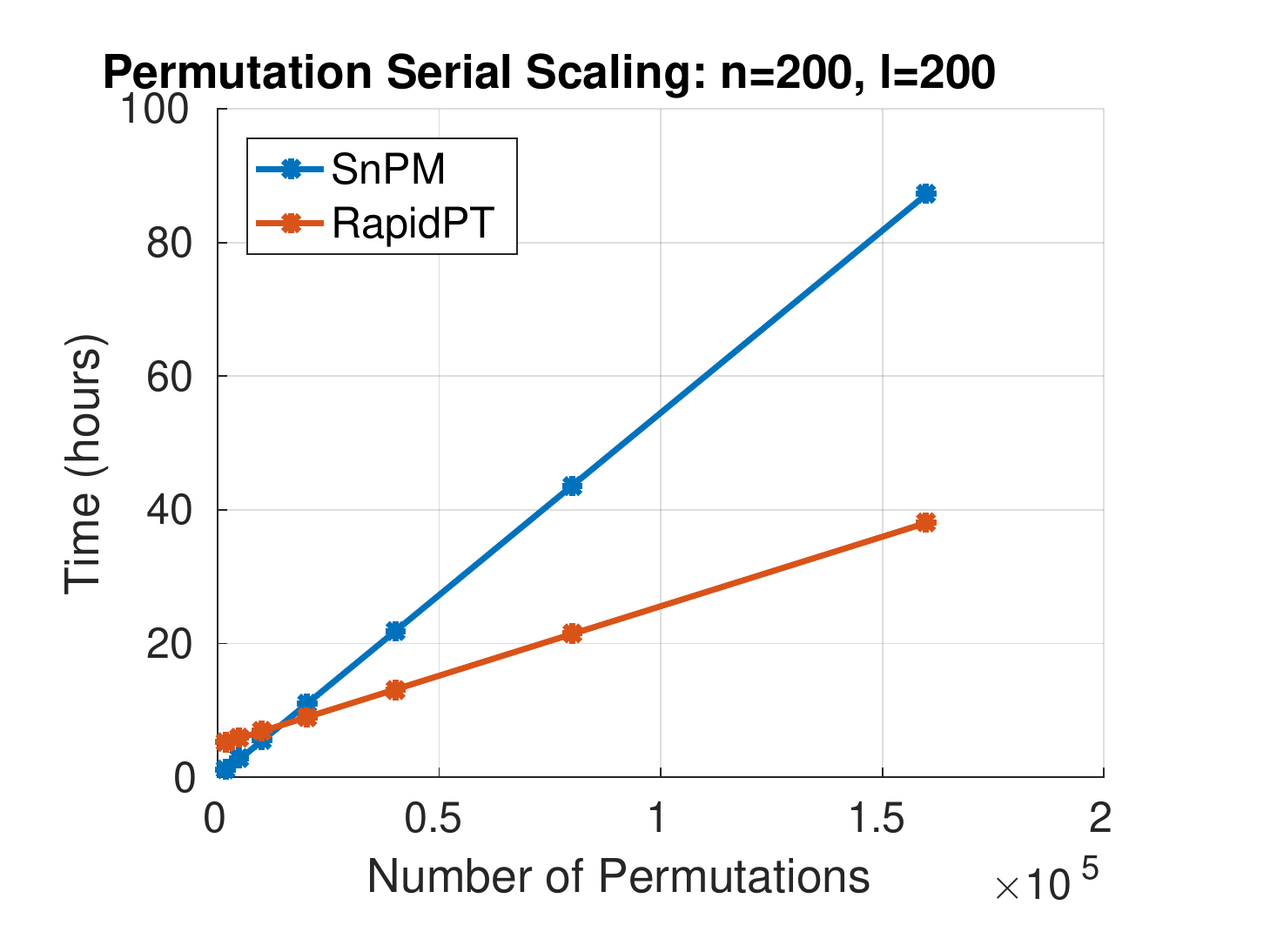}
	\includegraphics[width=0.25\textwidth]{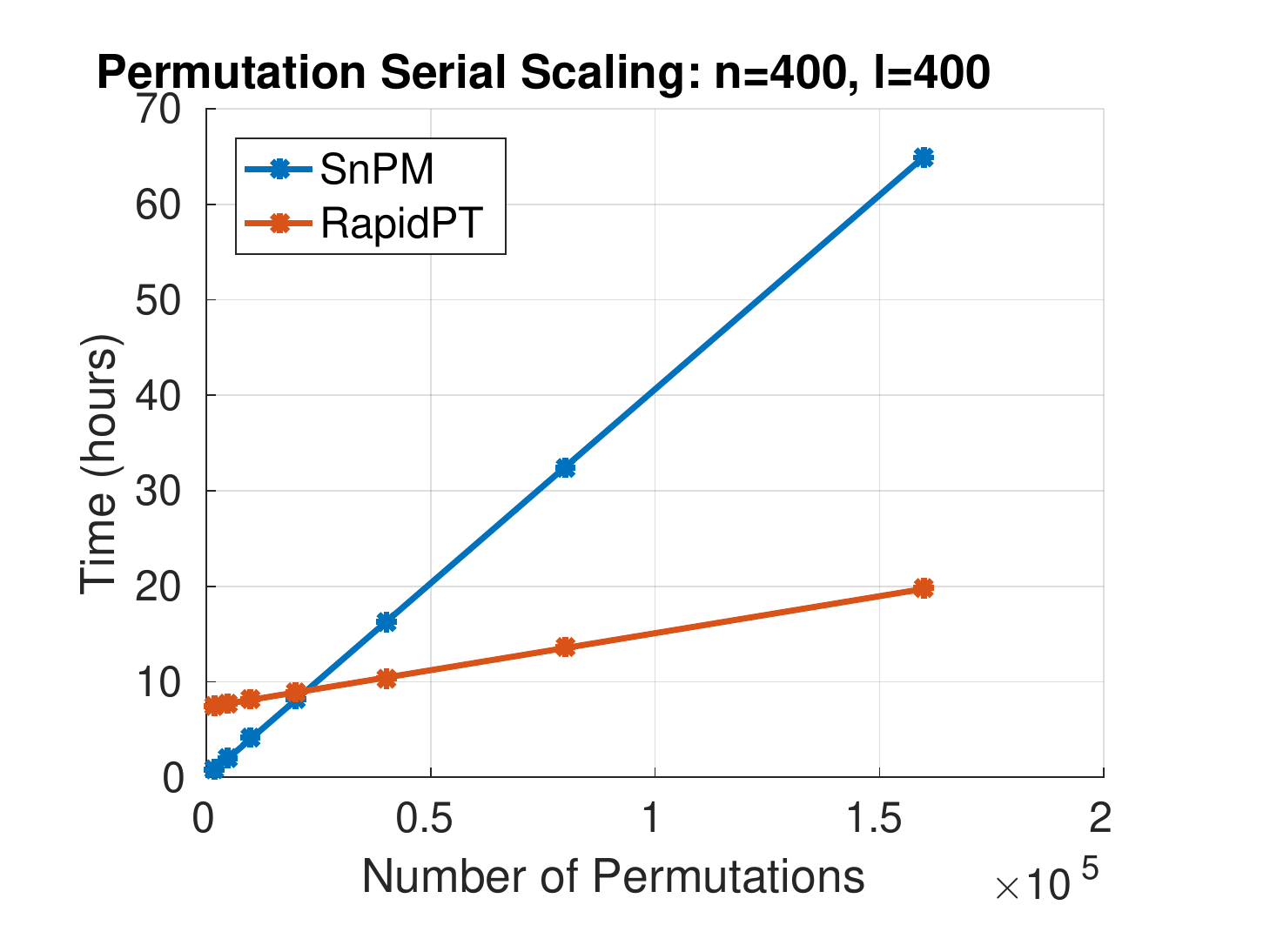}
}
\caption{Effect of the number of permutations on the runtime performance of RapidPT and SnPM on 16 cores (first row) and on a single core (second row). The hyperparameters used were: $\eta = 0.35\%$ and $l = n$.}
\label{fig:PermutationScaling}
\end{figure}

\begin{figure}[H]
\centerline{%
\includegraphics[width=0.4\textwidth]{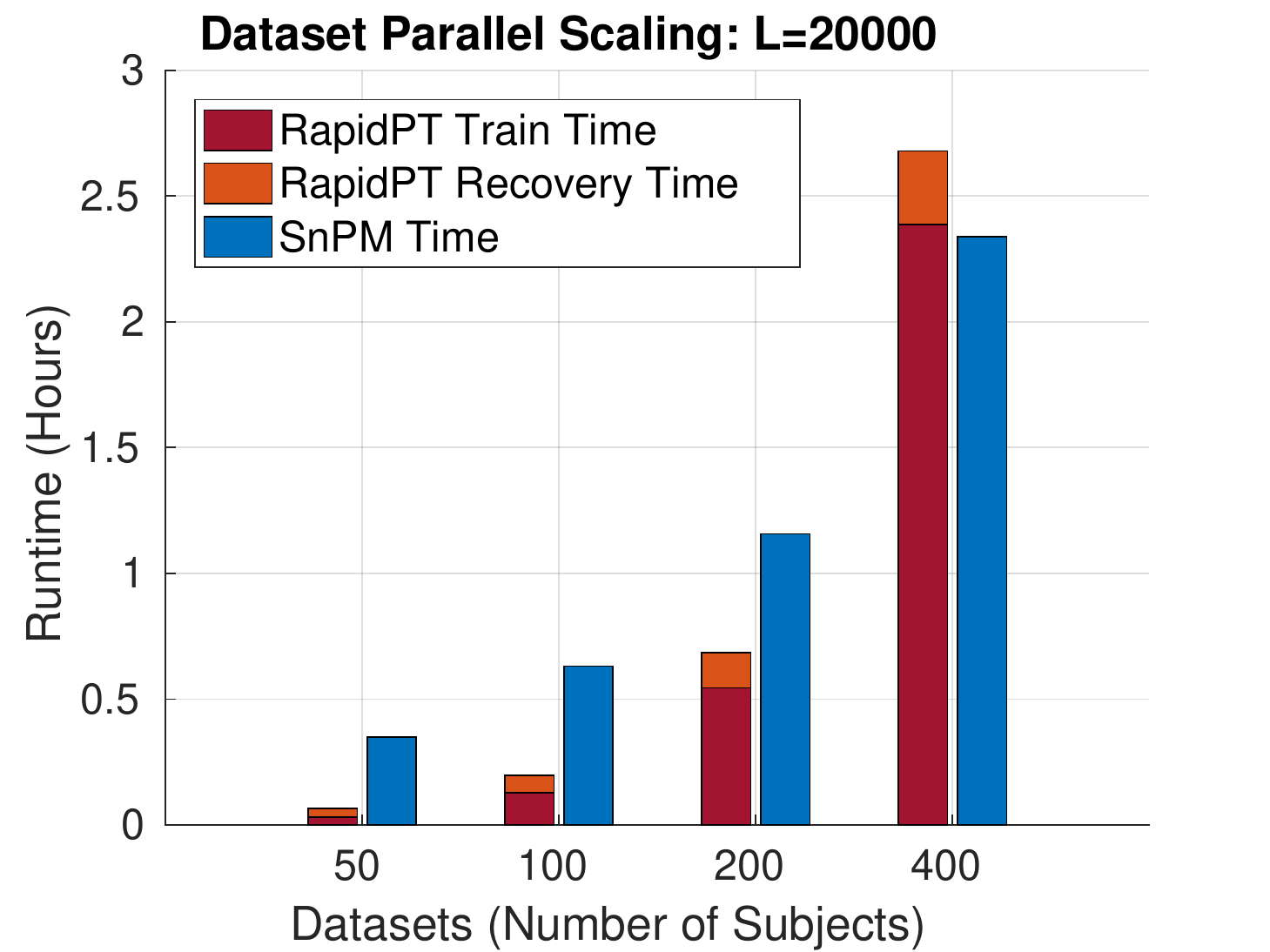}
\includegraphics[width=0.4\textwidth]{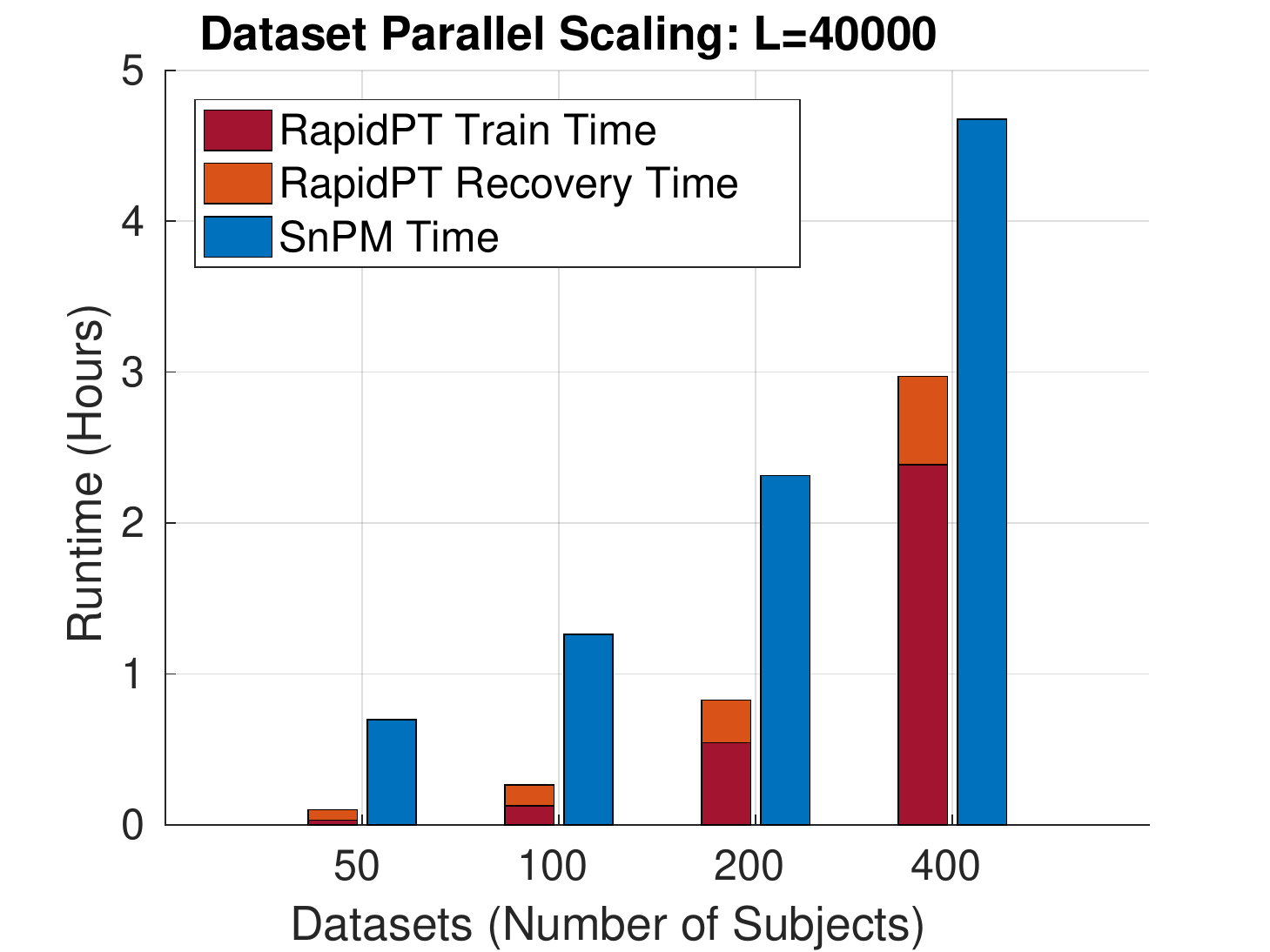}
}%
\centerline{%
\includegraphics[width=0.4\textwidth]{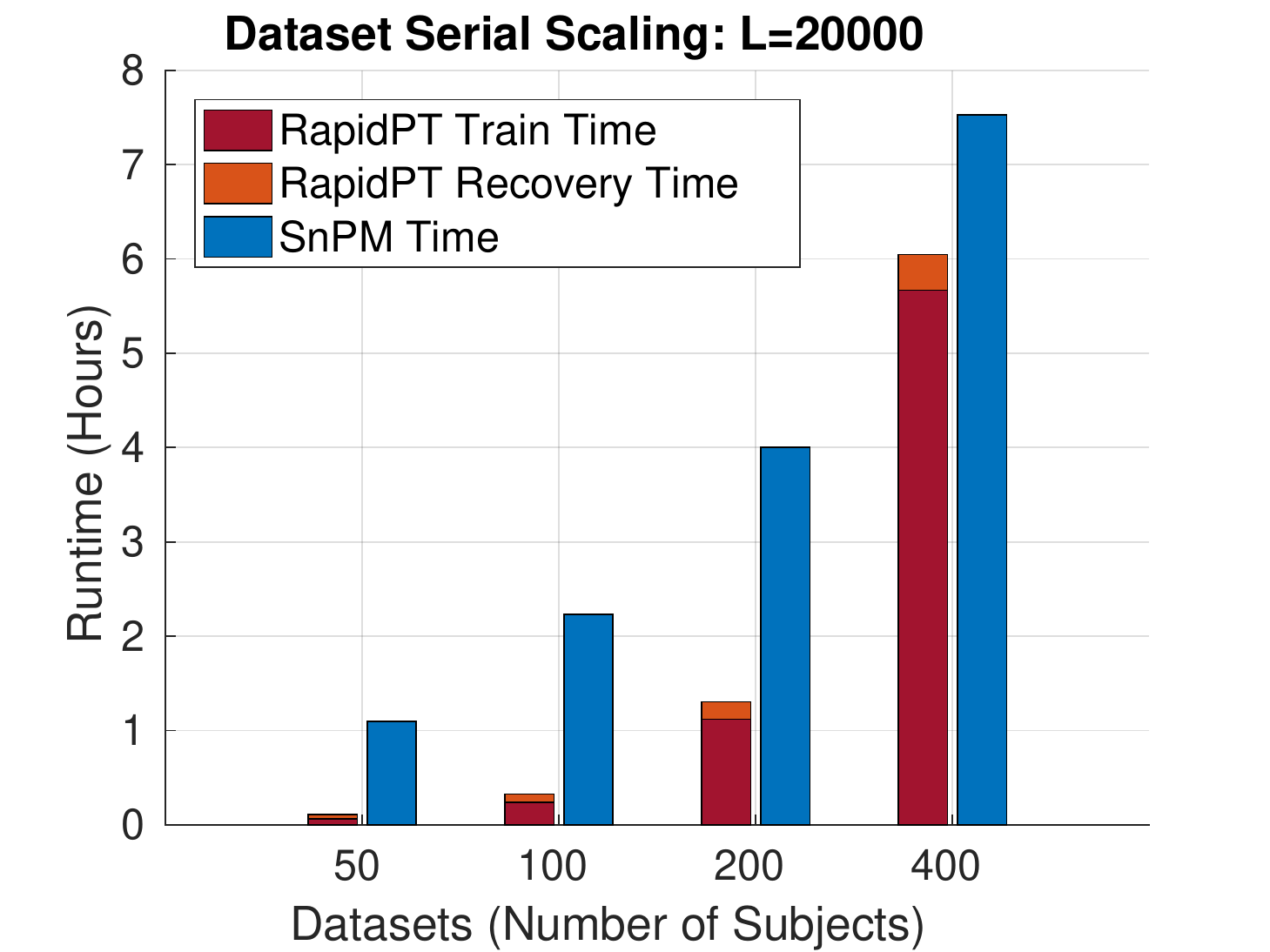}
\includegraphics[width=0.4\textwidth]{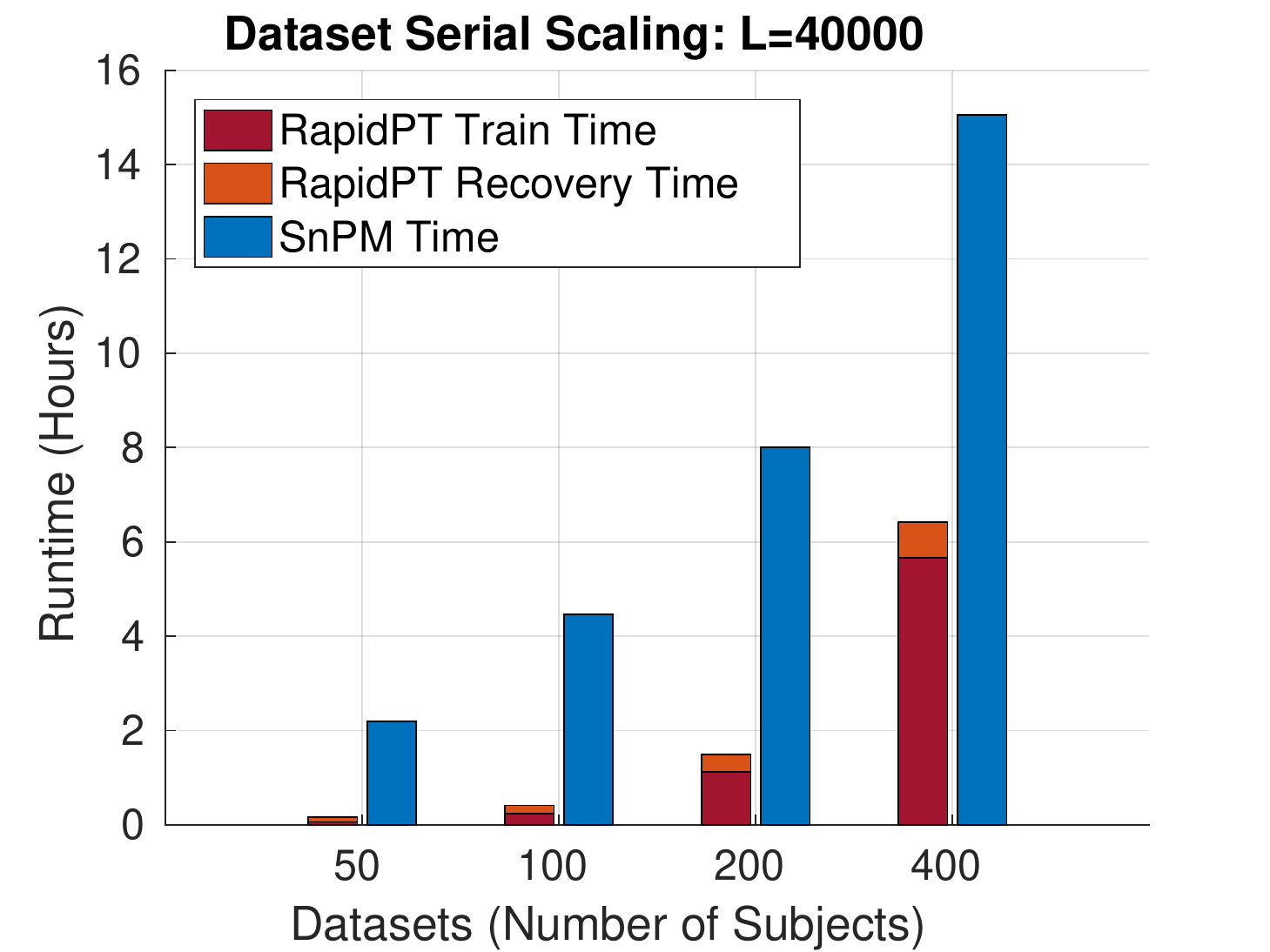}
}%

\caption{Effect of the dataset size on the performance of RapidPT and SnPM on 16 cores (first row) and on a single core (second row). The overall measured time of RapidPT is the result of the total time spent on the training phase and the recovery phase. The hyperparameters used were: $\eta = 0.35\%$, $L = 10000$, and $l = n$. }
\label{fig:DatasetScaling}
\end{figure}

As opposed to $\eta$ and $l$ which only seem to have an impact on the runtime performance of RapidPT, the number of permutations and the size of the dataset 
have an impact on the runtime of both RapidPT and SnPM. In this section, we compare how RapidPT and SnPM scale as we vary these two parameters. 
Figures \ref{fig:PermutationScaling} and \ref{fig:DatasetScaling} show the runtime performance of RapidPT for $\eta = 0.35\%$ and $l = n$.

\textit{Number of Permutations: }Figure \ref{fig:PermutationScaling} shows the super linear scaling of SnPM compared to RapidPT for all datasets as 
the number of permutation increases. Doubling the number of permutations in SnPM leads to an increase in the runtime by a factor of about two. 
On the other hand, doubling the number of permutations for RapidPT only affects the runtime of the recovery phase leading to small increases 
in the timing performance. The performance of both implementations is only comparable if we focus on the lower range of the number of permutations (~5000) across datasets. 
As this number increases, as was shown in Figure \ref{fig:SpeedupsSnPM}, RapidPT outperforms the permutation testing implementation within SnPM.

\textit{Dataset Size: }Figure \ref{fig:DatasetScaling} shows the effect of the dataset size on the runtime of RapidPT and SnPM. In SnPM, as expected, increasing the dataset by a factor of two leads to an increase on the runtime by a factor of two. In RapidPT, however, increasing the dataset size has a variable effect on the runtime. 
The training phase ends up contributing more to the runtime as the dataset size increases, while the recovery phase runtime increases at much slower rate. 

Overall, scaling the number of permutations have a stronger impact on the runtime performance of SnPM than RapidPT. On the other hand, scaling the dataset size 
has a more negative effect on the timing of RapidPT than in SnPM. Furthermore, if both parameters are increased at the same time the runtime of SnPM increases at a much 
faster rate than the runtime of RapidPT. 

\section{Discussion} \label{sec:discussion}

\subsection{Accuracy} \label{sec:accuracy_discussion}

\subsubsection{Recovered Max null Distribution}
Monte carlo permutation tests perform a randomization step where a random subset of the total number of permutations is chosen. 
This means that the constructed max null distribution from one run might slightly differ from another run, and as the number of permutations 
increases, this difference will decrease. In terms of KL-Divergence, this means that the KL-Divergence between two permutation testing runs on the 
same data will be small, but {\em not exactly zero}. The results show that given a good set of hyperparameters the KL-Divergence between 
a run of RapidPT versus a regular permutation testing run leads to a very low KL-Divergence which is the expected result, even if we run the 
{\em same permutation testing program twice}. The evaluated scenarios show that a sensible set of hyperparameters can be easily defined as long 
as the sub-sampling rate is sufficient for the recovery of the permutation testing matrix. The minimum number of sub-sampled entries needed to accurately recover $\T$ depends on the rank and dimensions of $\T$ as discussed in section \ref{sec:matcomp} and in the supplement. For the simulation study with $n=30$, the minimum $\eta$ required was $~1.6\%$ of all entries as shown in Figures \ref{fig:KLDivSimData} and \ref{fig:TThreshSimData}. The experiments on the other 48 synthetic datasets (Figure \ref{fig:KLDivBreadthSimData}) used $2\eta_{min}$ as the sub-sampling rate. In our experiments on the ADNI dataset, $\eta \geq 0.35\%$ led to a large enough set of sub-sampled entries to obtain an accurate estimate of the max null distribution. For a brief discussion on how to pick a sensible $\eta$ and the minimum sub-sampling rate ($\eta_{min}$), please refer to section 3 of the supplementary material. 
Overall, once we have a sensible $\eta$, the resulting max null distribution constructed by RapidPT is consistent to the one recovered by regular permutation testing. 

The number of permutations also has a significant impact on the KL-Divergence. As $L$ increases, the KL-Divergence decreases. However, this is also true between any two permutation testing runs on the same dataset. 

\subsubsection{Evaluating $p$-values}

As seen in the $p$-value spectrum plots, the $p$-values drawn from each max null distribution agree (given a good set of hyperparameters as discussed above). 
This follows from the low KL-Divergence, since the KL-Divergence is a measure of the overall difference between the distributions. The lowest differences, 
however, are located in the tails of the distributions. This means that the derived thresholds are expected to accept/reject almost the same null hypotheses. 

\textit{The extremely low $p$-value regime:} As we will discuss in section \ref{sec:accuracy_performance} the largest speedups from RapidPT are obtained as the number of permutations increases. The main reason to increase $L$ is to obtain smaller $p$-values. Figures \ref{fig:TThreshSimData} and \ref{fig:PVals} show that RapidPT recovers extremely accurate t-threshold (percent differences $ < 0.1\%$). Figures \ref{fig:pmaps400} and \ref{fig:pmaps200} demonstrate this in a practical scenario where a large number of permutation ($L = 100,000$) and a high percentage of the brain regions that rejected the null had a $p$-value $< 0.001$. Therefore, a user interested in using small $p$-values can benefit from large computational speedups when using RapidPT, with a negligible loss in accuracy.

\subsubsection{Resampling Risk}

\textit{Small signal datasets: }Although the $p$-values agree across methods, a small difference may lead to slightly more or less null hypotheses to 
be rejected by a given method. This has a direct impact on the resampling risk between RapidPT and regular permutation testing. In datasets where there is a 
small signal difference between groups, we may see an elevated resampling risk. The reason is because a very small number of null hypotheses will be rejected. 
Therefore, at a given $p$-value, $p$, one of the methods will reject a small number of null hypotheses which the second method will not reject until the $p$-value is 
$p + \delta$, where $\delta << 1$. This slight difference in the number of rejected null hypotheses may have a significant impact in the resampling risk. 
For instance in Figure \ref{fig:ResamplingRisk} ($N = 200$ at $p = 0.05$), RapidPT rejected $59$ (out of $\sim 568$k statistics) null hypotheses and SnPM rejected $71$: 
this led to a resampling risk of $8.45\%$. On the other hand, for $N = 400$ at $p = 0.05$ RapidPT rejected $2158$ (out of $\sim 570$k statistics) and SnPM rejected $2241$ null hypotheses, 
resulting in a lower resampling risk of $1.85\%$ even when there is a larger difference in the number of rejected hypotheses. Nonetheless, as we can see in the brain maps in Figures \ref{fig:pmaps400} and \ref{fig:pmaps200} that this difference of RapidPT's and SnPM's rejections were among the boundary voxels of the rejection region (i.e., the mismatch is not at the center of the rejection/significant region). Note that once a reasonable small smoothing filter is applied to nullify noise (e.g., stand-alone voxels) this apparent small difference will vanish visually. A similar situation is encountered in the $50$ and $100$ subject dataset (see supplement), where the number of null hypotheses rejected was extremely low ($< 10$) and single mismatch led to an elevated resampling risk. Therefore, the resampling risk is a useful measure if it is presented together with the number of null hypothesis being rejected. Hence, RapidPT yields a slightly more 
conservative test, primarily because it is based on sub-sampling. Nevertheless, this sub-sampling is robust to locating the same voxel clusters that displayed group differences as the other methods.


\subsection{Runtime Performance} \label{sec:accuracy_performance}

Our results show that under certain scenarios RapidPT provides substantial speedups over SnPM (state of the art) and NaivePT (simple implementation). 
From the runtime performance of NaivePT, it is evident that in practice it is more beneficial for the user to rely on a permutation testing toolbox such as SnPM. 
In the remainder of the discussion we will just consider the runtime performance of RapidPT and SnPM.

\textit{Dataset Size: }Overall, the largest speedups in both the serial and parallel setups were obtained in the runs for the smallest dataset ($N = 50$). 
The number of training samples used to estimate the basis $\U$ is smaller and consequently the training phase time decreases. 
As the dataset size increases, the training time introduces a considerable overhead which negatively impacts the speedup of RapidPT over SnPM when less than 20000 permutations are being performed.

\textit{Number of Permutations: }The number of permutations have a linear impact on the runtime of both RapidPT and SnPM. But as shown in Figure \ref{fig:PermutationScaling}, 
RapidPT runtime increases at a lower rate than SnPM's. This is expected since the number of permutations only impacts the runtime of the recovery phase. 
Therefore, the training phase time for a given setup is constant as the number of permutations changes. The results show that for datasets with less than or equal to 
$400$ subjects between $5000-10000$ permutations is the threshold where despite the training overhead in RapidPT, the expected speedup is considerable to justify 
its use. In the larger datasets, when performing less than $5000$ permutations, the training overhead becomes too large, as shown in the supplementary material.

\subsubsection{Serial vs. Parallel Performance}

The serial and parallel runs show very similar speedup trends across hyperparameters as shown in the results and supplementary material.
However, in terms of actual runtime, both RapidPT and SnPM benefit from being able to run on a parallel environment. This is an essential feature 
for any software toolbox that will be running on modern workstations because multiple cores are available in nearly all computers shipped today.

\subsection{Hyperparameter Recommendations} \label{sec:hyperreco}

\textit{Sub-sampling Rate: }The KL-Divergence results, Figures \ref{fig:KLDivSimData} and \ref{fig:KLDivSnPM}, show that as long as the sub-sampling rate is greater than or equal to a certain threshold, 
  RapidPT is able to accurately recover the max null distribution. For the simulation study this threshold was $~1.6\%$ and for the real data experiments $0.35\%$ was a high enough sub-sampling rate. A minimum $\eta_{min}$ can be calculated using LRMC theory as shown in Section 3 of the supplementary material. This minimum value is simply a function of the number of voxels and the number of data instances. The simulation results in Figures \ref{fig:KLDivSimData} and \ref{fig:TThreshSimData} explicitly show this $\eta_{min}$. The toolbox itself sets it to a default conservative value of $2*\eta_{min}$, which is also shown in Figures \ref{fig:KLDivSimData} and \ref{fig:TThreshSimData}.
  This slightly larger choice ensures the error in recovering the null is almost negligible.
  Overall, the accuracy of RapidPT will not significantly improve as the sub-sampling rate increases (pass $\eta_{min}$). 
  On the other hand, a low sub-sampling rate can significantly reduce the runtime, in particular in the large $L$ regime, and is therefore preferable.
  A user does not need to change this rate in practice (beyond what is given by the toolbox).
  However, if he/she is willing to be even more flexible by sacrificing some accuracy to achieve an even higher speed-up, it can be reduced appropriately.

\textit{Number of Training Samples: } The number of training samples will ideally be the exact rank of $\T$. This is usually not known, however, we know that 
the rank is bounded by the number of subjects in the data matrix used to generate $\T$. Therefore, in our evaluations, we are able to accurately recover 
the max null distribution even when using as low as $\frac{n}{2}$ training samples. The recommended number of training samples in practice is be $n$ (which follows from rank structure of the testing matrix) and is the default setting within RapidPT toolbox.

\textit{Number of Permutations: }When a large number of permutations is desired, RapidPT should be strongly considered due to its runtime gains. Not only RapidPT provides considerable runtime gains in the large $L$ regime, but also its accuracy will improve as the KL-Divergence results show. 
Note, however, that the user should also take into account the dataset size and look at the speedup colormaps to see the expected speedups.  

\textit{Dataset size: }Although not a hyperparameter, as discussed above, the dataset size should play a role when the user selects which method to use. 
Large datasets ($n \geq 200$), RapidPT will be a good option if the user is planning to perform a large number of permutations. 
For medium-sized datasets ($50 \leq n \leq 200$), RapidPT will most likely lead to good speedup gains. For small datasets ($n \leq 50$), although RapidPT might 
lead to speedup gains over regular permutation testing, the total runtime will be on the order of minutes anyway. 


  As we briefly discussed in Section \ref{sec:intro}, \cite{Winkler16} provides several strategies for reducing the runtime of permutation testing. 
  But the authors do not report significant speedup gains against regular permutation testing on a $50$ subject dataset 
  with $\approx 200000$ voxels. 
  On the other hand, RapidPT is able to consistently outperform a state of the art permutation testing implementation (SnPM) on a 
  $50$ subject dataset with $\approx 540$k voxels. 
  This boost in performance is, in large part, due to our low sub-sampling rates. 
  Our subspace tracking algorithm is, nonetheless, able to perform recovery in this sparse sampling setting. 

\subsection{SnPM Integration}

{SnPM is a toolbox that can be used within the software Statistical Parametric Mapping (SPM) \cite{SPM}. RapidPT has been integrated into the development version of SnPM \cite{SnPM}. This enables users to leverage the pre and post processing capabilities of SnPM. Through the graphical user interface (GUI) of SnPM the user can simply specify if they want to use RapidPT or not. Alternatively, the experienced user can also toggle a flag called \textit{RapidPT} inside the \textit{snpm\_defaults.m}. Once this flag is set the user can simply proceed with their normal SnPM workflow. The SnPM GUI does not allow the user to set RapidPT's hyperparameters ($\eta$, $l$), however, the online documentation walks the user through the process of setting them manually. A preview of the online documentation can be seen in Figure \ref{fig:SnPMPatch}. Further discussion and walkthroughs of how to use SnPM and RapidPT within SnPM can be found in the documentation of both toolboxes \cite{RapidPT,SnPM}.
The RapidPT library webpage is at \href{http://felipegb94.github.io/RapidPT/}{http://felipegb94.github.io/RapidPT/}. 

\begin{figure}[H]
\centerline{%
	\includegraphics[width=0.4\textwidth]{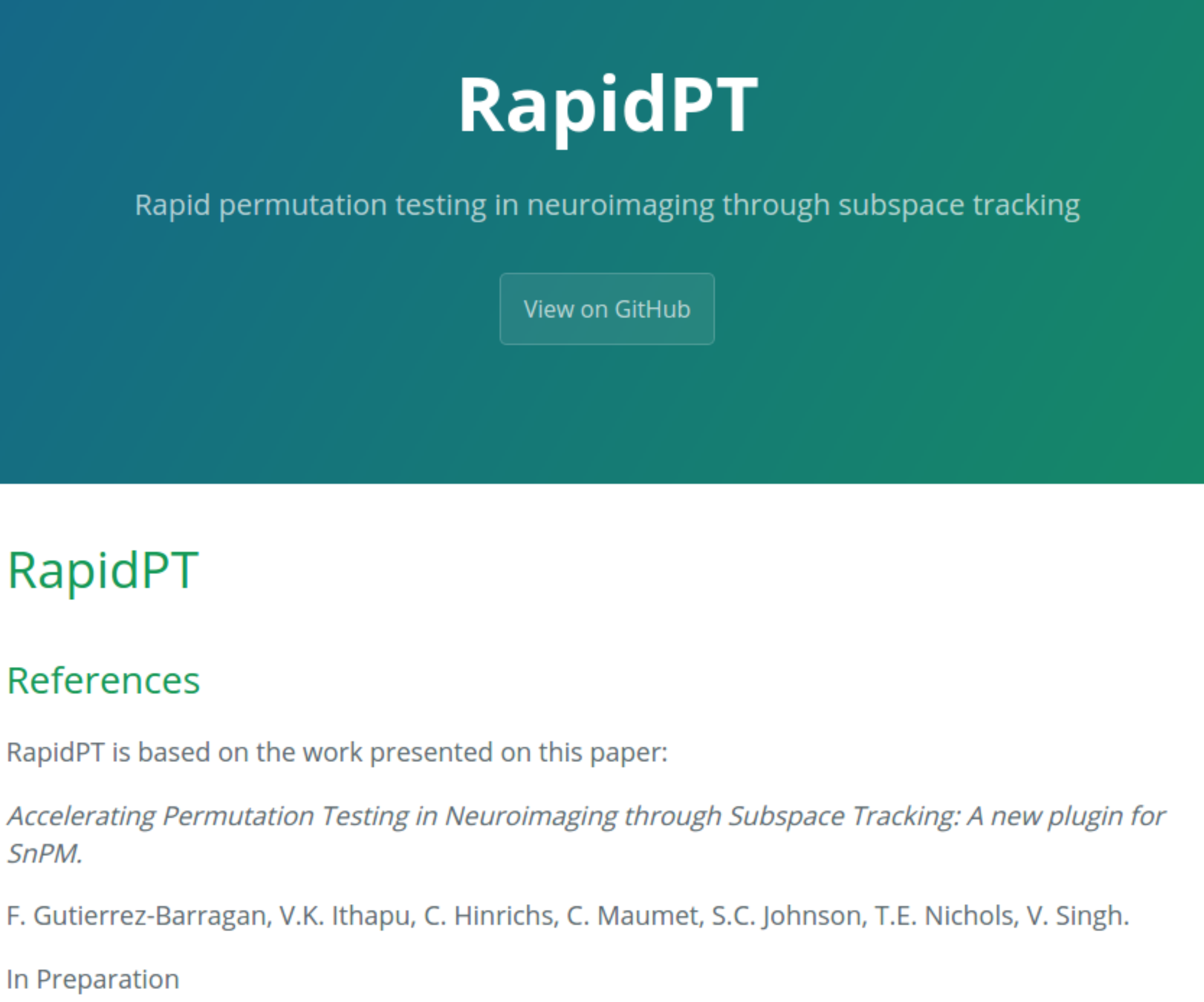}
	\includegraphics[width=0.4\textwidth]{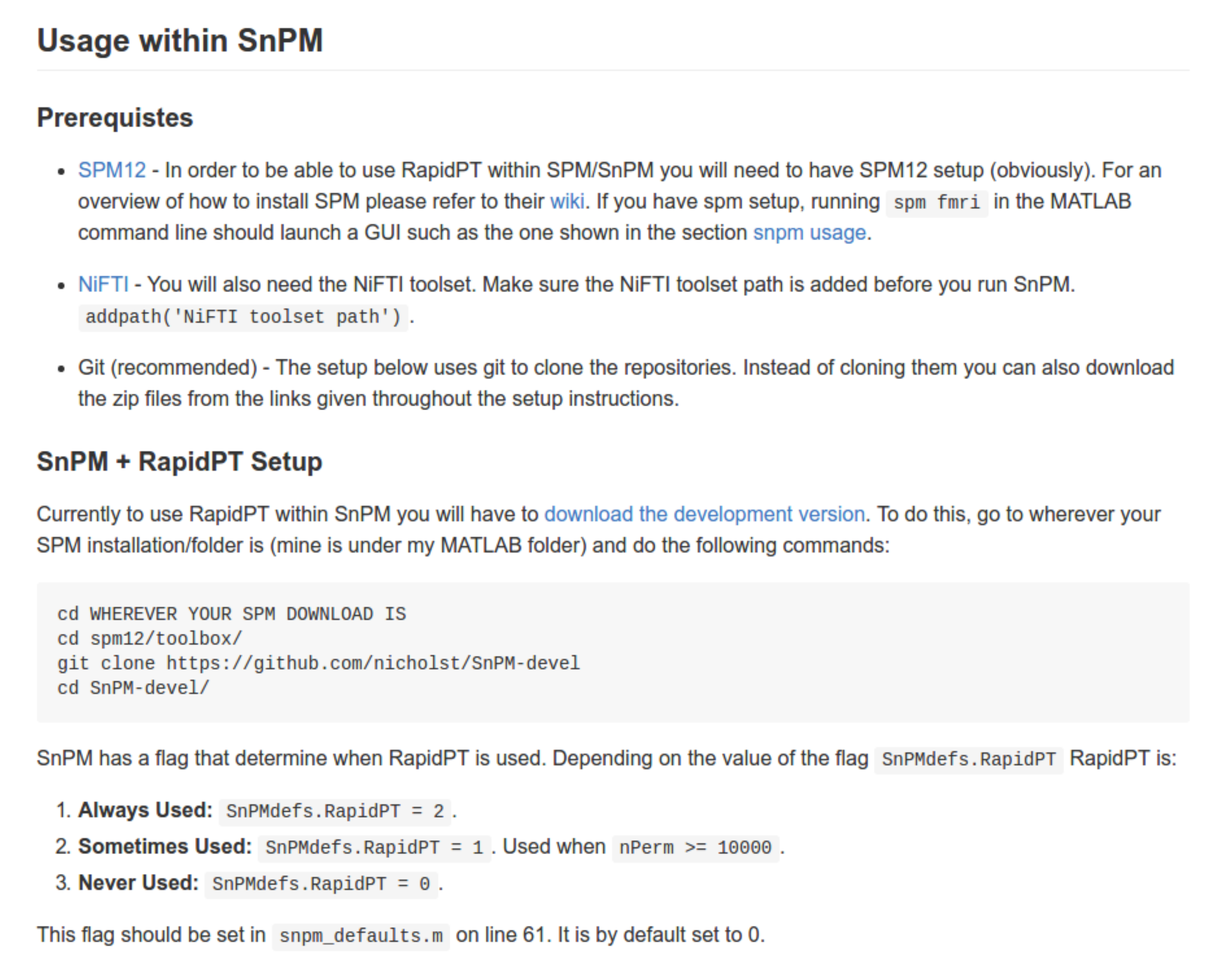}%
}%
\caption{Screenshots of the RapidPT website (left) and SnPM integration documentation (right).}
\label{fig:SnPMPatch}
\end{figure}

\section{Conclusion} \label{sec:conclusion}

In this paper, we have presented a new algorithmic framework that is able to efficiently approximate the max null distribution commonly obtained through permutation testing. By exploiting the structure of the permutation testing matrix, $\T$, and applying recent ideas from online matrix completion we show through our theoretical analysis and experimental evaluations that one can subsample entries of $\T$ at extremely low-rates and still construct a good estimate of $\T$. The algorithm first goes through a training phase where the basis and the distribution of the residual of $\T$ are estimated. Then it continues into the recovery phase where a small number of entries of each column of $\T$ are calculated and the rest are estimated through matrix completion. Finally, we obtain the max null distribution from the maximum value of each column in $\T$. Experiments on four varying sized datasets derived from the ADNI2 dataset showed that if we sub-sample at a high enough rate we can accurately recover the max null distribution at a fraction of the time in many scenarios. The implementation is available as a stand-alone open-source toolbox as well as a plugin for SnPM13 \cite{SnPM}, and is able to leverage multi-core architectures.

\paragraph{\bf Acknowledgements} The authors thank
Jia Xu for helping out with a preliminary implementation of the model.
This work was supported in part by NIH R01 AG040396; NIH R01 EB022883; NIH R01 AG021155; NSF CAREER grant 1252725; NSF RI 1116584; 
UW ADRC P50 AG033514 and UW ICTR 1UL1RR025011. 
Hinrichs was supported by a CIBM post-doctoral fellowship at Wisconsin via NLM grant 2T15LM007359. Nichols and Maumet are supported by the  Wellcome Trust (100309/Z/12/Z). The contents do not represent views of the Department of Veterans Affairs or the United States Government.

\section{Supplementary Material}\label{sec:supplement}
\href{http://pages.cs.wisc.edu/~felipe/assets/pdf/rpt-nimg2017/SupplMaterial.pdf}{http://pages.cs.wisc.edu/~felipe/assets/pdf/rpt-nimg2017/SupplMaterial.pdf}

\section*{References}
\bibliographystyle{elsarticle-harv}
\bibliography{permtest}

\clearpage

\end{document}